\documentclass[aps,prd,showpacs,amsfonts,notitlepage,amssymb,amsmath,nofootinbib,superscriptaddress]{revtex4-1}
\usepackage[utf8]{inputenc}
\usepackage[english]{babel}
\usepackage[titletoc,toc,title]{appendix}

\newcommand{\eq}[1]{Eq.~(\ref{#1})}
\newcommand{\fig}[1]{fig.~\ref{#1}}

\newcommand{\bsub}{\begin{subequations}}
\newcommand{\esub}{\end{subequations}}
 
\newcommand{\be}{\begin{eqnarray}}
\newcommand{\ee}{\end{eqnarray}}
\newcommand{\om}{\ensuremath{\omega}}

\newcommand{\pd}{\ensuremath{\partial}}

\newcommand{\lp}{\ensuremath{\left(}}
\newcommand{\rp}{\ensuremath{\right)}}
\newcommand{\abs}[1]{\ensuremath{\left \lvert #1 \right\rvert}} 
\newcommand{\bi} {\begin{itemize}}
\newcommand{\ei} {\end{itemize}}
\newcommand{\ben}{\begin{enumerate}}
\newcommand{\een}{\end{enumerate}}
\newcommand{\bmat}{\begin{pmatrix}}
\newcommand{\emat}{\end{pmatrix}}
\newcommand{\symba}{\phi_\om^{\rightarrow,d}}
\newcommand{\symbAt}{\phi_\om^{\leftarrow}}
\newcommand{\symbA}{\phi_\om^{\rightarrow}}
\newcommand{\symbb}{\lp \phi_{-\om}^{\rightarrow,d} \rp^*} 
\newcommand{\symbbno}{\phi_{-\om}^{\rightarrow,d}}

\newcommand{\symbain}{\phi_\om^{\rightarrow,d,{\rm in}}}
\newcommand{\symbAtin}{\phi_\om^{\leftarrow,{\rm in}}}
\newcommand{\symbAin}{\phi_\om^{\rightarrow,{\rm in}}}
\newcommand{\symbbin}{\lp \phi_{-\om}^{\rightarrow,d,{\rm in}} \rp^*} 
\newcommand{\symbaout}{\phi_\om^{\rightarrow,d,{\rm out}}}
\newcommand{\symbAtout}{\phi_\om^{\leftarrow,{\rm out}}}
\newcommand{\symbAout}{\phi_\om^{\rightarrow,{\rm out}}}
\newcommand{\symbbout}{\lp \phi_{-\om}^{\rightarrow,d,{\rm out}} \rp^*} 
\newcommand{\ommax}{\om_{\rm max}}

\usepackage[colorlinks=true,linkcolor=blue,citecolor=blue,urlcolor=blue]{hyperref}
\usepackage{graphicx}
\usepackage{amsmath}
\usepackage{slashed}     
\usepackage{xcolor}
\usepackage{amssymb}
\usepackage{mathrsfs}
\usepackage{multido}

\begin{document}

\author{Scott Robertson}
\affiliation{Institut Pprime, UPR 3346, CNRS-Universit\'{e} de Poitiers-ISAE ENSMA, 11 Boulevard Marie et Pierre Curie-T\'{e}l\'{e}port 2, BP 30179, 86962 Futuroscope Cedex, France}

\author{Florent Michel}
\affiliation{Laboratoire de Physique Th\'{e}orique, CNRS, Univ. Paris-Sud, Universit\'{e} Paris-Saclay, 91405 Orsay, France}

\author{Renaud Parentani}
\affiliation{Laboratoire de Physique Th\'{e}orique, CNRS, Univ. Paris-Sud, Universit\'{e} Paris-Saclay, 91405 Orsay, France}

\title{Scattering of gravity waves in subcritical flows over an obstacle} 
\date{\today}

\begin{abstract}
We numerically study the scattering coefficients of linear water waves on stationary flows above a localized obstacle. We compare the scattering on trans- and subcritical flows, and then focus on the latter which have been used in recent analog gravity experiments. The main difference concerns the magnitude of the mode amplification: whereas transcritical flows display a large amplification (which is generally in good agreement with the Hawking prediction), this effect is heavily suppressed in subcritical flows. This is due to the transmission across the obstacle for frequencies less than some critical value. As a result, subcritical flows display high- and low-frequency behaviors separated by a narrow band around the critical frequency. In the low-frequency regime, transmission of long wavelengths is accompanied by non-adiabatic scattering into short wavelengths, whose spectrum is approximately linear in frequency. By contrast, in the high-frequency regime, no simple description seems to exist. In particular, for obstacles similar to those recently used, we observe that the upstream slope still affects the scattering on the downstream side because of some residual transmission. 
\end{abstract}

\maketitle

\section{Introduction}

In 1981, Unruh pointed out that one might conceive of experiments where the analog version of black hole radiation could be observed in a moving medium~\cite{Unruh:1980cg}. Indeed, when the flow is stationary and transcritical, i.e., when the flow speed crosses the wave velocity, the propagation of linear density perturbations is governed by an equation which has the form of a d'Alembertian in a black hole geometry. As a result, the scattering coefficients should be identical to those encoding the Hawking effect. However, it was then realized that this ceases to be exact when taking into account the dispersive effects that occur at short wavelengths in condensed matter media~\cite{Jacobson:1991gr,Unruh:1994je}. 

As a result, to be able to predict what should be seen in experiments, one should compute the scattering coefficients taking into account the specific dispersive effects characterizing the medium. It was first understood that the spectrum is robust~\cite{Unruh:1994je,Brout:1995wp,Corley:1996ar,Corley:1997pr,Unruh:2004zk,Balbinot:2006ua},  i.e., that the spectral deviations from the standard thermal distribution are suppressed by positive powers of $\kappa/\Lambda$, where $\kappa$ and $\Lambda$ are respectively the analog version of the surface gravity and the dispersive scale above which dispersion effects are significant. Hence when $\kappa/\Lambda\ll 1$, the emitted spectrum closely follows a Planck distribution with a temperature given by $\kappa/2\pi$ in units $\hbar = c = k_B = 1$. When $\kappa/\Lambda$ is not negligible, the spectrum is no longer Planckian~\cite{Macher:2009tw,Macher:2009nz}, yet its main properties can be understood in terms of two parameters: $\kappa$ and a critical high frequency $\ommax$, which is linearly related to $\Lambda$ but which also depends on the parameters of the background flow~\cite{Finazzi:2012iu}. In particular, there is a smooth transition from the standard relativistic regime $\kappa/\ommax \ll 1$, to a dispersive regime $\kappa/\ommax \gg 1$ where $\kappa$ no longer plays any role.  It is fair to say that the scattering coefficients in transcritical flows are now well understood, see~\cite{Robertson:2012ku,Coutant:2011in} for reviews.

When considering the experiments based on surface waves in water tanks~\cite{Rousseaux:2007is,Rousseaux:2010md,Weinfurtner:2010nu,Euve:2014aga,Euve:2015vml}, one encounters two novel effects. Firstly, the background flows investigated up to now have been subcritical rather than transcritical. Since there is no analog Killing horizon in such flows, the link with the Hawking effect is {\it a priori} unclear. In fact, the spectral properties are not well understood, and have so far received much less attention than their counterparts in transcritical flows. Preliminary studies indicate that several regimes are found, and that various parameters are relevant in each regime~\cite{Michel:2014zsa,Michel:2015aga,Coutant16}. Secondly, downstream from the obstacle, the free surface is modulated by a zero-frequency undulation with a macroscopic amplitude and a long extension~\cite{Lawrence87,Coutant:2012mf}. Typically, the undulation is longer than a meter and its amplitude is of the order of $1$ cm, larger that the typical amplitude of the waves sent by the wavemaker which is of the order of a few mm. The extra scattering on such an extended modulation is poorly understood. Numerical simulations indicate that it might play a significant role in experiments~\cite{Euve:2015vml}, see also~\cite{Busch:2014hla} for a study in the context of atomic Bose-Einstein condensates. 

We shall study these two aspects in turn. In this first paper, we focus on the scattering coefficients in subcritical flows with no undulation downstream from the obstacle. Our principal aim is to characterize the main properties of these coefficients, and to show how they depend on the background flow parameters.  We hope our predictions can be tested in forthcoming experiments. In a future paper we shall study the scattering on the undulation itself. 

The present work is organized as follows. In Sec.~\ref{Sts} we present the simplified wave equation for linear perturbations and the particular parametrization of background flows over an obstacle used in our analysis. Then, we identify the four modes involved in the scattering, and compare the behaviors of the 16 scattering coefficients in a typical transcritical and a subcritical flow. We end the section by studying the evolution of the scattering coefficients when gradually replacing a transcritical flow by a subcritical one. In Sec.~\ref{sec:systematic_analysis} we focus on sub- and near-critical flows. We show that the scattering on such flows should be analyzed separately in three different frequency regimes, in each of which we identify the relevant flow parameters. We conclude in Sec.~\ref{Conc}. In Appendix~\ref{3reg} we show how the three regimes appear when studying the effective temperature as a function of the upstream and downstream slopes of the flow, and in Appendix~\ref{app:slope_asymmetry} we examine more closely the respective roles played by these two slopes when the flow is asymmetrical. 

\section{Scattering in trans- and subcritical flows} 
\label{Sts}

\subsection{The simplified wave equation}

We shall study linear surface waves propagating in inhomogeneous flows of an ideal, inviscid, incompressible fluid. Following~\cite{Schutzhold:2002rf,Unruh:2013gga,Michel:2014zsa} the flows are assumed to be stationary, irrotational, and laminar. We assume they take place in an elongated flume and neglect any dependence on the directions orthogonal to the mean velocity. In addition, we neglect capillary effects, which means that the wavelengths we consider are significantly larger that the typical capillary length ($\sim$ a few mm for water). Finally, we assume that the inhomogeneity of the flow is due to an obstacle put on the bottom of the flume.

Under these assumptions and considering waves which are homogeneous in the transverse direction, the dispersion relation between the (conserved) angular frequency $\omega$ and the wave number $k$ in the longitudinal direction is 
\be \label{eq:disprel}
\Omega^2 \equiv (\omega - vk)^2 = g k \tanh(hk), 
\ee
where $v$ is the horizontal flow velocity, $h$ the water depth, and $g$ the gravitational acceleration. In inhomogeneous flows, $v$, $h$, $k$, and $\Omega$ depend on $x$, the position in the longitudinal direction. The quantity $\Omega = \omega - vk$ gives the frequency in the frame co-moving with the fluid. Although it is not constant, its sign plays a crucial role in the analysis of the scattering.  

Despite the simplicity of \eq{eq:disprel}, the linear equation governing the propagation of waves is rather complicated. The explicit expression can be found in~\cite{Unruh:2013gga,Coutant:2012mf}. In particular, because of the term in $\tanh(hk)$, it contains operators with arbitrarily high orders of $\partial_x$. To simplify the numerical resolution, as in~\cite{Michel:2014zsa,Euve:2014aga,Michel:2015aga}, we consider a quartic truncation of this equation keeping the ordering of $v(x)$, $h(x)$ and $\pd_x$. Namely, we work with 
\be \label{eq:quartic}
\left[ \lp \pd_t + \pd_x v(x) \rp \lp \pd_t + v(x) \pd_x \rp - g \lp \pd_x h(x) \pd_x + \frac{1}{3} \pd_x \lp h(x) \pd_x \rp^3 \rp \right] \phi = 0, 
\ee
where $\phi$ is the perturbation of the velocity potential. It is related to the linear variation of the water depth $\delta h$ through
\be \label{eq:dh}
\delta h(t,x) = -\frac{1}{g}\left(\partial_t+ v \partial_x\right)\phi.
\ee
The truncated dispersion relation associated with \eq{eq:quartic} is 
\be \label{eq:disprelq} 
(\om - vk)^2 = g h \, k^2 \lp 1 - \frac{(h k)^2}{3} \rp.
\ee
In the hydrodynamical limit $h k \ll 1$, the (local value of the) speed of propagation of shallow waves becomes $c(x) = \sqrt{g h(x)}$. 

When the Froude number $F = v/c$ is close to 1, \eq{eq:quartic} becomes equivalent to the full wave equation in the range of frequencies we are interested in.  It is thus sufficient to characterize the main properties of the scattering for near-critical flows. We refer to~\cite{Coutant:2011in} for an analytical calculation of the scattering coefficients based on \eq{eq:quartic} when the flow is transcritical. In these flows, the link with the Hawking effect, and the first deviations due to dispersion, are both clear. For the low frequency behavior in subcritical flows, we refer to~\cite{Coutant16} which appeared while we were finishing the present work.

\eq{eq:quartic} has a conserved scalar product with the same structure as that of the complete equation. It is given by 
\be 
\lp \phi_1 \vert \phi_2 \rp \equiv i \int \lp \phi_1^*(t,x) \lp \pd_t + v(x) \pd_x \rp \phi_2(t,x) - \phi_2(t,x) \lp \pd_t + v(x) \pd_x \rp \phi_1^*(t,x) \rp dx, 
\label{eq:norm}
\ee
where $\phi_1$ and $\phi_2$ are two complex solutions. We refer to~\cite{Coutant:2012mf} for the relation between \eq{eq:norm} and the wave energy, and for the fact that the norm $\lp \phi_1 \vert \phi_1 \rp$ is not positive definite.
In fact, the sign of the norm is that of $\Omega$, the frequency in the co-moving frame (see Eq. (\ref{eq:disprel})).~\footnote{\label{ftnWA} The conservation of the norm should not be confused with that of the wave action~\cite{Bretherton1968}, although these notions are closely related. While the former is exact, the conservation of the wave action is an approximate (adiabatic) law which only applies to flows with low temporal and spatial gradients. The link is clear when restricting attention to stationary inhomogeneous flows. In this case, the validity of the WKB approximation of \eq{eq:quartic}, see~\cite{Coutant:2012mf}, guarantees that the wave action is constant. Considering a stationary mode $\varphi_\om = e^{- i \om t} \phi_\om(x)$ solution of \eq{eq:quartic}, the wave action is given by $W = |\Omega v_{g}| |\phi_\om|^2 = g^2 |v_{g}| |\delta h_\om|^2/|\Omega|$ where $\Omega = \om - v k_\om$, $v_{g} = d\om/dk=d\Omega/dk+v$, and $g \,\delta h_\om = i \Omega \phi_\om$, see \eq{eq:dh}. The scattering coefficients we shall compute encode non-adiabatic effects~\cite{Massar:1997en,Coutant16}, i.e., violations of the conservation of the wave action.} 

\subsection{The parametrization of inhomogeneous flows}

\begin{figure} 
\begin{center}
\includegraphics[width=0.44 \linewidth]{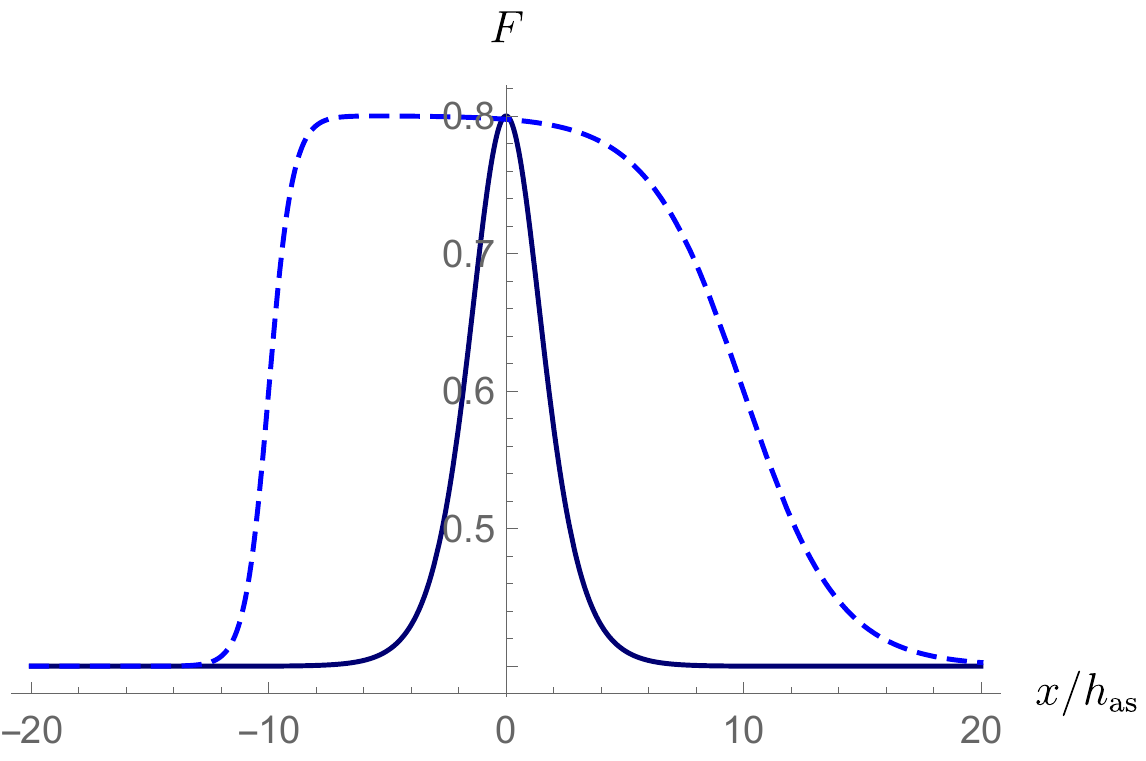} 
\includegraphics[width=0.44 \linewidth]{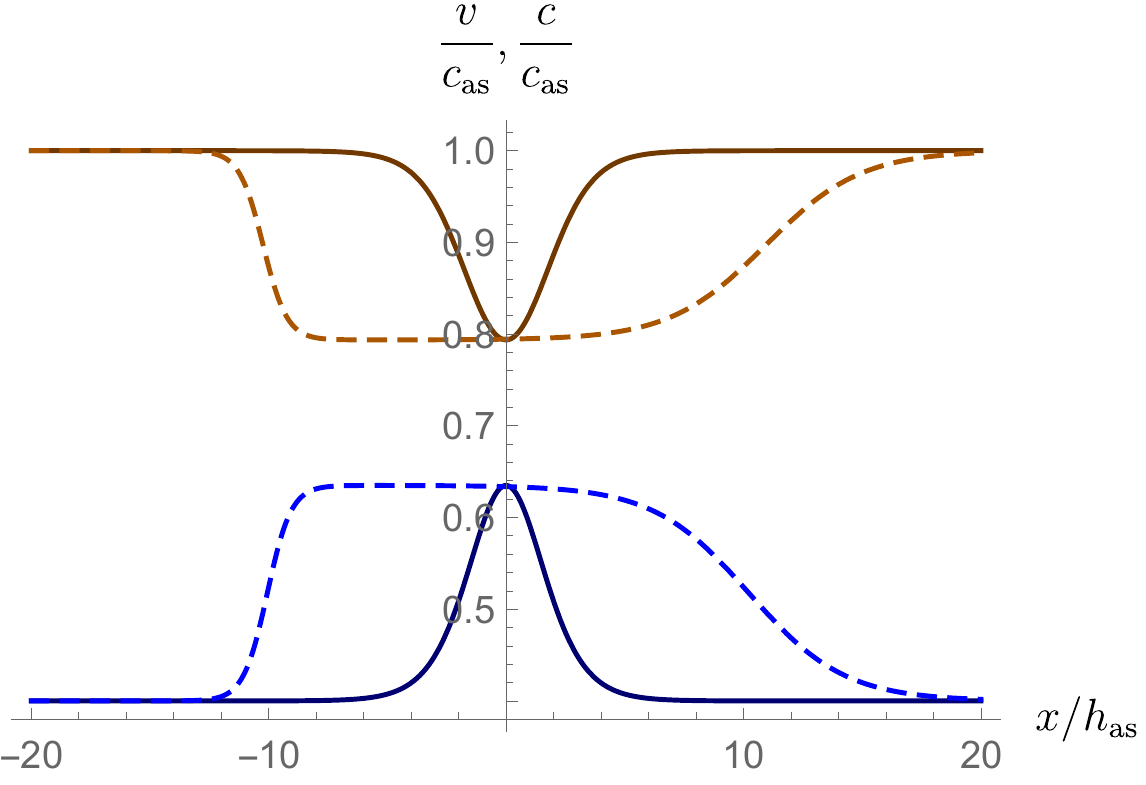} 
\end{center}
\caption{We show the profiles of $F(x)$, $v(x)$, and $c(x)$ for two subcritical flows with the same values of $F_{\rm as} = 0.4$ and $F_{\rm max}= 0.8$, see \eq{eq:f}. The continuous curves describe a narrow symmetrical obstacle with $a_L h_{\rm as} = a_R h_{\rm as} = 0.5$ and $L/h_{\rm as} = 1$, whereas the dotted lines show a long asymmetric obstacle with $a_L  h_{\rm as} = 4 a_R h_{\rm as} = 1$ and $L/h_{\rm as} = 20$. 
The horizontal axis gives the distance from the center of the narrow obstacle, in the adimensional unit $x/h_{\rm as}$. On the right plot, $c$ is in orange and $v$ in blue in units of $c_{\rm as} = \sqrt{g h_{\rm as}}$, the asymptotic value of $c$. For the narrow obstacle, the effective values of \eq{eq:eff} are $L_{\rm eff}/h_{\rm as} \approx  2.8 $, $\sigma_{R} h_{\rm as} = \sigma_{L} h_{\rm as} \approx 0.38$. For the long obstacle, as expected, one gets $L_{\rm eff} \approx L$, $\sigma_{R}\approx a_R / 2$, and $\sigma_{L} \approx a_L / 2$.
} \label{fig:flow} 
\end{figure}

Assuming the flow is homogeneous in the vertical direction, the local low-frequency wave speed and background flow velocity are respectively given by $c(x) = \sqrt{g h(x)}$ and $v(x) = J / h(x)$, where $J$ is the conserved water current. The local value of the Froude number is thus 
\be 
F(x) = \frac{J}{g^{1/2}h(x)^{3/2}}.
\ee
In this paper, we work with $J > 0$, that is, the flow goes from left to right. We phenomenologically describe the properties of the flow on top of a localized obstacle using the following parametrization of $F(x)$:~\footnote{An alternative approach would be to consider background flows that are solutions of the hydrodynamical equations 
over known obstacles.  This approach has been presented in Appendix A of~\cite{Michel:2014zsa}. We verified that the behavior of the scattering coefficients is similar to that presented here.} 
\be \label{eq:F}
F(x) = F_{\rm as} + \lp F_{\rm max} - F_{\rm as} \rp f(x),
\ee
where 
\be \label{eq:f}
f(x) = \mathcal{N} \left[ 1 - \tanh \lp a_L (x+L/2) \rp \tanh \lp a_R (x-L/2) \rp \right].
\ee
The constant $\mathcal{N}$ is chosen so that $\mathop{{\rm max}}_{x \in \mathbb{R}} f(x) = 1$, and the parameters $a_L$, $a_R$, and $L$ are strictly positive. $F_{\rm max}$ is the maximum value of $F(x)$ reached on top of the obstacle, see \fig{fig:flow}. $F_{\rm as}$ is its asymptotic value, and is smaller than 1 so that the flows we consider are all asymptotically subcritical. By analogy with the transcritical case where $F$ crosses $1$, we will often refer to the upstream slope ($x \approx -L/2$) as the {\it black hole}, and to the downstream slope ($x \approx L/2$) as the {\it white hole} (even though there is no analogue Killing horizon if $F_{\rm max} < 1$). 

When $L$ is smaller than or of the same order as $1/a_L + 1/a_R$, $L$, $a_L$, and $a_R$ do not individually give accurate estimations of the length and slopes of the obstacle. It will thus be convenient to define effective values in the following way. We call $x_R$ (resp. $x_L$) the value of $x$ where $-\partial_x f(x)$ (resp. $\partial_x f(x)$) is largest. For large values of $L$, one obtains $x_R \approx - x_L \approx L/2$, but these can differ significantly for smaller lengths, see \fig{fig:flow}. We thus define the effective length $L_{\rm eff}$ and slopes $\sigma_{R/L}$ by 
\be 
& L_{\rm eff} \equiv x_R - x_L, \nonumber
\\ 
& \sigma_{R/L} \equiv \abs{\partial_x f(x_{R/L})}.  
\label{eq:eff}
\ee

It should be noticed that Eqs.~(\ref{eq:F}) and~(\ref{eq:f}) involve only dimensionless quantities when expressing $x$, $L$, $a_L$, and $a_R$ in units of the asymptotic water depth $h_{\rm as}$. As a result, each set of parameters effectively corresponds to a one-parameter family of water depth profiles $h(x)$ related to each other by a rescaling of all lengths. Moreover, this transformation does not change the behavior of the scattering coefficients. Indeed, the non-linear fluid equations~\cite{Unruh:2013gga,Coutant:2012mf} contain only one dimensionful parameter when surface tension and viscosity are neglected: the gravitational acceleration $g$. They are thus invariant under multiplication of all lengths by a positive number $\eta$ and all times by $\sqrt{\eta}$. This implies that the scattering coefficients extracted from the linear wave equation \eqref{eq:quartic} are also left invariant. 

\subsection{The $4\times 4$ $S$-matrix}

\begin{figure}
\begin{center}
\includegraphics[scale=0.8]{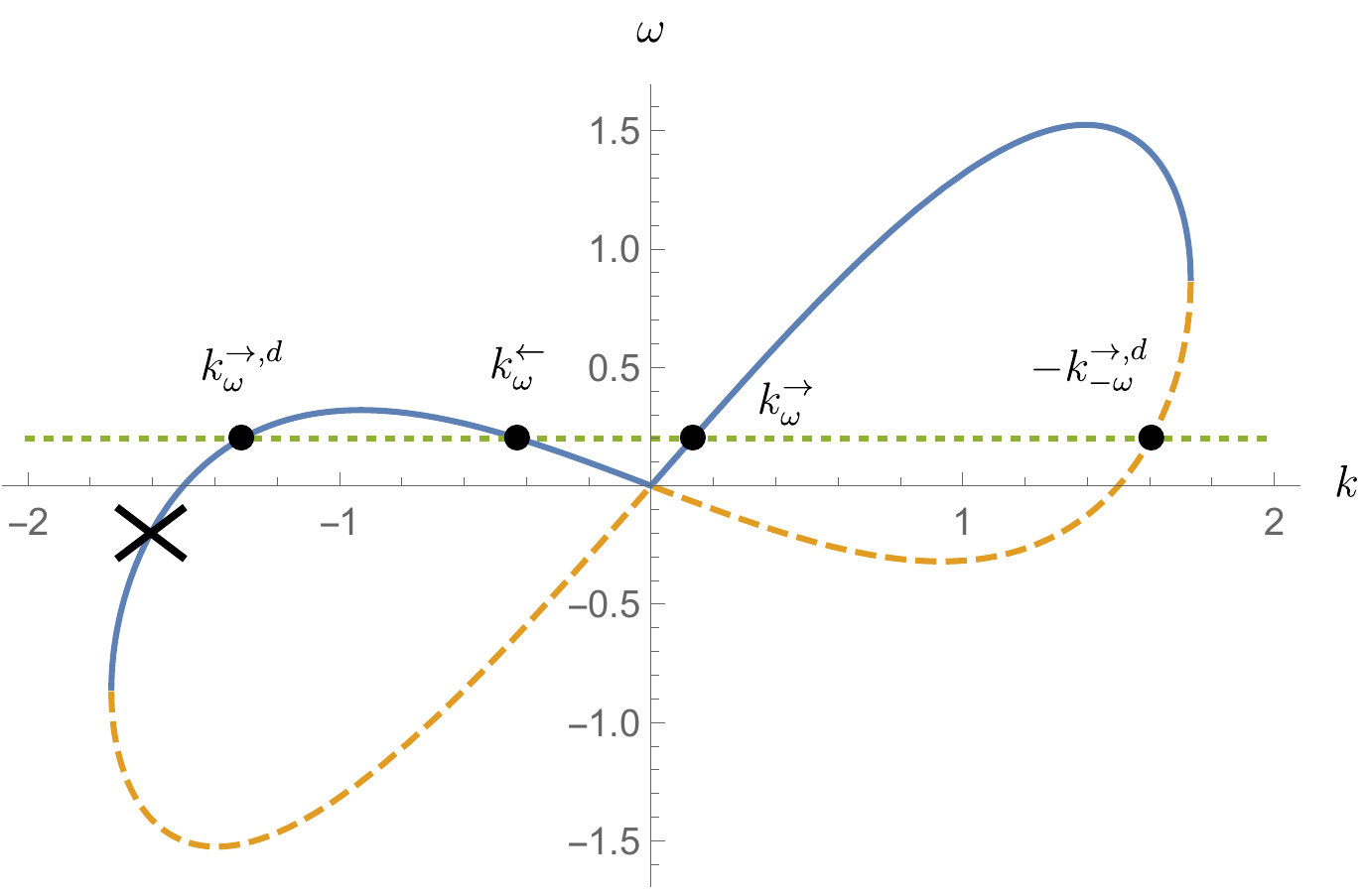} 
\end{center}
\caption{Dispersion relation \eq{eq:disprelq} for a subcritical flow with $F = 0.5$. $k$ is expressed in units of $h_{\rm as}^{-1}$, and $\om$ in units of $\sqrt{g/h_{\rm as}}$. The blue, continuous curve shows roots with $\Omega = \omega - vk > 0$, whereas $\Omega$ is negative along the orange, dashed curve. The green, dotted line shows $\om = 0.2$. It is smaller than $\ommax \approx 0.32$ where the two roots on the upper left quadrant merge. Large dots show the 4 wave vectors for $\om = 0.2$. The symbols are the same as those carried by the corresponding asymptotic modes which are listed in the text. Notice that the root $-k_{-\om}^{\rightarrow,d}$, for which $\Omega < 0$, is the opposite of that represented by a cross which has conserved frequency $-\omega < 0$ but co-moving frequency $-\Omega > 0$.} 
\label{fig:disprel}
\end{figure}
\begin{figure}
\begin{center}
\includegraphics[scale=0.8]{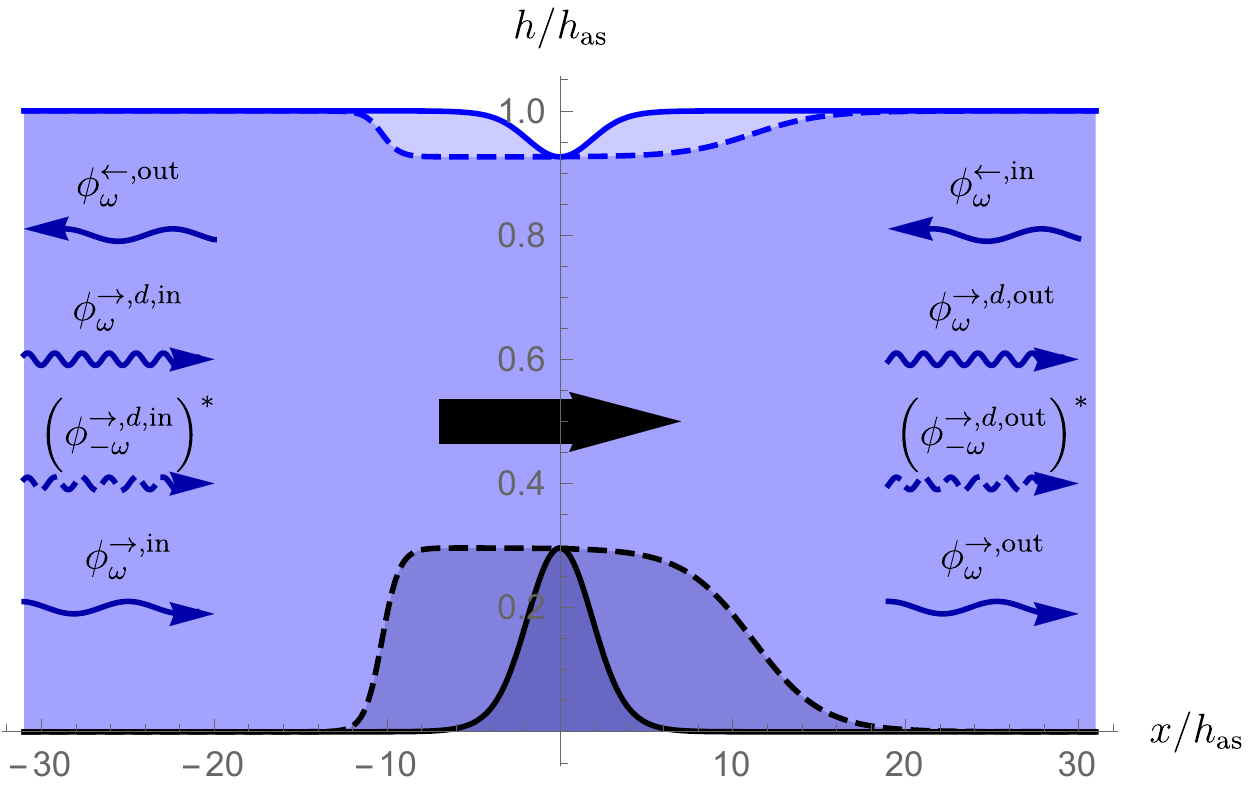} 
\end{center}
\caption{This figure shows schematically the bottom and the free surface of the flows associated with the two profiles of \fig{fig:flow}. The thick, black arrow shows the direction of the current. The eight wavy arrows indicate the asymptotic incoming and outgoing modes entering in the $4\times 4$ $S$-matrix discussed in the text. Dispersive short wavelength modes are indicated by rapid oscillations, while long wavelength modes are represented by longer oscillations. 
Dashed arrows indicate negative-energy waves.  The incoming mode sent by a wavemaker in the experiments is $\symbAtin$, top right side. When scattered on the obstacle, it produces the four outward-pointing arrows.
} \label{fig:fww}
\end{figure}

Since \eq{eq:quartic} does not depend explicitly on time, one can decompose any solution in terms of modes with fixed angular frequency $\om$. Moreover, in the asymptotic regions where $h$ is constant, these stationary modes are superpositions of plane waves $\phi_\omega \propto e^{i (k x - \om t)}$, where $k$ is related to $\om$ by \eq{eq:disprelq}. In the present work, we only consider frequencies in the interval $0 < \om < \om_{\rm max}$, where $\om_{\rm max}$ is the frequency at which two roots in the upper left quadrant of \fig{fig:disprel} merge (at Froude number $F$ equal to its asymptotic value $F_{\rm as}$). Using the quartic dispersion relation of \eq{eq:disprelq}, and $c_{\rm as} = \sqrt{g h_{\rm as}}$, it is given by  
\be 
\om_{\rm max} &=& 
\frac{c_{\rm as}}{h_{\rm as}} \sqrt{
6 \frac
{F_{\rm as} + \sqrt{F_{\rm as}^2+8}}
{\lp 3 F_{\rm as} + \sqrt{F_{\rm as}^2+8} \rp^3}
\lp 1-F_{\rm as}^2 \rp^3
}, \nonumber \\
& \simeq & \frac{c_{\rm as}}{3 h_{\rm as}}
(1 - F_{\rm as}^2)^{3/2} ,
\label{eq:ommax}
\ee
where the second equation is valid for $1 - F_{\rm as} \ll 1$, for more details we refer to Eqs.~(9) and~(10) of~\cite{Michel:2014zsa}. In the domain $0 < \om < \om_{\rm max}$, there are four real roots of \eq{eq:disprelq}, and thus four plane waves satisfying \eq{eq:quartic}. Explicitly, these are the following: 
\begin{itemize}
\item $\symbAt$ is hydrodynamic (in that its wave vector vanishes as $\om \to 0$) and left-moving. This is the mode sent by a wave-maker against the flow from the right side~\cite{Rousseaux:2007is,Weinfurtner:2010nu,Euve:2014aga,Euve:2015vml}, see \fig{fig:fww}. 
\item $\symba$ is a dispersive mode (in that its wave vector does not vanish when $\om \to 0$) and right-moving. 
\item $\symbb$ is also dispersive and right-moving. 
\item $\symbA$ is hydrodynamic and right-moving.
\end{itemize} 
The third mode has been complex-conjugated because its norm is negative, see \eq{eq:norm}, while the other three modes have positive norms.  We adopt the standard notation such that all modes without complex conjugation have scalar product $\delta(\om - \om')$, and hence, according to the definition (\ref{eq:norm}), the complex conjugated modes have scalar product $-\delta(\om - \om')$. It should be noticed that $\symbbno$ carries a negative energy. Hence, when increasing the amplitude of this mode, the wave energy is reduced, see~\cite{Coutant:2012mf} for more details. The arrow in the superscript gives the sign of the group velocity in the laboratory frame, i.e., the sign of $\partial_\om k$. The first 3 modes are counter-propagating with respect to the fluid. In transcritical flows, their mixing through scattering on the obstacle encodes the analog Hawking effect~\cite{Unruh:1994je,Brout:1995wp}. 
The last mode instead is co-propagating (with respect to the fluid) and plays no significant role in this regard. In fact, to obtain a good analogy with the standard Hawking prediction, one should minimize the coefficients governing its mixing with the three other modes~\cite{Macher:2009tw,Busch:2014bza}. 
 
We now consider two bases of globally defined modes, that is, solutions of \eq{eq:quartic} defined for all $x$. The {\it in} basis contains four modes with only one incoming wave, i.e., one asymptotic wave with group velocity oriented towards the horizon.  Similarly, the {\it out} basis comprises those modes with only one outgoing wave. The aim of the present work is to determine numerically the properties of the scattering matrix relating these two bases. We shall denote by a superscript ``${\rm in}$'' (resp. ``${\rm out}$'') the in (resp. out) modes, so that, for instance, $\symbAtin$ is the (global) mode which asymptotically contains only $\symbAt$ as incoming wave. 

Generalizing the notation used for the $3 \times 3$ $S$-matrix of Ref.~\cite{Macher:2009tw}, we write the relationship between the two bases as
\be \label{eq:16coeffs} 
\begin{pmatrix}
\symbAtin \\ 
\symbain \\ 
\symbbin \\
\symbAin \\ 
\end{pmatrix} =
\begin{pmatrix}
\tilde{A}_\om & \alpha_\om & \beta_\om &  A_\om^{(v)}  \\ 
\bar{\alpha}_\om &  A_\om & B_\om & \alpha_\om^{(v)} \\  
\bar{\beta}_\om & \bar{B}_\om & \bar{A}_\om 
& \beta_\om^{(v)}  \\
 \bar{A}_\om^{(v)} & \bar{\alpha}_\om^{(v)} & \bar{\beta}_\om^{(v)} & 
A_\om^{(vv)} 
\end{pmatrix}
\begin{pmatrix}
\symbAtout \\ 
\symbaout \\  
\symbbout\\
\symbAout 
\end{pmatrix}.
\ee
The superscript $(v)$ has been added to ease the identification of the coefficients involving the co-propagating mode $\symbA$. The $S$-matrix is an element of $U(3,1)$. This is a direct consequence of the fact that the scalar product of \eq{eq:norm} is conserved, and that the norm of $\symbb$ is the opposite of that of the three other modes. As a result, the squared absolute values of the coefficients of the first line satisfy
\be \label{eq:unit1}
|\tilde{A}_\om |^2 +  |\alpha_\om |^2 - | \beta_\om |^2 + |A_\om^{(v)}  |^2 = 1.
\ee
(When the transmission $|\tilde{A}_\om|^2$ and the reflection $|A_\om^{(v)}|^2 $ channels can be neglected, one recovers the standard $2 \times 2$ mode mixing which gives $|\alpha_\om |^2 - | \beta_\om |^2 = 1$.) Similar equations apply to the other lines, and to the columns. In these 8 relations, the squared absolute values of the 4 $\beta$ coefficients and the 2 $B$ coefficients are all multiplied by a minus sign. These 6 coefficients encode some mode amplification compensated for by excitation of the negative energy mode. 

\subsection{The behavior of the 16 scattering coefficients} 

\subsubsection{Transcritical flows}

To prepare the analysis of the scattering in subcritical flows, we first show how the 16 coefficients of \eq{eq:16coeffs} behave in a transcritical flow with $F_{\rm max} = 1.4$ and $F_{\rm as} = 0.6$. For simplicity, we choose a symmetric flow. We also choose to work with a narrow obstacle, as this eases the observation of the transmission occuring at very low frequency. Explicitly, we work with $a_R h_{\rm as} = a_{L} h_{\rm as} = 2$ and $L/h_{\rm as} = 2$. Since the flow is transcritical it has two analog horizons where $F = 1$. The analog Hawking temperature $T_H = \vert \partial_x (v-c) \vert /2\pi = \vert c\, \partial_x F \vert /2\pi$ evaluated on the horizons is $T_H \approx 0.111 \sqrt{g/h_{\rm as}}$. To give an example, if one chooses $h_{\rm as} = 0.140 {\rm m}$, the white (black) hole horizon is at $x \approx (-) 0.142 {\rm m}$, $T_H \approx 0.93 {\rm Hz}$ and $\ommax \approx 1.93 {\rm Hz}$.

\begin{figure} 
\begin{center} 
\includegraphics[width=0.49 \linewidth]{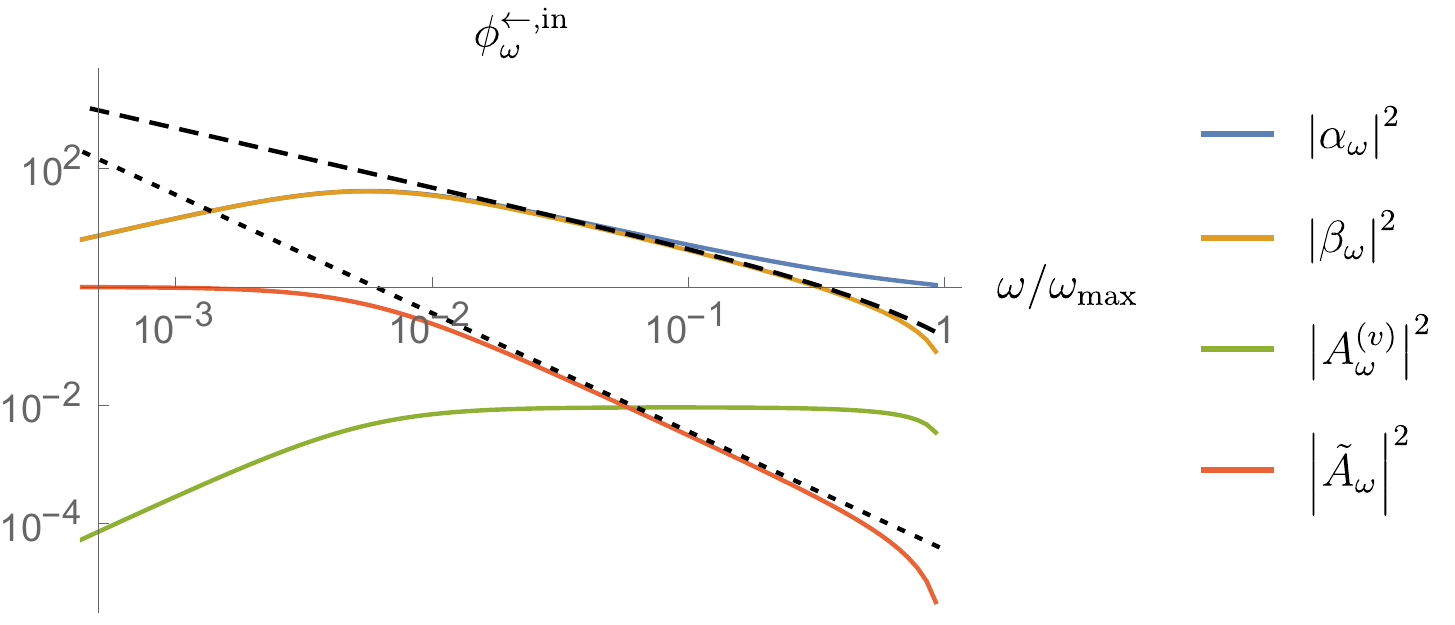}
\includegraphics[width=0.49 \linewidth]{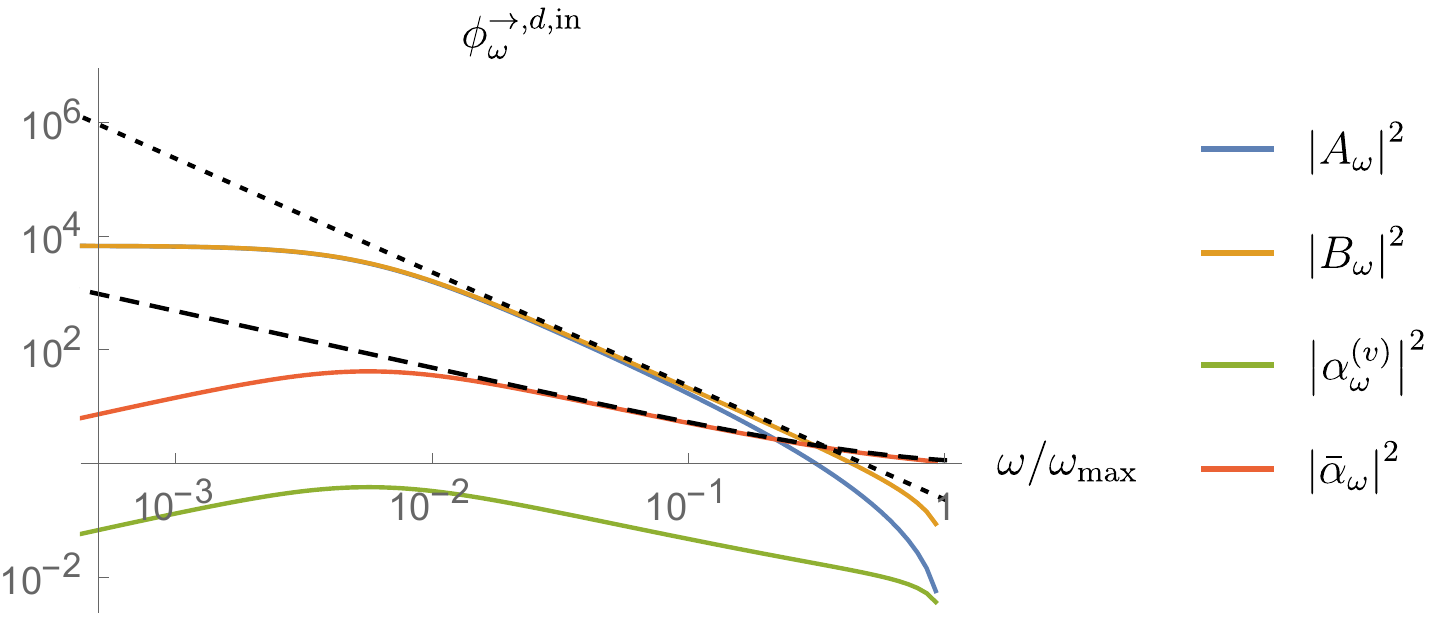}
\includegraphics[width=0.49 \linewidth]{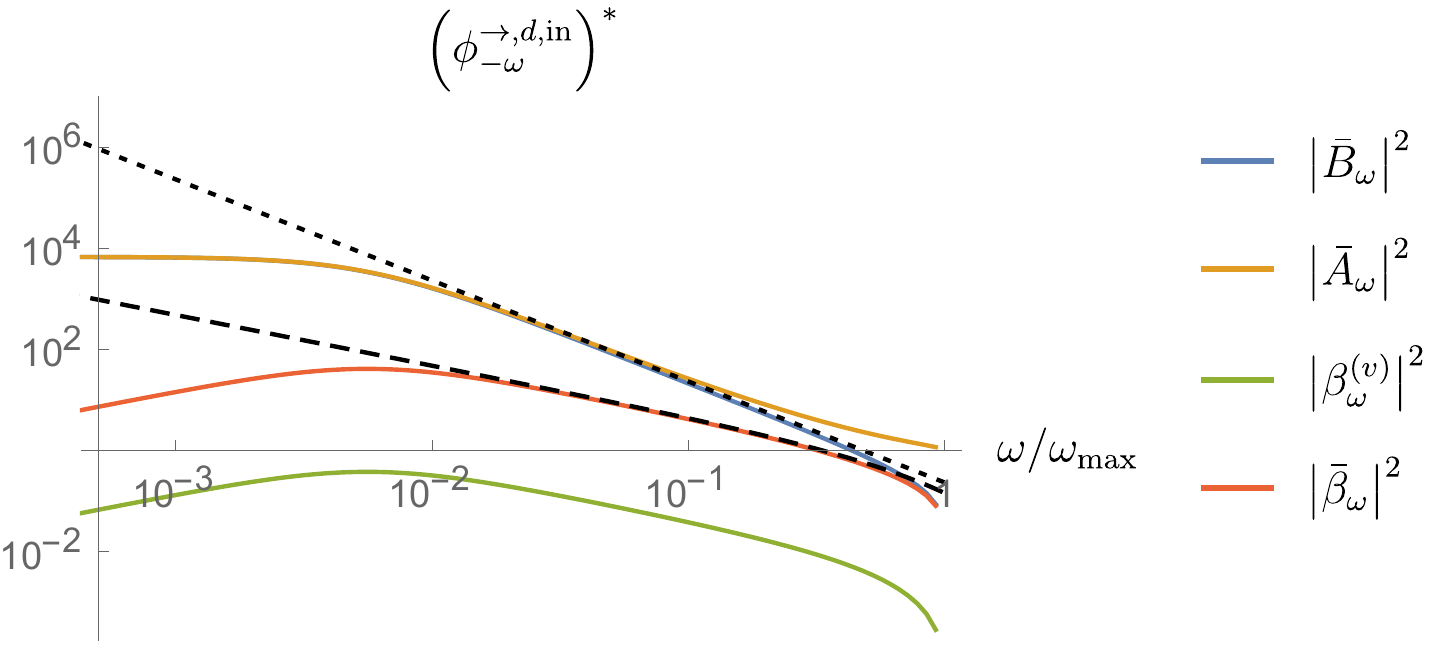}
\includegraphics[width=0.49 \linewidth]{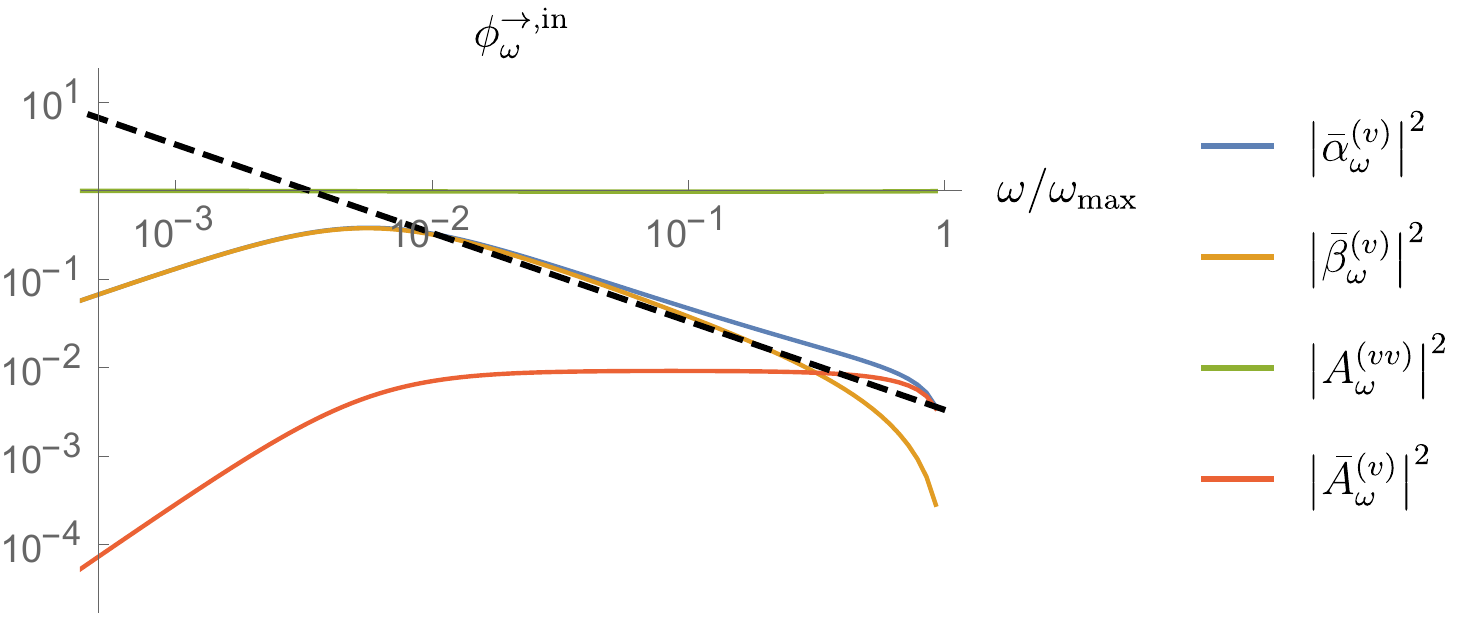}
\end{center}
\caption{In each panel, as a function of $\om/\ommax$, we show the squared absolute values of the 4 scattering coefficients associated with the incoming mode indicated above the plot. All plots are shown on a log-log scale. We work with a transcritical flow given by \eq{eq:F} with $F_{\rm as} = 0.6$, $F_{\rm max} = 1.4$, $a_R = a_L = 2\,h_{\rm as}^{-1}$, $L=2\,h_{\rm as}$, $h_{\rm min} \approx 0.57\,h_{\rm as}$. The dashed lines show the Planck spectrum $1/(e^{\om / T_H}-1)$ (top left and bottom left) $1/(1 - e^{-\om / T_H})$  (top right) and $T_H / (144 \om)$ (bottom right), where $T_H$ is the analog Hawking temperature, see text. Dotted straight lines show $(T_H / \om)^2$ (top right and bottom left), and $(T_H / (1200 \om))^2$ (top left). One clearly sees that these thermal curves are no longer followed for $\omega/\om_{\rm max} \lesssim 6.\, 10^{-3}$.}
\label{fig:16_trans}
\end{figure}

In each panel of \fig{fig:16_trans}, as a function of $\omega/\omega_{\rm max}$, we represent in log-log plots the squared absolute values of the four coefficients when sending each of the incoming waves of the left-hand side of \eq{eq:16coeffs}.  The symbol of the incoming mode is given on top of the panel, while each color always indicates the same outgoing mode, namely blue for $\symbaout$, orange for $\symbbout$, green for the co-propagating mode $\symbAout$, and red for $\symbAtout$.

The most important observation is that the absolute values of some scattering coefficients are significantly larger than 1. This indicates that the mode amplification (pair creation in quantum terms) induced by the scattering on this transcritical flow is large. Since the Hawking prediction is $\abs{\beta_\om}^2 = 1/ \lp e^{\om / T_H}-1 \rp$, one should look for curves which grow like $T_H / \om$ for $\om \to 0$. 

On the first panel, when sending $\symbAtin$ from the downstream right side, this growth characterizes the modes $\symbaout$ and $\symbbout$. This is to be expected from the Hawking radiation taking place in a white hole flow: in this case, the outgoing radiation is carried by the two dispersive modes emitted on the same side of the horizon. To show the quality of the agreement between the numerical outcome and the Hawking spectrum, the dotted black line follows the Planck law with the temperature $T_H$ evaluated on the white hole horizon. We can see that the agreement is excellent in a wide domain of frequencies containing $\om = T_H$ even though we work in a rather dispersive regime since $T_H/ \ommax \sim 0.48$~\cite{Finazzi:2012iu}. The upper limit of the domain is near $\ommax$, while its lower limit where the growth stops is here $\om_c \sim 6 \times 10^{-3} \, \ommax$.  This is due to the transmission across the obstacle of ultra low-frequency modes.  In fact, in the ultra low-frequency regime, we notice that $|\alpha_{\omega}|^{2}$ and $|\beta_{\omega}|^{2}$ agree with each other, and decrease linearly in $\om$ for $\om \to 0$.  As a result, the zero-frequency limit is fully characterized by the frequency $\sigma_{\beta}$ defined by
\begin{alignat}{2}
|\beta_{\omega}|^{2} \sim \frac{\omega}{\sigma_{\beta}} \,, & \qquad \omega \rightarrow 0 \,.
\label{eq:sigma_beta_defn}
\end{alignat} 
The critical frequency $\om_c$ is then given by $\om_c = (T_H \, \sigma_{\beta})^{1/2}$. This simple relation follows from matching the two behaviors of $|\beta_{\omega}|^{2}$ above and below $\om_c$, namely $|\beta_{\omega}|^{2} \sim T_H/\om$ and $|\beta_{\omega}|^{2} \sim \om/\sigma_\beta$, respectively.  When working in the limit of steep slopes, an approximate expression for $\om_c$ is
\be \label{eq:omc}
\frac{\omega_c}{\sqrt{g/h_{\rm as}}} 
\sim 
\sqrt{\frac{h_{\rm as}}{h_{\rm min}}} 
\lp F_{\rm max} - 1 \rp 
\sqrt{3 \left(F_{\rm max}^2-1\right)} \, e^{-k_{\text{dec}} L}. 
\ee
see  Eq.~(20) of~\cite{Michel:2014zsa}. Here $k_{\rm dec}$ is the imaginary part of the root of the dispersion relation at $\om = 0$ in the upper complex plane. In the present flow, one gets $k_{\rm dec}h_{\rm as} = 3.0$.  Equation (\ref{eq:omc}) gives a reliable estimation of $\om_c$ for sufficiently long obstacles, i.e. for $k_{\rm dec} L \gg 1$.  We shall see in Sec.~\ref{sec:systematic_analysis} that the damping of the evanescent mode also plays a crucial role in the characterization of subcritical flows.  

On the second panel, when sending the short wavelength mode $\symbain$ from the left side, one observes that the growth in $1/\om$ characterizes the mode $\symbAtout$ in red, as expected from the Hawking effect taking place on the black hole side. To underline the agreement the dashed black line here follows the theoretical prediction $\abs{\bar{\alpha}_\om}^2 \approx \abs{\beta_\om}^2 +1 = 1/(1- e^{-\om / T_H})$. Again the agreement is excellent down to the low-frequency cut-off $\sim \om_c$ where the growth stops. We also notice the presence of two curves which grow like $(1/\om)^2$. This behavior is indicated by a dotted straight line which gives $(T_H/\om)^2$. This growth is due to the fact that these modes have been scattered on {\it both} horizons. As a result their scattering coefficients essentially grow like the product of the amplification associated with each horizon, as was discussed in \cite{Coutant:2011fz}. The same observations apply to the first two coefficients of the third plot which are obtained when sending the dispersive negative norm mode from the left. On the third panel, we also see that the mode $\symbAtout$ in red closely follows the Planck law indicated by a dashed line.

On the last panel, irrespective of the frequency, we see that the co-propagating mode $\symbA$ is essentially transmitted. This indicates that the mode $\symbA$ nearly decouples from the three other modes, which are counter-propagating with respect to the fluid. In addition, when considering the green curves on the three other panels, one verifies that their values are always subdominant. These observations establish that (in transcritical flows at least) the scattering coefficients involving the co-propagating mode can be neglected, to a good approximation.

\subsubsection{Subcritical flows}

\begin{figure} 
\begin{center}
\includegraphics[width=0.49 \linewidth]{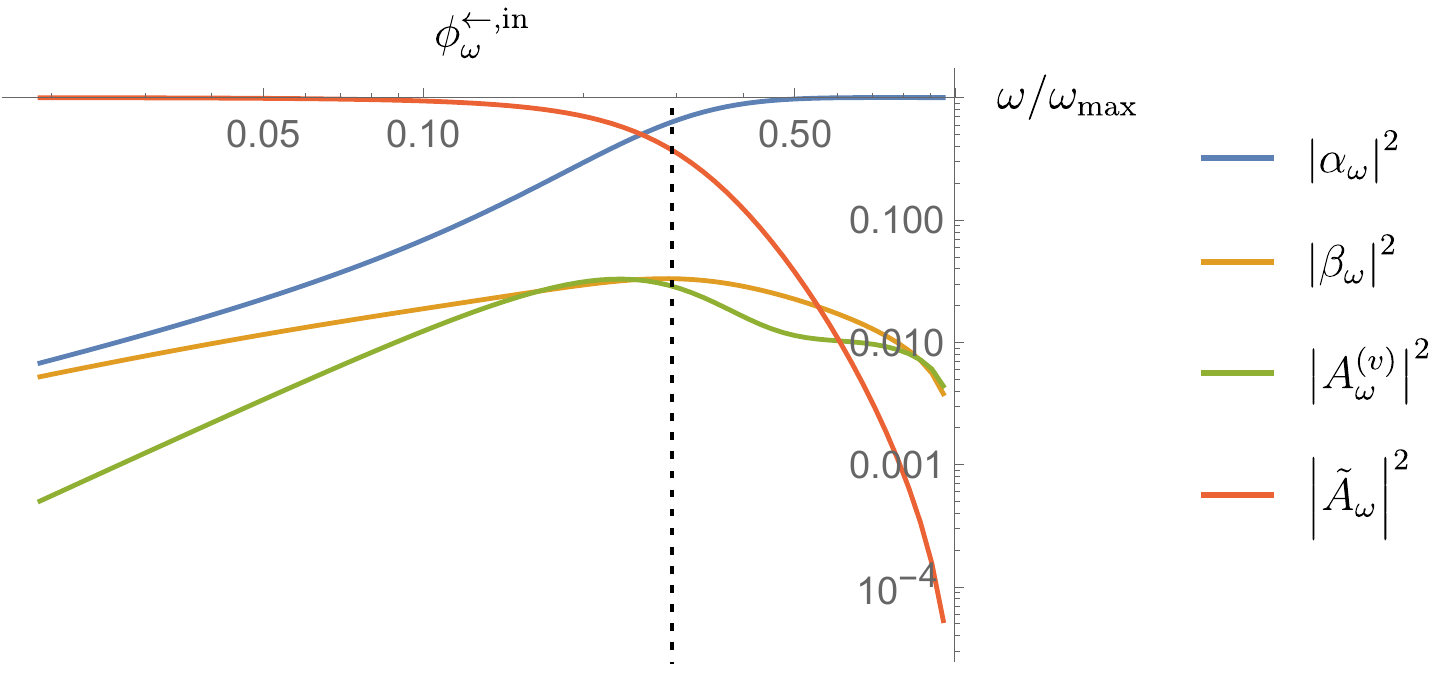}
\includegraphics[width=0.49 \linewidth]{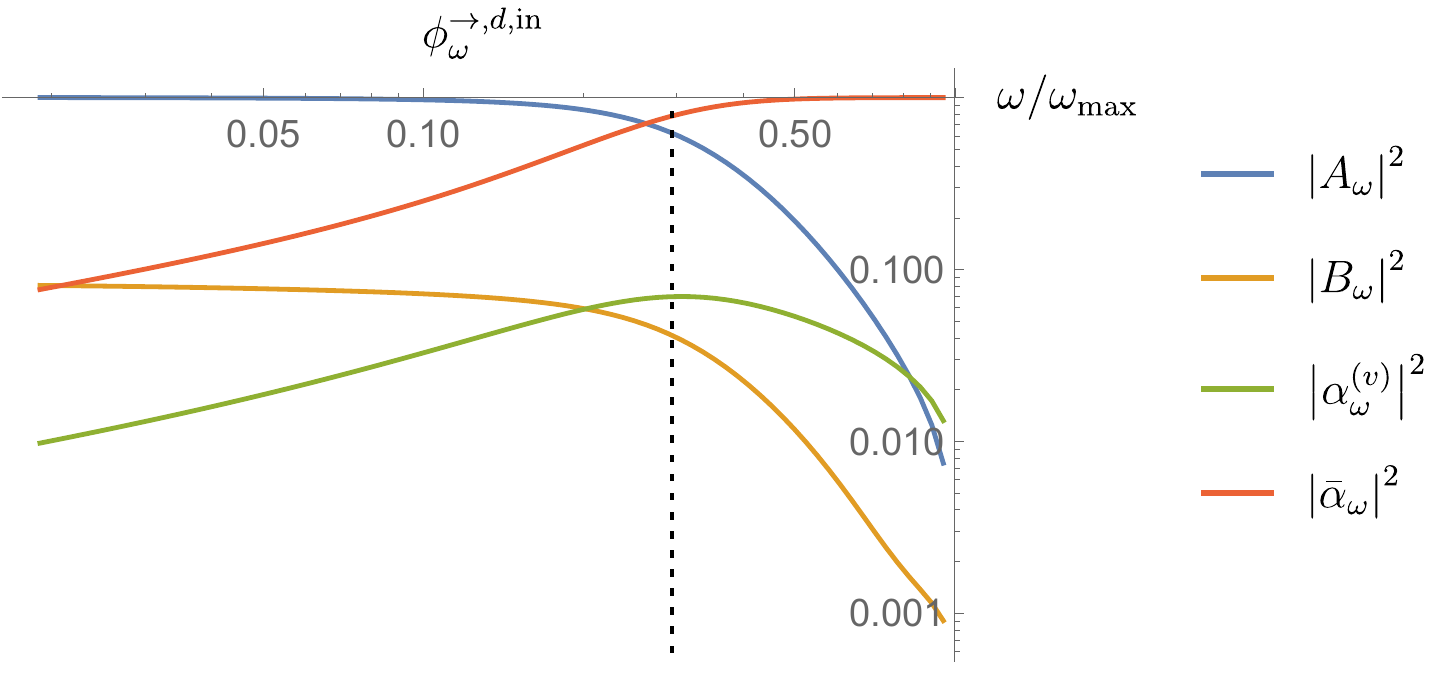}
\includegraphics[width=0.49 \linewidth]{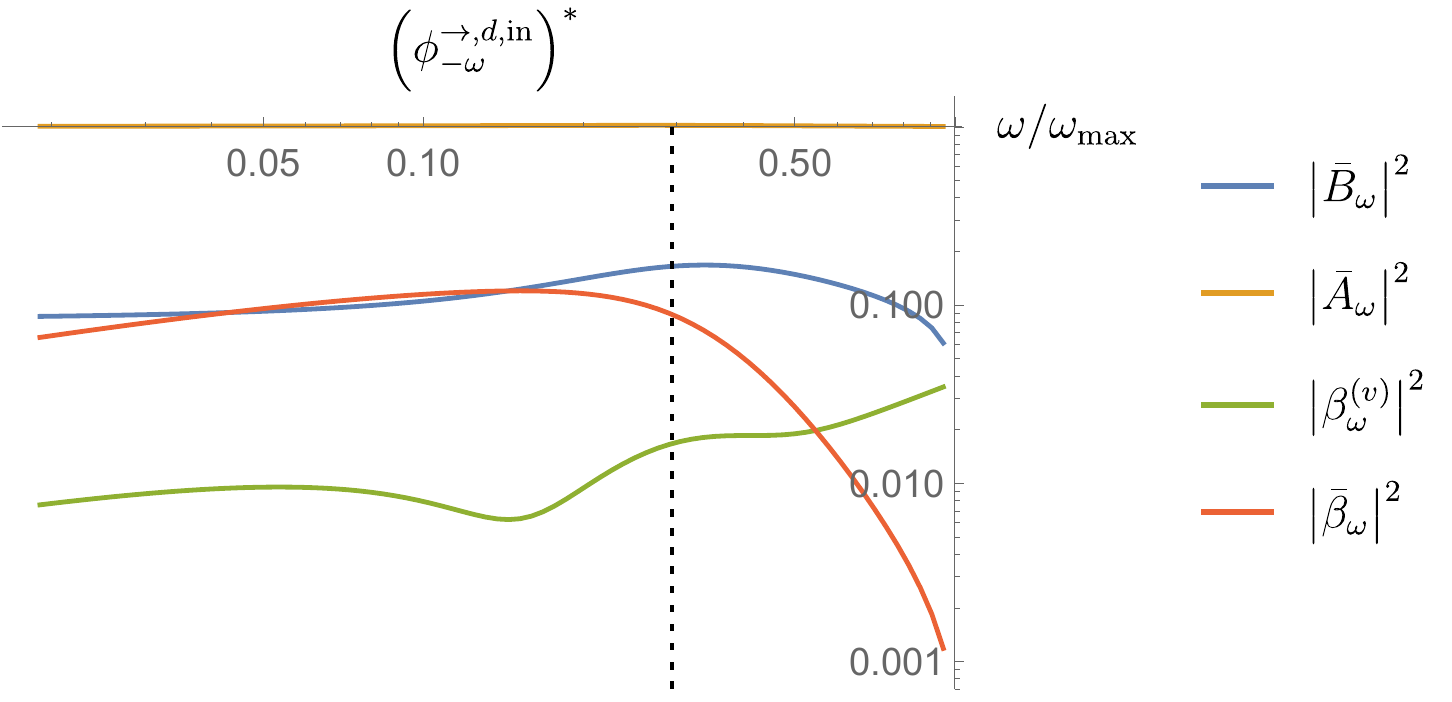}
\includegraphics[width=0.49 \linewidth]{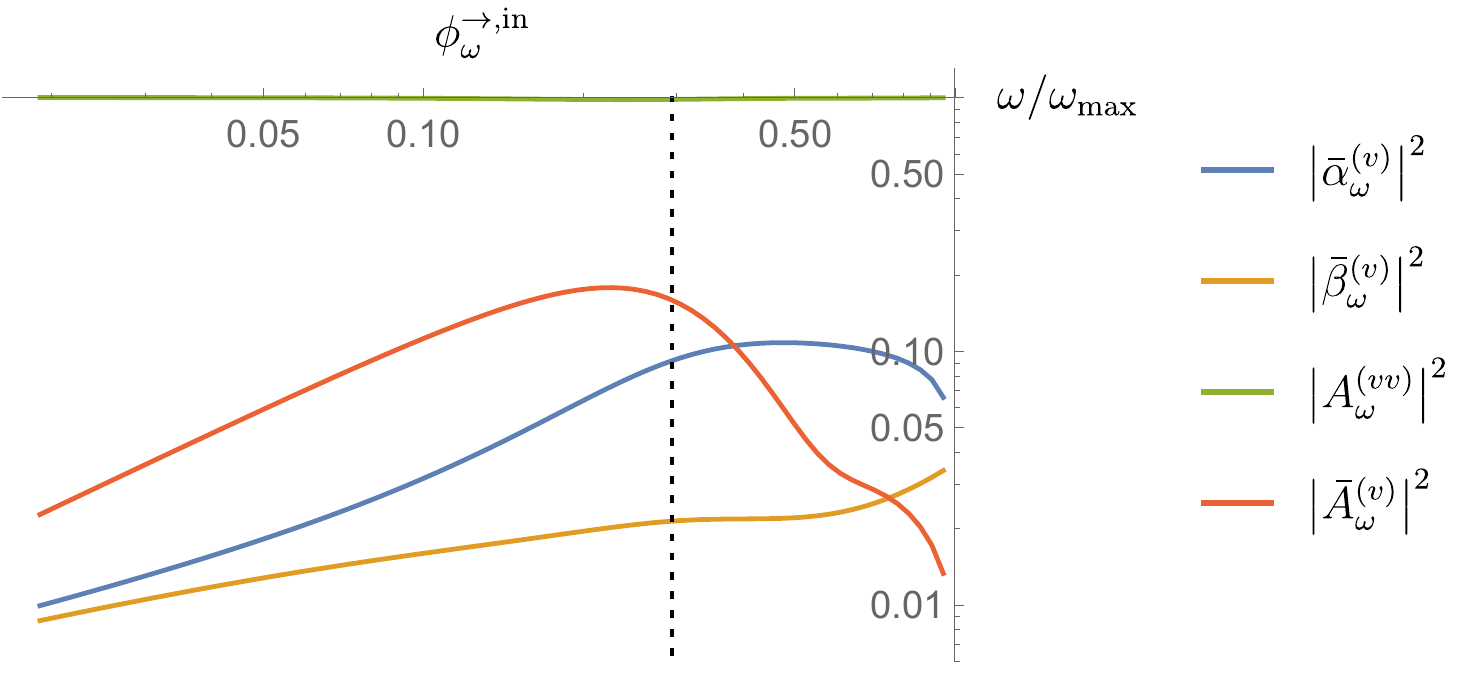}
\end{center}
\caption{As a function of $\om/\ommax$, we show the squared absolute values of the 16 scattering coefficients for a subcritical flow with $F_{\rm as} = 0.4$, $F_{\rm max} = 0.8$, and $a_R = 4 a_L = 2\,h_{\rm as}^{-1}$, $L=4\,h_{\rm as}$. All plots are shown on a log-log scale. The dotted vertical line shows $\om_{\rm min} = 0.28 \ommax$. On the two upper plots, the large transmission for frequencies lower than $\om_{\rm min}$ is clearly visible. On the left upper plot, one also sees that $|\alpha_\om|^2$ and $|\beta_\om|^2$ go to 0 as $\om$ for $\om \to 0$.
} \label{fig:16_sub}
\end{figure}

 We now consider the scattering coefficients in a subcritical flow with $F_{\rm max} = 0.8$, $F_{\rm as} = 0.4$, $a_R h_{\rm as} = 4 a_L h_{\rm as} = 2$ and $L/h_{\rm as} =4$. The effective values of \eq{eq:eff} are $\sigma_L h_{\rm as} \approx 0.27$, $\sigma_R h_{\rm as} \approx 1.03$, and $L_{\rm eff}/h_{\rm as} \approx 4$. In the four panels of \fig{fig:16_sub}, as a function of $\om/\ommax$, we show the log-log plots representing the squared absolute values of the same scattering coefficients as in \fig{fig:16_trans}, following the same notational conventions. 

The main difference one immediately sees is that the scattering coefficients are suppressed with respect to the transcritical case, never becoming appreciably larger than 1. This reveals that in subcritical flows there is no significant mode amplification. In other words, the 6 anomalous coefficients mixing modes with opposite norms all remain much smaller than 1. For instance, in the first panel, the squared norm of the $\beta_\om$ coefficient (encoding the scattering on the ``white hole'' side) is always smaller than $0.04$. The same observation applies to the $\bar \beta_\om$ coefficient encoding the scattering on the ``black hole'' side, see the red curve of the third panel. The lesson here is very clear: when the Froude number remains smaller than 1, the typical growth of the $|\beta_\om|^2$ coefficients in ${\om}^{-1}$ is no longer found. This could be understood from the absence of any Killing horizon in the associated effective metric $ds^2 = - c^2 dt^2 + (dx - v dt)^2$~\cite{Unruh:1980cg}. 

The absence of horizons in subcritical flows introduces a new critical frequency, which we shall call $\om_{\rm min}$, and which is indicated by a vertical line in the four panels of Figure \ref{fig:16_sub}.  It is the frequency at which the dispersion relation has a double root for $F = F_{\rm max}$, vanishing as $F_{\rm max} \to 1$. In the quartic approximation of \eq{eq:disprelq}, it is thus  given by the same expression of \eq{eq:ommax} but now evaluated on top of the obstacle where $h$ and $c$ reach their minimal values: 
\be
\om_{\rm min} \simeq \frac{c_{\rm min}}{3 h_{\rm min}} (1 - F_{\rm max}^2)^{3/2}.
\label{eq:ommin}
\ee 
For $\om > \om_{\rm min}$, the two upper panels show that the hydrodynamical mode is blocked and reflected onto the dispersive mode, and vice versa, see the red and blue curves. This can be understood from the fact that the corresponding characteristics have a turning point for $\om > \om_{\rm min}$~\cite{Michel:2014zsa}. Similarly, the absence of significant scattering experienced by the negative norm mode and the co-propagating modes (see the two lower panels) can also be understood from the validity of the WKB approximation for the propagation of both of these modes.

For $\om < \om_{\rm min}$, the situation is even simpler as the four incident modes are essentially transmitted above the obstacle. In fact, the mode mixing coefficients are all small, as can be understood from the fact that they encode non-adiabatic corrections in a domain where the WKB approximation is reliable~\cite{Coutant16}. In the limit $\om \to 0$, the squared norms of the coefficients relating a dispersive mode and a hydrodynamic one go to zero as $O(\om)$~\cite{Michel:2014zsa}, while those relating the two hydrodynamic modes decrease faster, as $O(\om^2)$. Notice however that $B_\om$ and $\bar{B}_\om$ go to non-vanishing values. This behavior is similar to the one found at very low frequency in transcritical flows, although the non-vanishing values are much smaller in subcritical flows because the growth $|\bar{B}_\om|^2 \sim |{B}_\om|^2 \sim 1/\omega^2$ found in~\fig{fig:16_trans} is no longer present. 

\subsection{Evolution of the scattering coefficients of $\symbAtin$ when varying $F_{\rm max}$}
\label{sub:Fmax}

We observed in the previous subsection that the behavior of the coefficients critically depends on whether the flow is sub- or transcritical. To display the transition between these two behaviors, we gradually lower $F_{\rm max}$ from $1.2$ to $0.8$, focussing on the left-moving incoming mode $\symbAtin$, which is most relevant for the experiments performed in Nice, Vancouver, and Poitiers~\cite{Rousseaux:2007is,Weinfurtner:2010nu,Euve:2014aga,Euve:2015vml}. 
Explicitly, the first line of \eq{eq:16coeffs} gives
\be
\symbAtin = \tilde{A}_\om \symbAtout + \alpha_\om \symbaout + \beta_\om \symbbout +  A_\om^{(v)} \symbAout,
\label{eq:incm}
\ee
where the four scattering coefficients satisfy \eq{eq:unit1}.

\begin{figure} 
\begin{center}
\includegraphics[width=0.49 \linewidth]{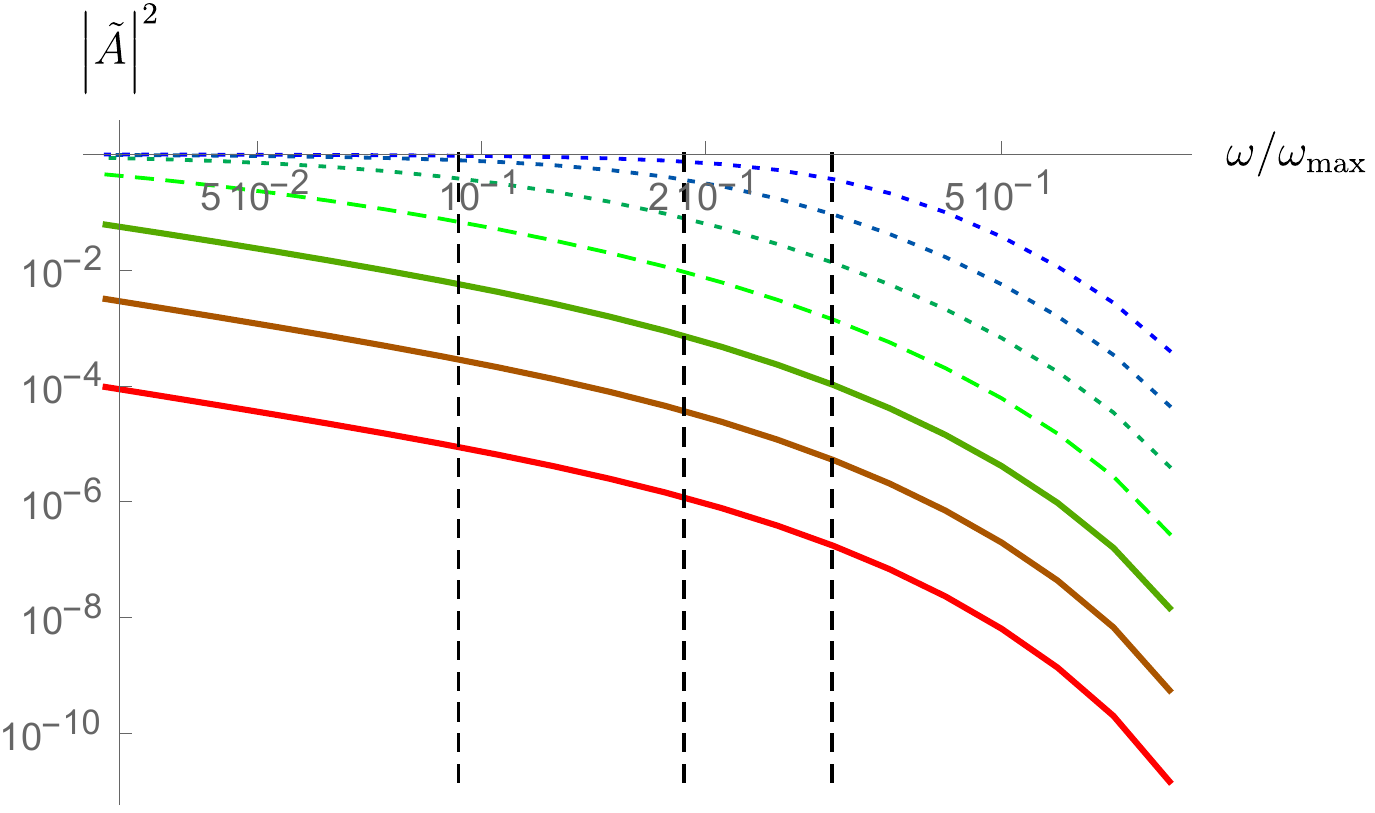}
\includegraphics[width=0.49 \linewidth]{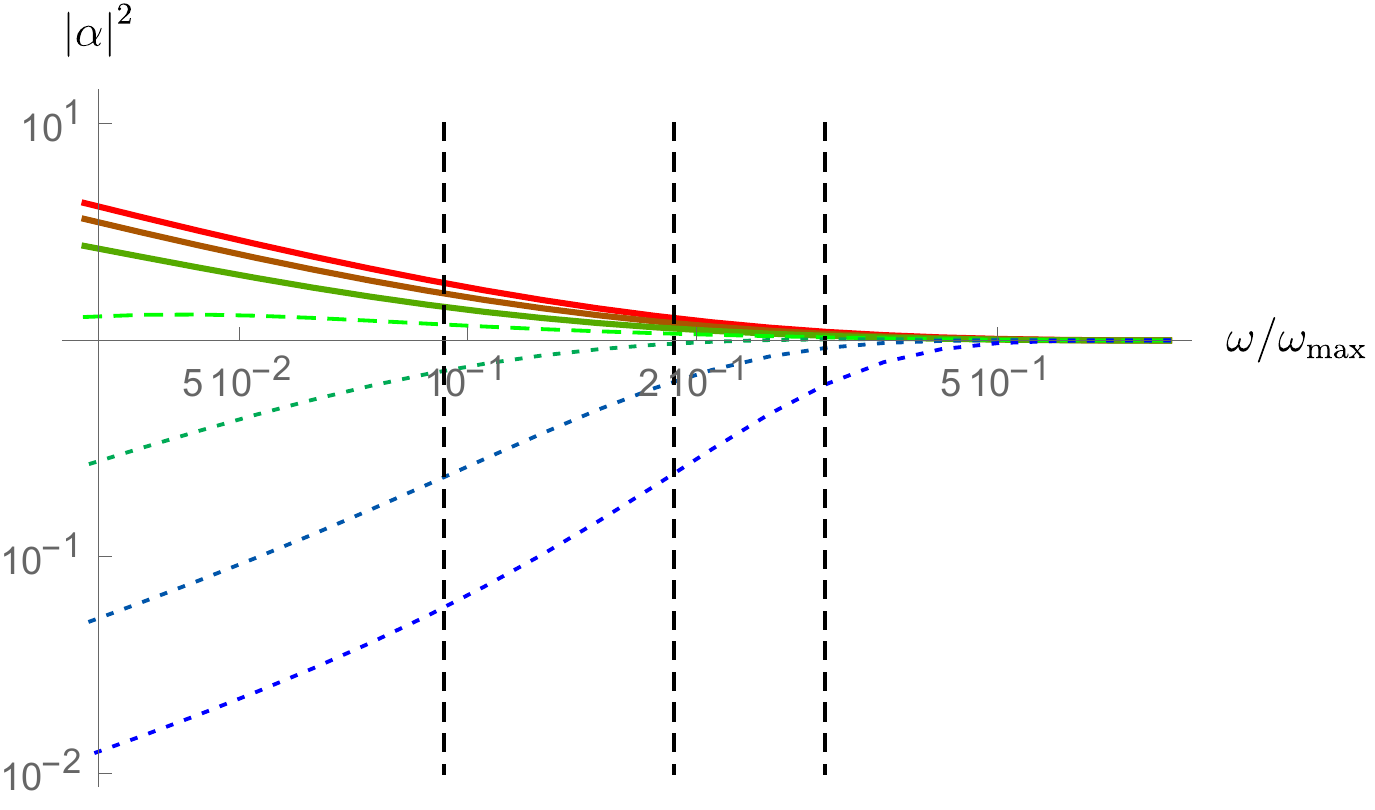}
\includegraphics[width=0.49 \linewidth]{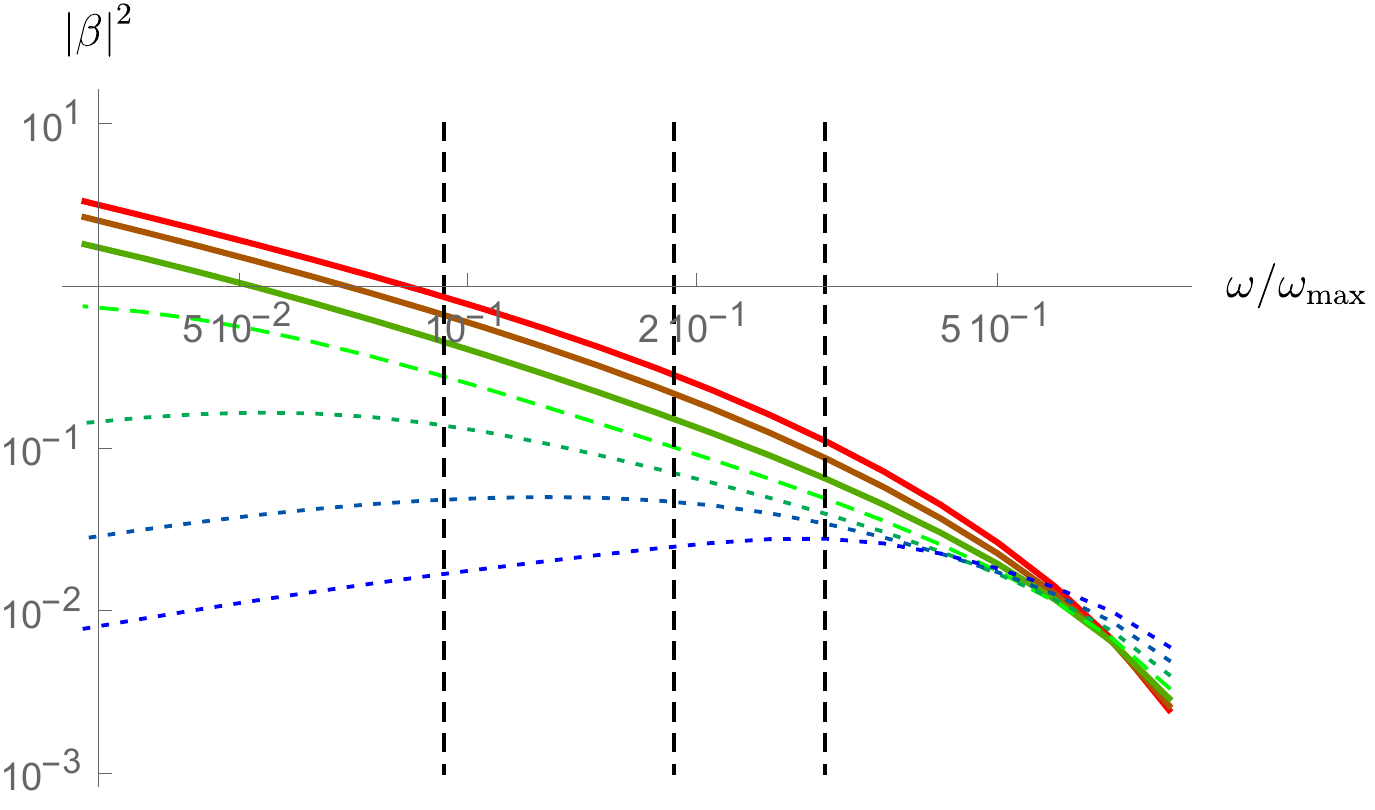}
\includegraphics[width=0.49 \linewidth]{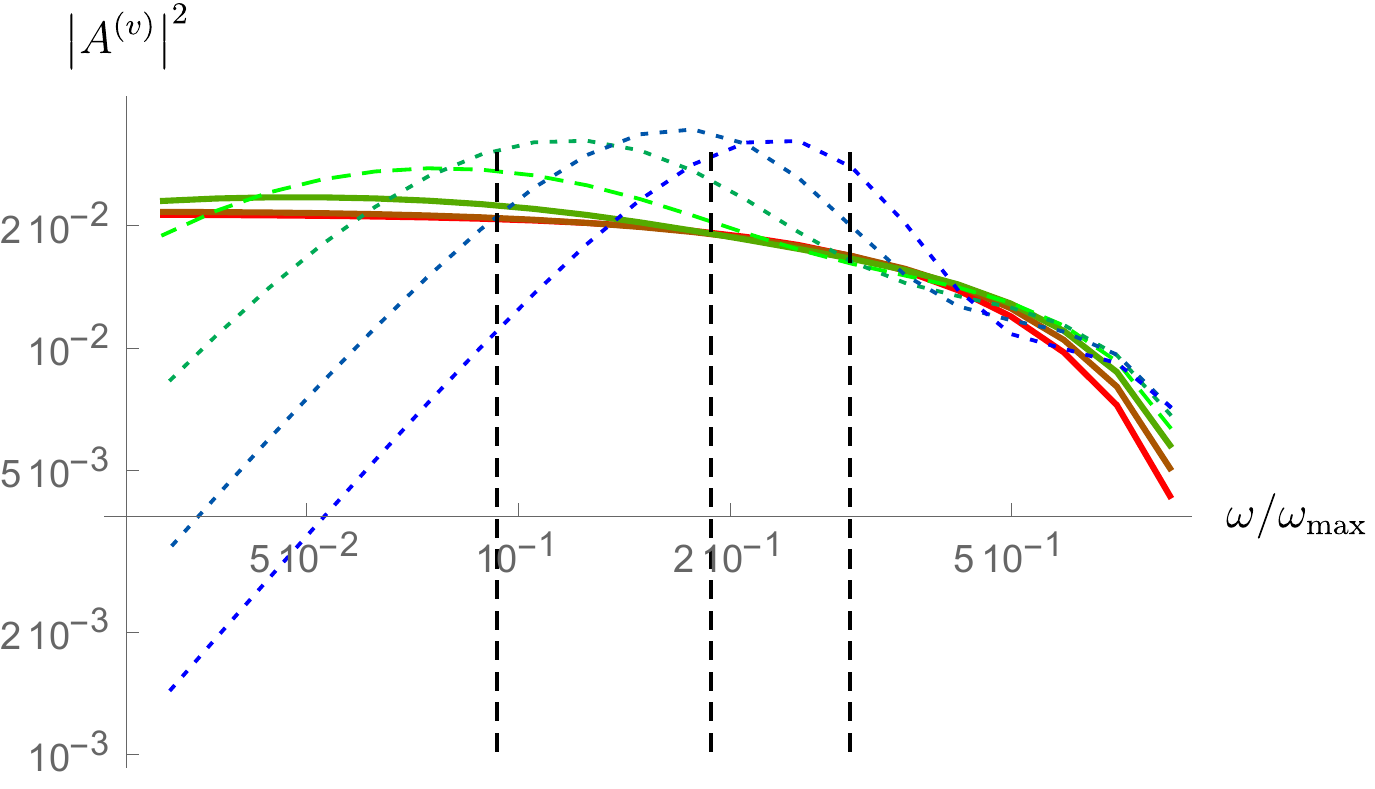}
\end{center}
\caption{As a function of $\om/\ommax$, we show the squared absolute values of the 4 scattering coefficients of the mode $\symbAtin$ for flows with 7 different values of $F_{\rm max}$ but the same values of $F_{\rm as}$ and $\kappa_{R/L}$ of \eq{eq:kappaeff} as those used in  \fig{fig:16_sub}. $F_{\rm max}$ takes equally spaced values from $1.2$ (red curves) to $0.8$ (blue curves). The green dashed curve corresponds to the critical case $F_{\rm max} = 1$. It separates the 3 subcritical flows (dotted lines) from the 3 transcritical ones (continuous lines). The three dotted vertical lines give the values of the critical frequency $\omega_{\rm min}$ of \eq{eq:ommin} below which the incident waves are essentially transmitted, i.e., $|\tilde A_\om|^2 \approx 1^-$. The most interesting panel is that of $\vert \beta_\om \vert^2$ representing the anomalous mode mixing. When decreasing $F_{\rm max}$, one clearly sees the replacement of the low frequency behavior in $1/\om$, by a behavior linear in $\om$. For the critical flow, one sees that $\vert \beta_\om \vert^2 \sim 1$ at low frequency.} \label{fig:flow4}
\end{figure}

The precise evolution of the scattering coefficients when decreasing $F_{\rm max}$ depends on the variations of the other flow parameters. Here, we work with fixed values of $F_{\rm as} = 0.4$ and $L/h_{\rm as} = 4$, which are the same as those used in \fig{fig:16_sub}, while we vary the parameters $a_{R/L}$ of \eq{eq:f} so that the generalized surface gravities
\be 
\kappa_{R/L} \equiv \abs{\partial_x (v-c)|_{x_{R/L}}}, 
\label{eq:kappaeff}
\ee
differ by less than $10 \%$ when varying $F_{\rm max}$ from $1.2$ to $0.8$. Explicitly, the values of $a_R$ and $a_L$ used in \fig{fig:flow4} are derived from those of \fig{fig:16_sub} by dividing by $\lp (F_{\rm max} - F_{\rm as})/0.4  \rp^{2/3}$, so that $F_{\rm max} = 0.8$ corresponds to exactly the same flow in both figures. 

The upper left plot of \fig{fig:flow4} shows the squared absolute value of the transmission coefficient $\tilde A_\om$ for 7 flows: three subcritical, one critical ($F_{\rm max} = 1$) and three transcritical. For the three subcritical flows (dotted curves), for $\om$ smaller than the corresponding values of $\om_{\rm min}$ which are indicated by three dotted vertical lines, the transmission coefficient is close to 1, i.e., there is no blocking of incident waves. For the critical flow (dotted line), one sees that $|\tilde A_\om|^2$ approaches $1$ for low frequency. Instead, for the three transcritical flows, it remains smaller than $0.1$ for the whole frequency range shown in the figure. (Because of the finite size of the obstacle, it nevertheless tends to 1 in the limit $\om \to 0$.) Interestingly, when increasing $F_{\rm max}$ at fixed $\om$, $|\tilde A_\om|^2$ decreases nearly exponentially in the region where it is small. Correspondingly, the critical frequency $\om_c$ of \eq{eq:omc} at which transmission becomes significant increases and becomes of the order of $\om_{\rm max}$ when $F_{\rm max} = 1$. 

The dichotomy between trans- and subcritical flows is more pronounced when considering the coefficients $\alpha_\om$ and $\beta_\om$. When the flow is significantly transcritical, i.e. $F_{\rm max} > 1.1$, there is a wide frequency domain between $\om_c$ and $\om_{\rm max}$ where $\vert \alpha_\om\vert^2$ and $\vert \beta_\om \vert^2$ are proportional to $1/\om$. This interval shrinks when decreasing $F_{\rm max}$ 
and vanishes on reaching the critical case $F_{\rm max} = 1$. For all subcritical flows, one clearly sees that $\vert \alpha_\om\vert^2$ and $\vert \beta_\om \vert^2$ go to zero linearly as $\om \to 0$~\cite{Michel:2014zsa}. 
As a result, in subcritical flows the maximal value of $\vert \beta_\om \vert^2$ is reached near $\om_{\rm min}$, 
and steadily decreases as $F_{\rm max}$ is decreased further. 

In the lower right panel, for subcritical flows, we notice that $|{A_\om^{(v)}}|^2$ decreases as $\om^2$ for $\om \to 0$. In transcritical flows, this decrease can only be seen for frequencies close to or smaller than $\om_c$, leaving a wide interval where $|{A_\om^{(v)}}|^2$ is nearly constant, but not significant as $|{A_\om^{(v)}}|^2 \lesssim 0.03$.
 
\begin{figure}
\begin{center}
\includegraphics[width=0.49 \linewidth]{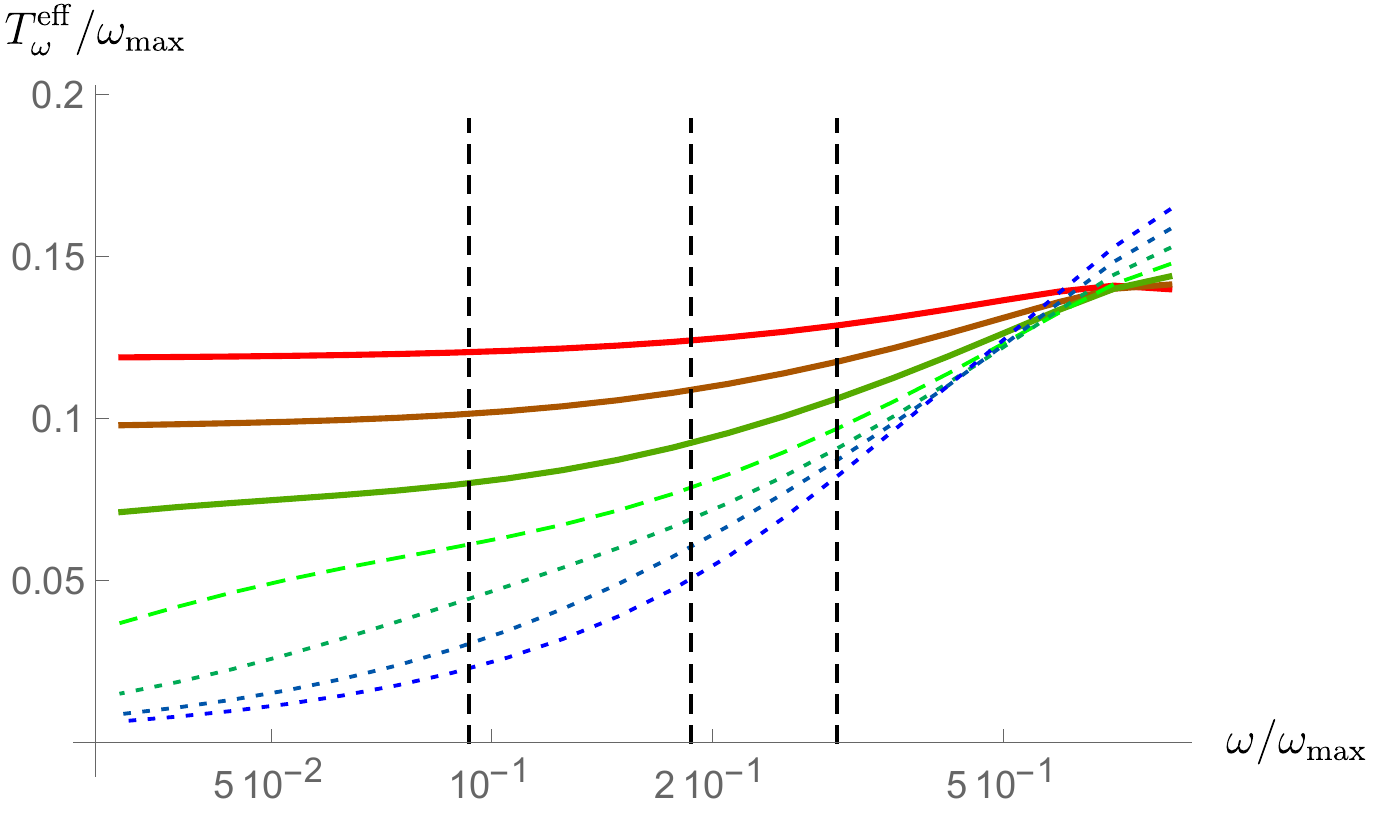}
\includegraphics[width=0.49 \linewidth]{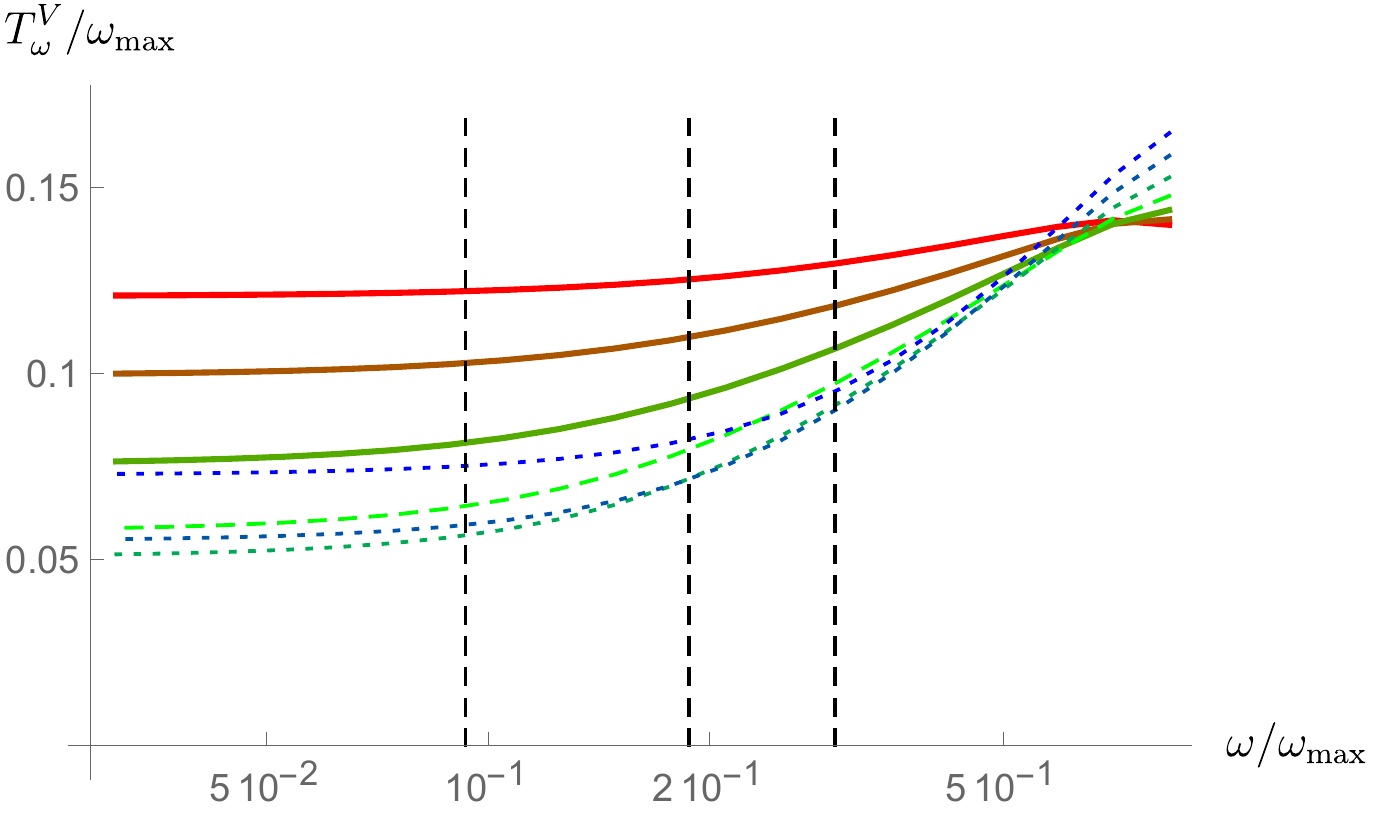}
\end{center}
\caption{On the left panel, as a function of $\om/\ommax$, we show $T_\omega^{\rm eff} / \ommax$ of \eq{eq:Teff} for the same 7 flows of \fig{fig:flow4}. Except near $\omega_{\rm max}$, at fixed $\om$, $T_\omega^{\rm eff}$ monotonically decreases when reducing $F_{\rm max}$. One also observes that $T_\omega^{\rm eff}$ ceases to be a (non-vanishing) constant at low frequency for critical and subcritical flows. The right panel shows $T_\omega^V$ of \eq{eq:TR} for the same flows. For the three transcritical flows, $T_\omega^V$ closely agrees with $T_\omega^{\rm eff}$ for all $\om$. Instead, for the critical and subcritical flows, its low frequency behavior radically differs from that of $T_\omega^{\rm eff}$. In particular, the constant value reached by $T_\omega^V$ for $\om \to 0$ increases when reducing $F_{\rm max} < 1$.} \label{fig:2temp}
\end{figure}
To complete this comparison, it is interesting to study the behaviors of two effective temperatures which have been used to characterize the spectrum. The first one is defined by $\abs{\beta_\om}^2 = 1/(e^{\omega / T_\om^{\rm eff}} - 1 )$, i.e.,
\be \label{eq:Teff} 
\ln \frac{\abs{\beta_\om}^2}{1+ \abs{\beta_\om}^2} =  - \frac{\om}{T_\om^{\rm eff}}.
\ee
Constancy of $T_\om^{\rm eff}$ is equivalent to $\abs{\beta_\om}^2$ following the Planck law with temperature $T_{\om}^{\rm eff}$, see~\cite{Macher:2009tw,Macher:2009nz,Finazzi:2012iu}. The second one is defined by~\cite{Weinfurtner:2010nu} 
\be \label{eq:TR} 
\ln \abs{ \frac{\beta_\om}{\alpha_\om} }^2 = - \frac{\om}{T_\om^{V}} \,. 
\ee
These coincide whenever ${\abs{\alpha_\om}^2} - {\abs{\beta_\om}^2}= 1$. 
In \fig{fig:2temp}, they are shown as functions of $\omega$ for the same flows as those of \fig{fig:flow4}. In transcritical flows and for $\om_c \ll \om \ll \om_{\rm max}$, they are both nearly constant and very close to each other, as can be understood from the fact that the transmission $\vert \tilde A_\om\vert^2$ and the ``gray body'' factor $\vert A^{(v)}_\om \vert^2 $ are both negligible. In this case, ${\abs{\alpha_\om}^2} - {\abs{\beta_\om}^2}= 1$ follows from unitarity, see \eq{eq:unit1}. 

However, they strongly differ in subcritical and near-critical flows. (In fact they also differ in transcritical flows but only at very low frequencies, for $\om < \om_c$.) In these cases, $T_\om^{\rm eff}$ goes to zero linearly when $\om \to 0$ because of the aforementioned behavior of $\abs{\beta_\om}^2$, i.e., the suppression of the amplification mechanism at low frequencies due to transmission. On the other hand, $T_\om^V$ approaches a finite value in that limit. This is because $\abs{\alpha_\om}^2$ and $\abs{\beta_\om}^2$ both go to zero linearly, so that their ratio goes to a finite, non-vanishing constant. Interestingly, we notice that this constant value {\it increases} when decreasing $F_{\rm \max}$, as can be seen in the crossing of the dotted lines occurring for $\om / \ommax \sim 0.2$ in the right plot of \fig{fig:2temp}. Our numerical simulations suggest that it goes to infinity in the limit $F_{\rm max} \to F_{\rm as}$, i.e., when approaching a homogeneous flow without obstacle.

\section{Influence of the background flow parameters} 
\label{sec:systematic_analysis}

Let us now focus our attention on subcritical and near-critical flows. As in Sec. \ref{sub:Fmax}, we again 
restrict our attention to the left-moving incoming mode of \eq{eq:incm}. Our aim is to identify the relevant parameters determining the spectral properties of the scattering coefficients. To this end, we consider three different phenomena characterized by the value of the frequency: 
\begin{itemize}
\item{When increasing $\omega$ near $\omega_{\rm min}$, the scattering of $\phi_{\omega}^{\leftarrow,{\rm in}}$ varies 
from near-total transmission across the obstacle to an essential reflection from the obstacle.
More precisely, the transmission coefficient $\vert \widetilde{A}_{\omega} \vert^{2}$ varies from near $1$ to near $0$, while $\vert \alpha_{\omega} \vert^{2}$ varies in the opposite manner.
The sharpness of the transition will be quantified by the derivative of $\vert \widetilde{A}_{\omega} \vert^{2}$ at $\omega_{\mathrm{min}}$.} 
\item{Below $\omega_{\mathrm{min}}$, $\left|\alpha_{\omega}\right|^{2}$ and $ \left|\beta_{\omega}\right|^{2}$, the squared absolute values of the coefficients multiplying the dispersive modes in \eq{eq:incm} become close to each other, and both vanish linearly in $\omega$ for $\omega \to 0$.}
\item{Above $\omega_{\mathrm{min}}$, so long as the obstacle is sufficiently long that tunnelling effects are negligible, we expect only the flow properties in the downstream (white hole) region to be relevant, just as if the flow were transcritical. It is in this frequency domain that one could hope to obtain a close relationship with the Hawking predictions. For narrow obstacles, however, the behavior in this regime can be rather complicated.} 
\end{itemize}

In Appendix~\ref{3reg}, it can be seen that these three behaviors are clearly present when considering the effective temperature of \eq{eq:Teff} in the $(\kappa_{R},\kappa_{L})$-plane. Here, we shall look separately at the three scenarios delineated above, picking out the relevant parameters of the flow which determine the main behavior of the scattering coefficients in each case. 

\subsection{Transition near $\omega_{\rm min}$}

For $\omega > \omega_{\rm min}$, the characteristics for the left-moving incident mode $k_{\omega}^{\leftarrow}$ are blocked: there is a turning point they cannot pass, instead continuously evolving into right-moving characteristics of the outgoing dispersive mode $k_{\omega}^{\rightarrow,d}$~\cite{Michel:2014zsa}. An entirely analogous blocking occurs for the right-moving incident mode $k_{\omega}^{\rightarrow,d}$ from the left side, which continuously evolves into the left-moving outgoing mode $k_{\omega}^{\leftarrow}$.  By contrast, for $\omega < \omega_{\rm min}$, no such blocking occurs, and the characteristics of both modes traverse the obstacle.  There is thus a significant change in behavior at $\omega_{\rm min}$, quite independent of the analogue Hawking effect, involving only the scattering coefficients $\vert \alpha_{\omega} \vert ^{2}$ and $\vert \widetilde{A}_{\omega} \vert ^{2}$ of \eq{eq:incm}. We shall consider $\vert \widetilde{A}_{\omega} \vert ^{2}$, and define the dimensionless parameter
\begin{equation}
S \equiv - \left. \frac{d \vert \widetilde{A}_{\omega} \vert ^{2}}{d\left({\rm ln} \, \omega\right)} \right| _{\omega_{\rm min}} = -\omega_{\rm min} \left. \frac{d \vert \widetilde{A}_{\omega} \vert ^{2}}{d\omega} \right| _{\omega_{\rm min}} \,.
\label{eq:Sdefn}
\end{equation}

\begin{figure}
\includegraphics[width=0.49\columnwidth]{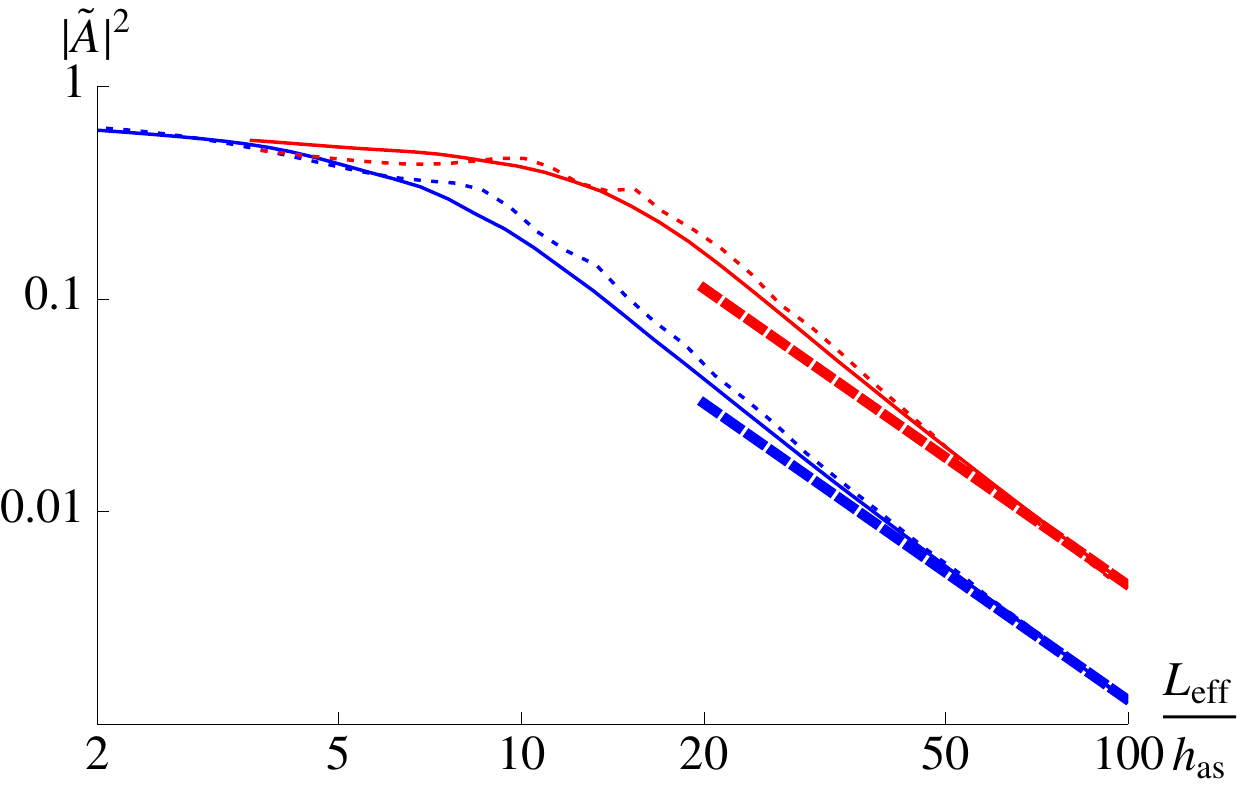} 
\includegraphics[width=0.49\columnwidth]{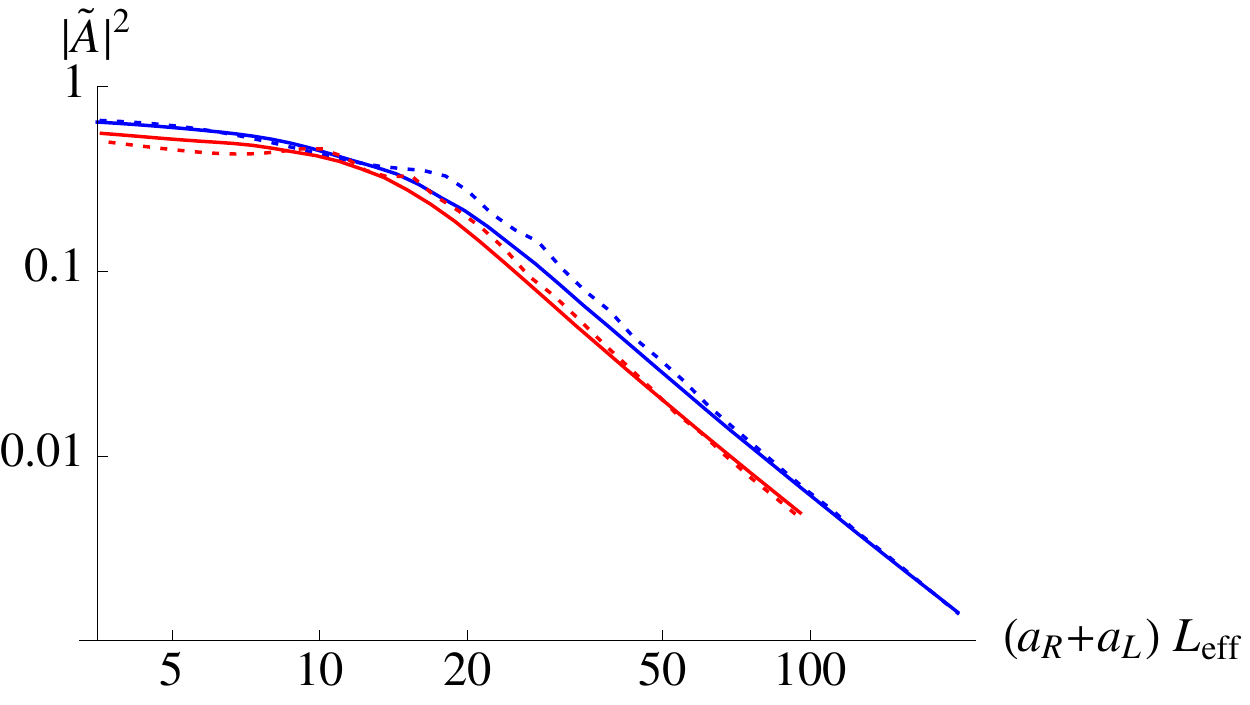}
\caption{Here is plotted the transmission coefficient $\vert \widetilde{A}_{\omega} \vert ^{2}$ at $\omega_{\rm min}$ for a variety of flows, in log-log scale as a function of the adimensionalised effective length $L_{\rm eff}/h_{\rm as}$. The different colors correspond to different slopes in the profile: blue represents $a_{R} h_{\rm as} = 0.5$ and $a_{L} h_{\rm as} = 1.6$ (close to those of the obstacle used in Refs.~\cite{Weinfurtner:2010nu,Euve:2014aga}), while red represents the symmetric obstacle with $a_{R} h_{\rm as} = a_{L} h_{\rm as} = 1.6$. The different styles of curve correspond to different values of $F_{\rm max}$: $0.6$ (solid) and $0.9$ (dotted).  The thick dashed curves indicate the behavior $\vert \widetilde{A}_{\omega} \vert^{2} \propto (L_{\rm eff}/h_{\rm as})^{-2}$. In the right plot, we attempt to account for the effect of the slopes by using $\left(a_{R}+a_{L}\right)\,L_{\rm eff}$ as a variable (rather than $L_{\rm eff}/h_{\rm as}$). It is clear that, unlike $L_{\rm eff}/h_{\rm as}$, the value of $F_{\rm max}$ does not play a crucial role. 
\label{fig:TransWmin}}
\end{figure}

In Figure \ref{fig:TransWmin} is shown the transmission coefficient $\vert \widetilde{A}_{\omega_{\rm min}} \vert ^{2}$ evaluated at $\omega_{\rm min}$ for a variety of flows, with particular emphasis on how it depends on the adimensionalised effective length $L_{\rm eff}/h_{\rm as}$ of the obstacle. For obstacles which are narrow enough, $\vert \widetilde{A}_{\omega_{\rm min}} \vert ^{2}$ is approximately constant and close to $0.5$, so that $\omega_{\mathrm{min}}$ marks the midpoint of the transition.  However, for longer obstacles, $\vert \widetilde{A}_{\omega_{\rm min}} \vert ^{2}$ scales as $(L_{\rm eff}/h_{\rm as})^{-2}$. We can make sense of this by noting that, in the limit where $L_{\mathrm{eff}}/h_{\rm as}$ becomes infinite and we are left with a single horizon, there can be no transmission at all for $\omega > \omega_{\rm min}$, so to maintain continuity of the scattering coefficients we must have $\vert \widetilde{A}_{\omega_{\rm min}} \vert ^{2}$ going to zero in this limit. Figure~\ref{fig:TransWmin} also indicates that the effects of the slope can be approximately accounted for by using $(a_{R}+a_{L})\,L_{\rm eff}$ as the variable rather than $L_{\rm eff}/h_{\rm as}$. (See Eq. (\ref{eq:f}) for the definition of $a_{R/L}$.) Finally, fig.~\ref{fig:TransWmin} shows that $F_{\rm max}$ has little bearing on $\vert \widetilde{A}_{\omega_{\rm min}} \vert ^{2}$, as was observed in Ref.~\cite{Euve:2014aga}. 

\begin{figure} 
\includegraphics[width=0.45\columnwidth]{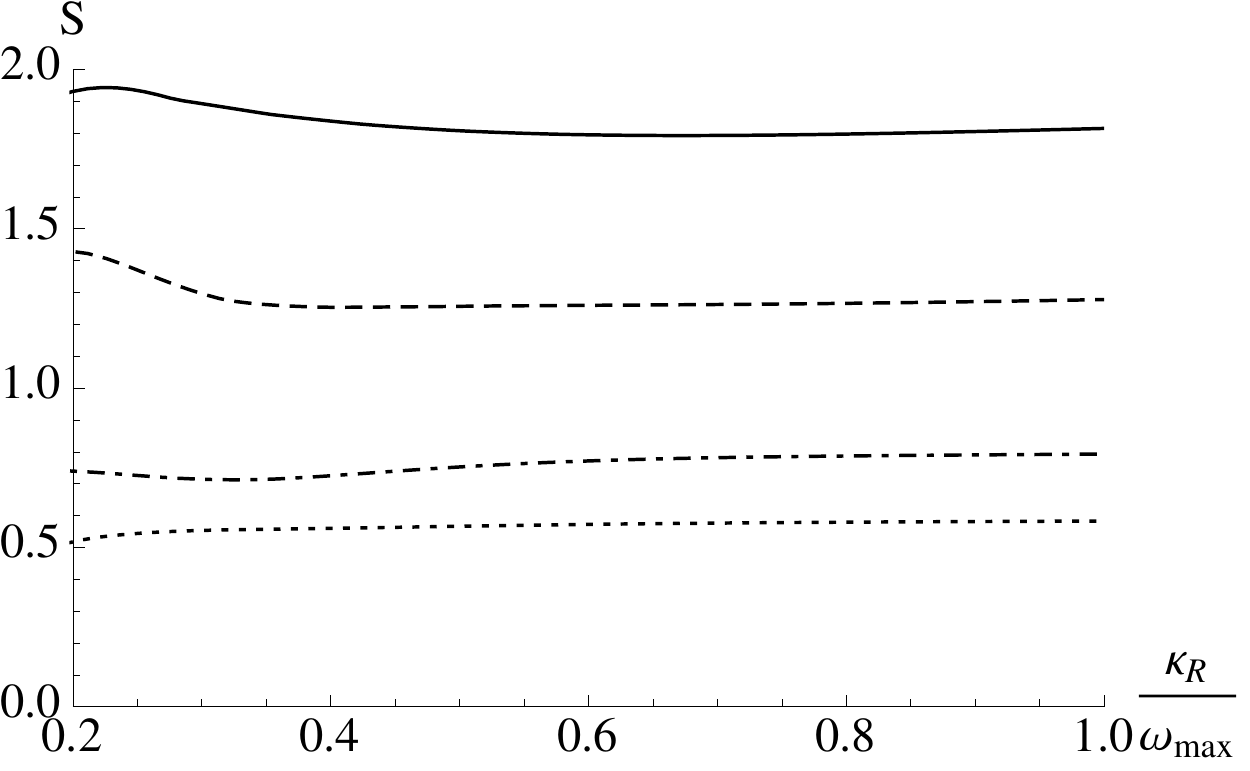} \, \includegraphics[width=0.45\columnwidth]{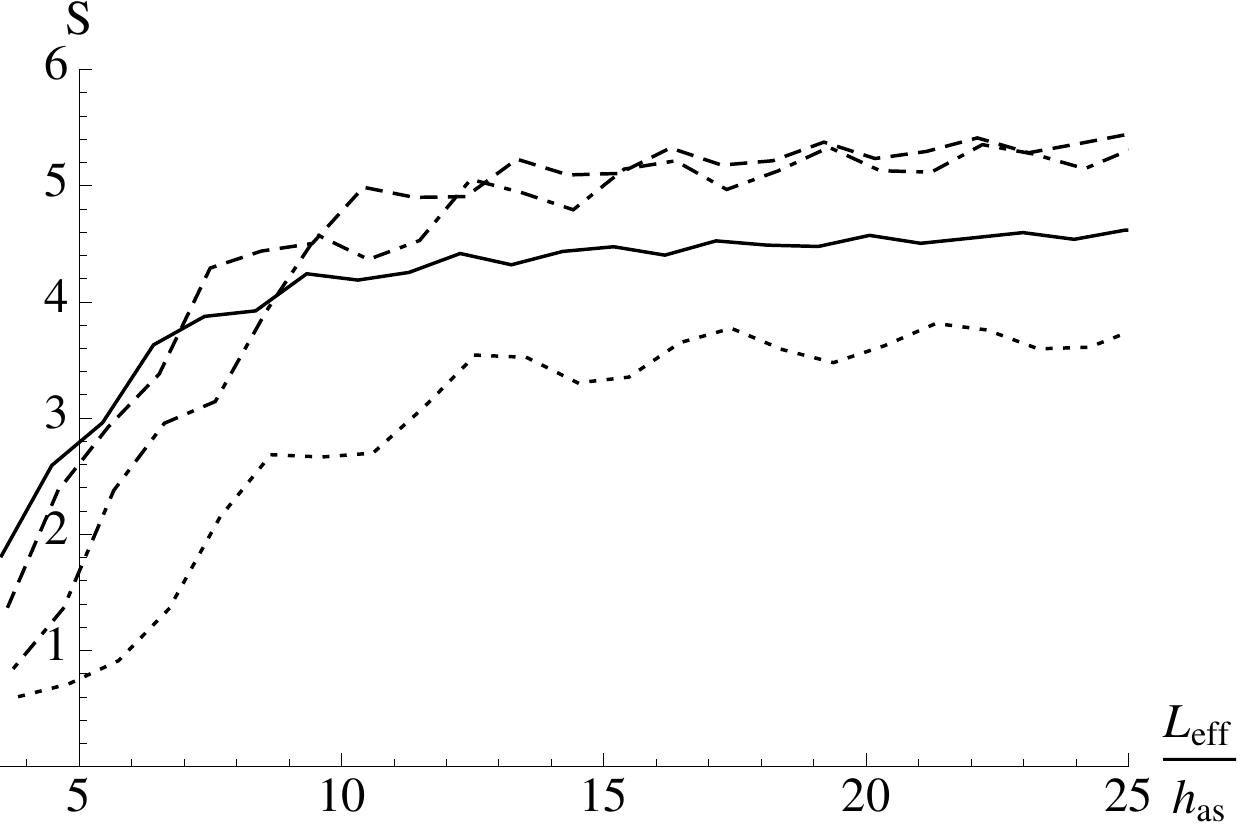}
\caption{On the left is plotted $S$ of Eq. (\ref{eq:Sdefn}) as a function of the downstream slope $\kappa_{R}$, defined in Eq. (\ref{eq:kappaeff}). We have taken $F_{\mathrm{as}}=0.16$, $L_{\rm eff}/h_{\rm as} = 3.5$, and $\kappa_{L}/\omega_{\rm max} = 0.4$. Meanwhile, on the right is plotted $S$ as a function of the adimensionalised effective length $L_{\rm eff}/h_{\rm as}$, with $F_{\mathrm{as}}=0.16$, $\kappa_{L}/\omega_{\rm max}=0.7$ and $\kappa_{R}/\omega_{\rm max}=0.4$. The different curves correspond to different values of $F_{\mathrm{max}}$: $0.6$ (solid), $0.7$ (dashed), $0.8$ (dot-dashed) and $0.9$ (dotted).
\label{fig:S_kappaR}}
\end{figure}

In Figure \ref{fig:S_kappaR} are shown plots of the adimensionalised derivative $S$ of Eq. (\ref{eq:Sdefn}) for several different values of $F_{\mathrm{max}}$. In the left panel, $\kappa_{R}$ is varied while $F_{\mathrm{as}}$ is fixed at $0.16$, the adimensionalized effective length $L_{\mathrm{eff}}/h_{\rm as}$ is fixed at $3.5$, and $\kappa_{L}/\omega_{\rm max}$ is fixed at $0.4$. We see that, while there is a dependence on the slope $\kappa_{R}$, this is not as important as the dependence on $F_{\mathrm{max}}$, with $S$ being systematically reduced as $F_{\mathrm{max}}$ is increased. In the right panel, $L_{\rm eff}$ is varied while $F_{\rm as}=0.16$, $\kappa_{R}/\omega_{\rm max}=0.4$ and $\kappa_{L}/\omega_{\rm max}=0.7$. We see there that $L_{\rm eff}/h_{\rm as}$ is an important quantity in determining $S$ when both are relatively small.  Indeed, when increasing $L_{\rm eff}/h_{\rm as}$ from $4$ to $15$, $S$ is seen to increase by a factor of between $2$ and $5$, depending on the value of $F_{\mathrm{max}}$.  At large $L_{\rm eff}/h_{\rm as}$, however, $S$ shows only small oscillations around some $F_{\mathrm{max}}$-dependent limiting value. Notice also that, unlike at small $L_{\rm eff}/h_{\rm as}$, the dependence on $F_{\mathrm{max}}$ is non-monotonic at large $L_{\rm eff}/h_{\rm as}$. 

There is a clear lesson here in the case of relatively narrow obstacles (i.e. $L_{\rm eff}/h_{\rm as} \lesssim 5$).  According to Figure \ref{fig:TransWmin}, the critical frequency $\omega_{\rm min}$ corresponds more or less to the midpoint of the transition region, and hence $S$ (which is defined at $\omega_{\rm min}$) serves as a good indication of the sharpness of the transition.  Turning to Figure \ref{fig:S_kappaR}, we find that in this regime, the sharpness of the transition increases with increasing $L_{\rm eff}$ and decreases with increasing $F_{\rm max}$, while it is essentially independent of $\kappa_{R}$.  The dependence on $L_{\rm eff}$ is particularly intuitive: the narrower the obstacle, the higher will be the rate of tunnelling across it, and so we need higher frequencies with more rapidly decaying evanescent modes in order to find a mode which is truly blocked.  It is less clear how to interpret the results for large $L_{\rm eff}/h_{\rm as}$, for then $S$ is no longer measured at the midpoint of the transition.  

\subsection{Low-frequency regime} 

For $\omega < \omega_{\mathrm{min}}$, there are no turning points according to geometrical optics, so that the incident wave is essentially transmitted, i.e., $|\widetilde{A}_{\omega}|^{2} \approx 1$. This is clearly seen in the top left panel of Figure \ref{fig:16_sub}. Furthermore, the same panel reveals that $|\alpha_{\omega}|^{2} \approx |\beta_{\omega}|^{2} \sim \omega/\sigma_{\beta}$ for $\omega \to 0$, in accordance with Eq. (\ref{eq:sigma_beta_defn}).

To characterize the zero-frequency limit, we study how the frequency $\sigma_{\beta}$ depends on the flow parameters. 
\begin{figure}
\begin{center} 
\includegraphics[width=0.45\columnwidth]{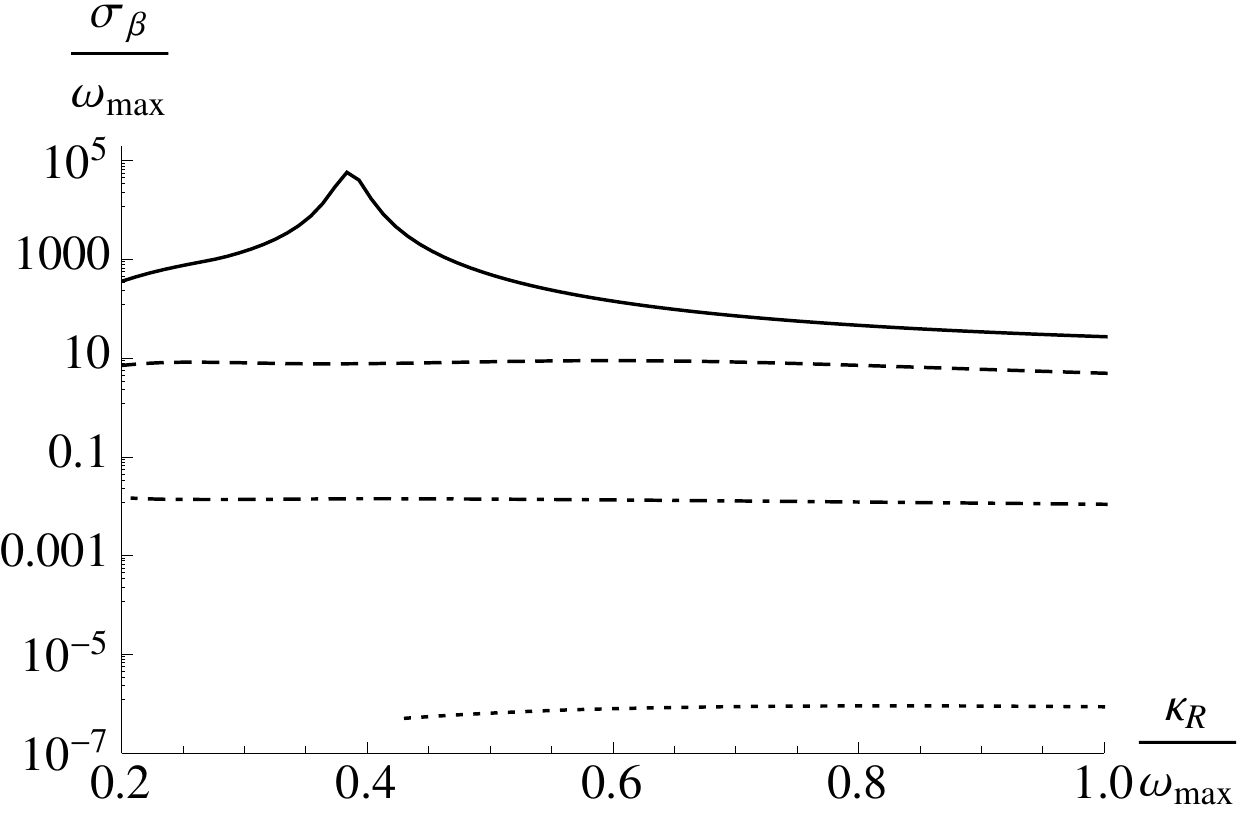} 
\end{center}
\caption{Here is plotted $\sigma_{\beta}/\omega_{\mathrm{max}}$, see Eq. (\ref{eq:sigma_beta_defn}), as a function of the downstream slope $\kappa_{R}$ adimensionalized by $\omega_{\mathrm{max}}$. We have fixed $F_{\mathrm{as}}=0.16$, $L_{\rm eff}/h_{\rm as}=3.5$ and $\kappa_{L}/\omega_{\rm max} = 0.4$. The various curves correspond to different values of $F_{\mathrm{max}}$: the subcritical cases $F_{\mathrm{max}}=0.6$ (solid) and $0.8$ (dashed), the critical case $F_{\mathrm{max}}=1$ (dot-dashed) and the transcritical case $F_{\mathrm{max}}=1.2$ (dotted). Note that the latter curve does not extend below $\kappa_{R}/\omega_{\rm max} \approx 0.4$, since this is the lowest value compatible with the fixed values of $\kappa_{L}$ and $L_{\rm eff}$. It is clear that the slope $\kappa_{R}$ plays a much weaker role than the value of $F_{\mathrm{max}}$. 
\label{fig:sb_kappaR} }
\end{figure}
\begin{figure}
\begin{center} 
\includegraphics[width=0.5\columnwidth]{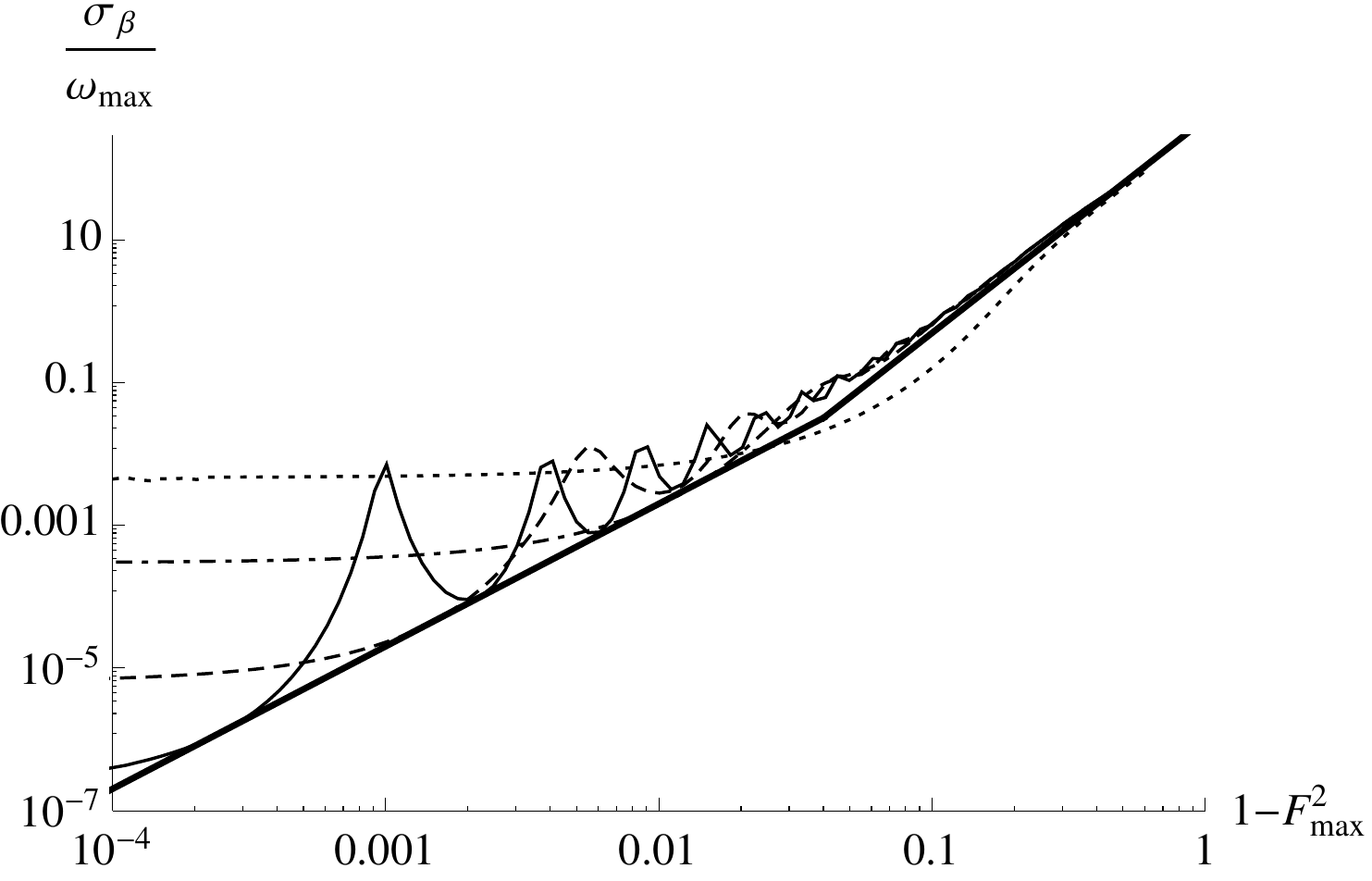} 
\end{center}
\caption{Here is plotted $\sigma_{\beta}/\omega_{\mathrm{max}}$ as a function of $1-F_{\mathrm{max}}^{2}$ on a logarithmic scale.  The parameters $a_{R} h_{\rm as}$ and $a_{L} h_{\rm as}$ are fixed at $0.5$ and $1.6$, respectively. The various curves correspond to different values of $L/h_{\rm as}$: $5$ (dotted), $10$ (dot-dashed), $20$ (dashed) and $40$ (solid). The thick line shows the limiting behaviour: for large $1-F_{\mathrm{max}}^{2}$ it is proportional to $\left(1-F_{\mathrm{max}}^{2}\right)^{3}$ and is seen to coincide with the $\sigma_{\beta}$ curves of larger $L$ in this regime, whereas for small $1-F_{\mathrm{max}}^{2}$ it is proportional to $\left(1-F_{\mathrm{max}}^{2}\right)^{2}$ and follows the bottom of the oscillations in $\sigma_{\beta}$. 
\label{fig:sbFmax}}
\end{figure}
In Figure \ref{fig:sb_kappaR} is plotted $\sigma_{\beta}$ as a function of the downstream slope $\kappa_{R}$, with fixed values of $F_{\mathrm{as}}=0.16$, $L_{\mathrm{eff}}/h_{\rm as} = 3.5$ and $\kappa_{L}/\omega_{\rm max}=0.4$. The various curves correspond to different values of $F_{\mathrm{max}}$, which we allow to vary from a subcritical to a supercritical value.  We note that, although there is a clear dependence on the slope $\kappa_{R}$, this is subdominant with respect to the dependence on $F_{\mathrm{max}}$, whose effect is much greater.~\footnote{An exception to this is the peak in the $F_{\mathrm{max}}=0.6$ curve centred around $\kappa_{R} \approx 0.4\,\omega_{\mathrm{max}} = \kappa_{L}$.  This is a resonant behavior in $\sigma_{\beta}$ due to the symmetry of the flow profile.} The rapid decrease of $\sigma_{\beta}$ with increasing $F_{\mathrm{max}}$ can be understood from the results presented in Figure \ref{fig:16_trans}: when the flow is transcritical, the scattering coefficients $\vert \alpha_{\omega} \vert^{2}$ and $\vert \beta_{\omega} \vert^{2}$ first increase as $1/\omega$ in some interval, in stark contrast to the linear behavior seen in the subcritical case.  Interpolating between these two different behaviors requires that $\sigma_{\beta}$ decrease when increasing $F_{\mathrm{max}}$, and indeed the window of validity of the linear behavior of \eq{eq:sigma_beta_defn} must shrink accordingly.  It does not vanish when $F_{\mathrm{max}}$ reaches 1, however; we recall from Figure \ref{fig:16_trans} that there exists an ultra-low frequency regime where tunnelling across the obstacle is significant, and where $\vert \beta_{\omega} \vert^{2} \approx \omega/\sigma_{\beta}$ even for transcritical flows.  This allows $\sigma_{\beta}$ to be well-defined even when $F_{\mathrm{max}} > 1$.

To further investigate the behavior of $\sigma_{\beta}$ with $F_{\mathrm{max}}$ as the latter approaches $1$, we fixed the values of $a_{R} h_{\rm as}$ and $a_{L} h_{\rm as}$ at $0.5$ and $1.6$, respectively, and plotted $\sigma_{\beta}$ for varying $F_{\mathrm{max}}$ and $L$.  The results are shown in Figure \ref{fig:sbFmax}.  Firstly, we notice that $\sigma_{\beta}$ does not vanish as $F_{\mathrm{max}} \rightarrow 1$ but approaches a finite value, which decreases with increasing $L$.  In this regime, $L$ has taken over as the relevant parameter.  Secondly, there is an interesting change of behaviour at $1-F_{\mathrm{max}}^{2} \sim 0.1$, a changeover point which is seemingly independent of $L$.  For $1-F_{\mathrm{max}}^{2}$ larger than this value, the curves converge to one which is proportional to $\left(1-F_{\mathrm{max}}^{2}\right)^{3} \propto\omega_{\mathrm{min}}^{2}$, with only the curve for the smallest value of $L$ showing significant deviations from the others.  In this regime, then, and so long as $L$ is not too small, $\omega_{\mathrm{min}}$ is the only relevant parameter in determining $\sigma_{\beta}$.  Finally, we note that there is also an intermediate regime where both $L$ and $\omega_{\rm min}$ are relevant parameters. In this third regime, there are significant oscillations in $\sigma_{\beta}$ with a period that depends on $L$. Interestingly, the troughs of these oscillations all follow a curve which is proportional to $\left(1-F_{\mathrm{max}}^{2}\right)^{2}$ (or $\omega_{\mathrm{min}}^{4/3}$, according to Eq. (\ref{eq:ommin})).~\footnote{While completing our numerical analysis, we became aware of~\cite{Coutant16} where the low-frequency regime is investigated in analytical terms.  We performed a few extra simulations which indicate good agreement with numerical integration of their Eq.~(B10). On the other hand, it is presently unclear to us if the various behaviors displayed in our Fig. \ref{fig:sbFmax} can be recovered from their Eq. (B14).  
We are thankful to Antonin Coutant for explanations about the expected validity domain of the equations of~\cite{Coutant16}.
}  

From an experimental perspective, however, it is quite unlikely for $F_{\rm max}$ to be so close to $1$ that we find ourselves in the region of Figure \ref{fig:sbFmax} where $L$ plays a significant role.  Generally speaking, then, and up to the possibility of resonant effects, $F_{\rm max}$ is by far the most relevant quantity in the determination of $\sigma_{\beta}$, the latter decreasing rapidly as $F_{\rm max}$ approaches $1$.  Sufficiently narrow obstacles constitute an exception, as we can begin to see from the $L/h_{\rm as}=5$ (dotted) curve of Fig. \ref{fig:sbFmax}.  But this effect is subdominant relative to the dependence on $F_{\rm max}$. 

\subsection{High-frequency regime}

It turns out that the high-frequency regime is the most complicated to describe.  For, while we might naively expect the spectrum here to be approximately thermal (since the wave is blocked much as in the transcritical case), it appears that this is only sometimes true. 
As we shall see, the difficulties come in part from the residual transmission across the obstacle. To get a flavor of the behavior in this regime, we shall study here the spectrum on a series of flows obtained by fixing one of the two slopes and letting the other vary. We shall examine the behavior of both $T_{\omega}^{\rm eff}$, the effective temperature at the mid-point of the high-frequency regime, i.e., at $\omega = (\omega_{\rm min} + \omega_{\rm max})/2$, and of its derivative $dT_{\omega}^{\rm eff}/d\omega$ evaluated at the same frequency. The latter quantity is very important in that it quantifies (at least approximately) the variation of the effective temperature, and thus the Planckianity of the spectrum, see \eq{eq:Teff}. 
 
\begin{figure}
\includegraphics[width=0.45\columnwidth]{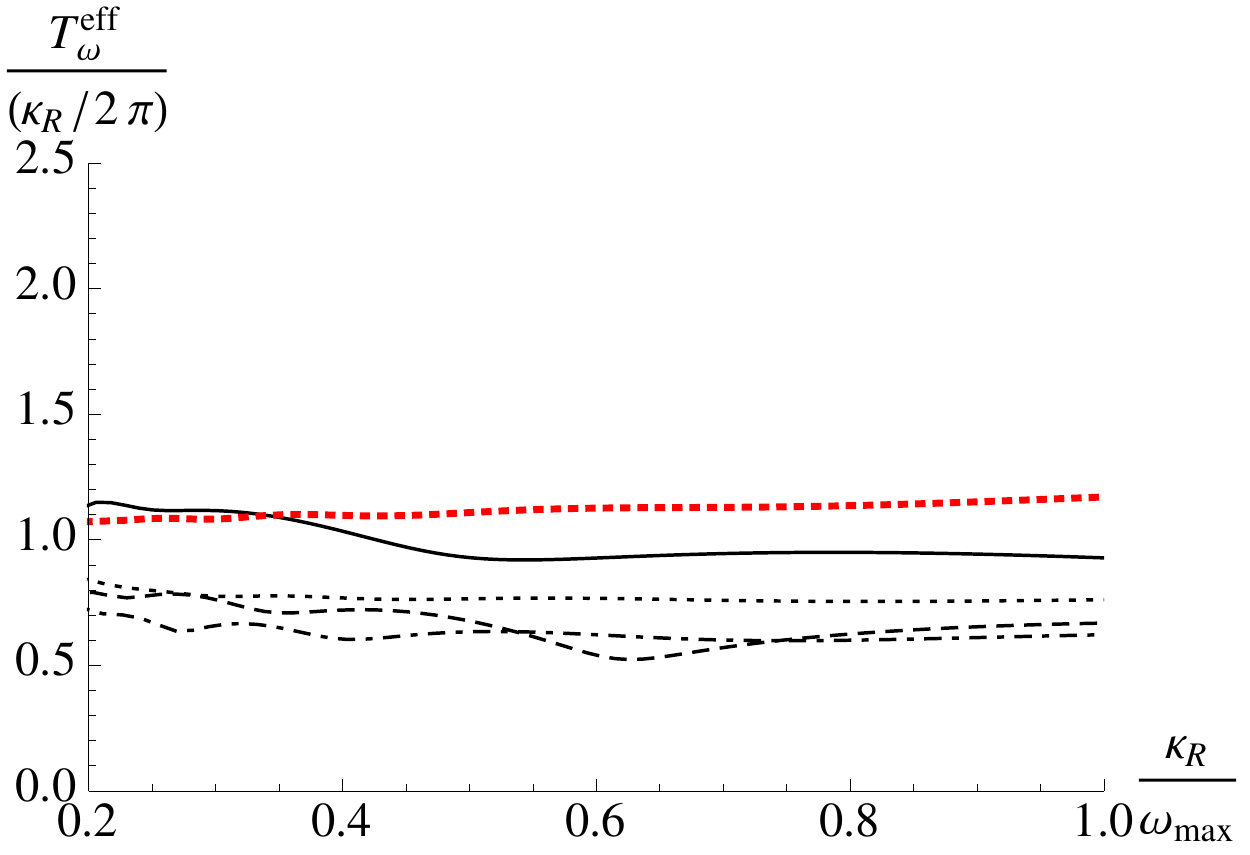} \, \includegraphics[width=0.45\columnwidth]{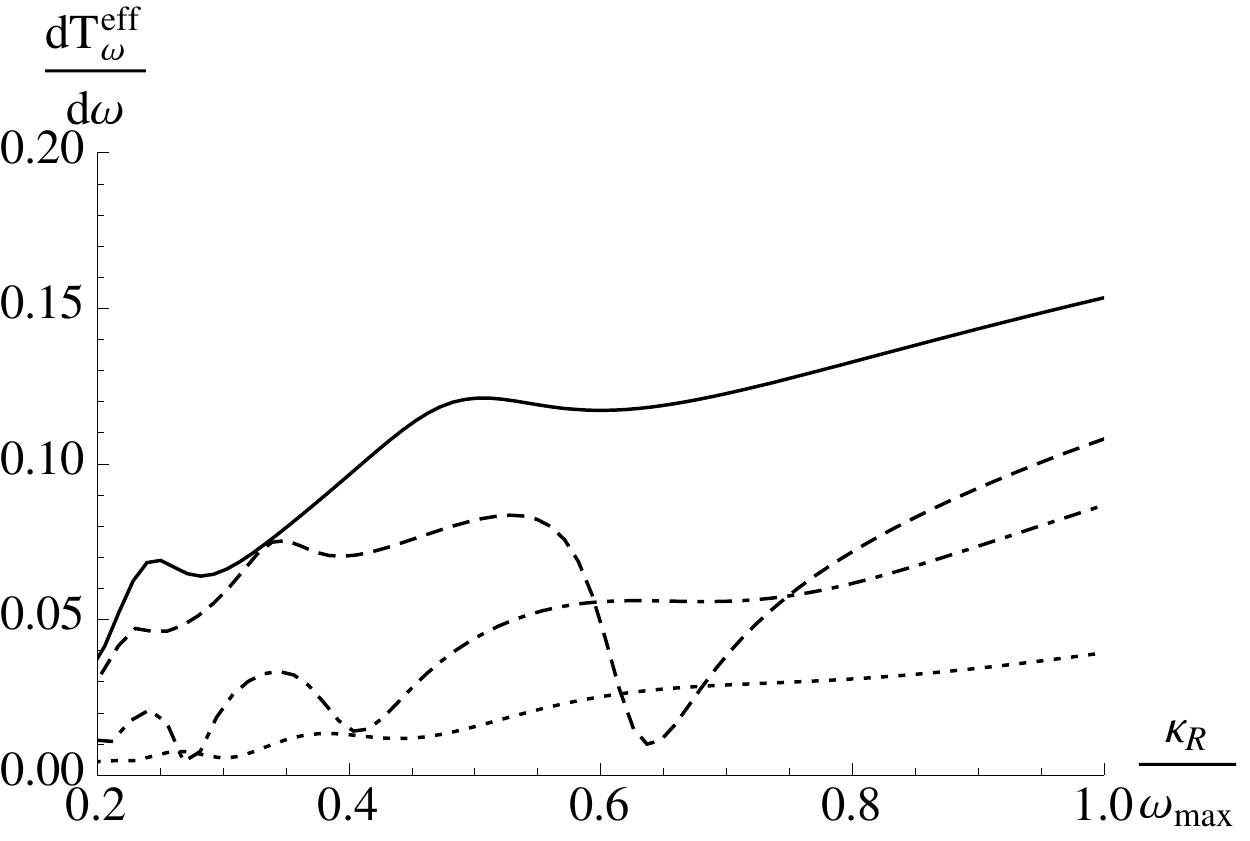} \\
\includegraphics[width=0.45\columnwidth]{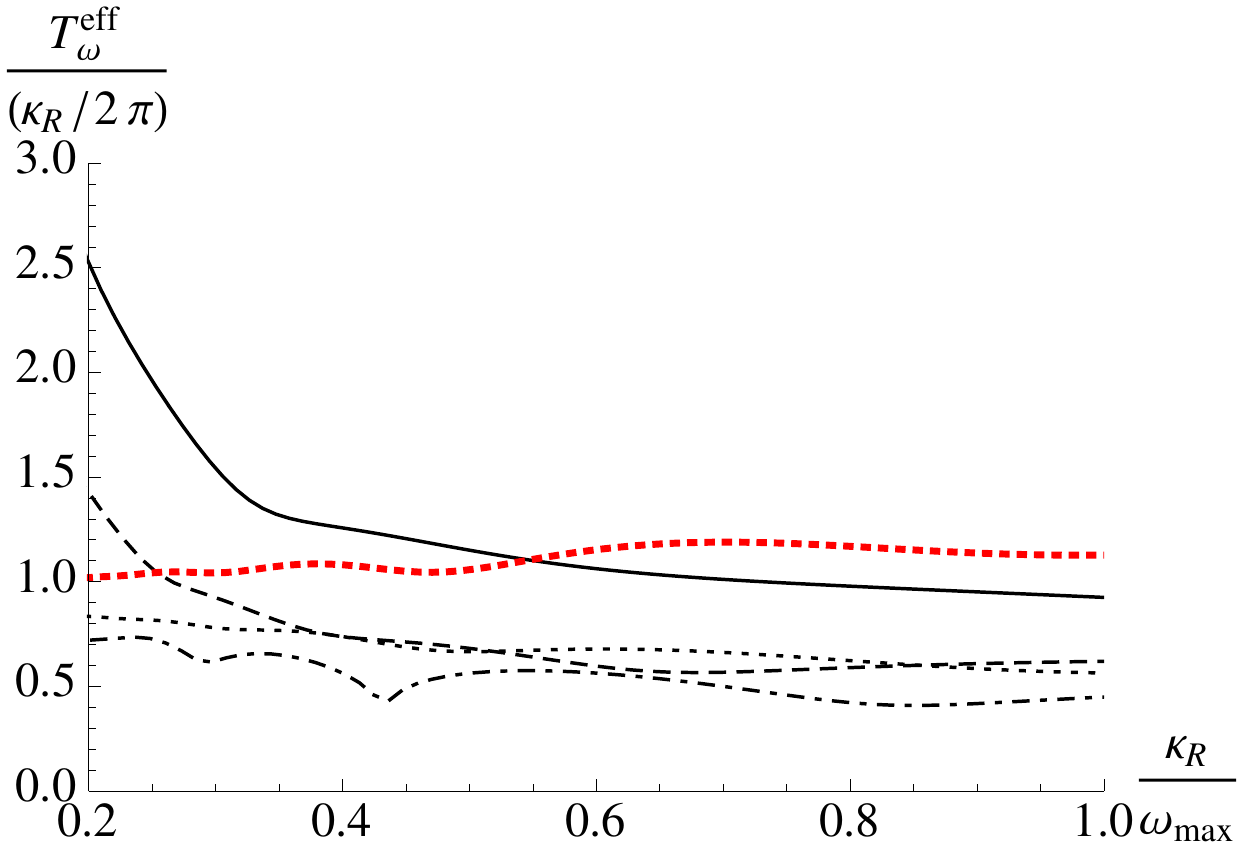} \, \includegraphics[width=0.45\columnwidth]{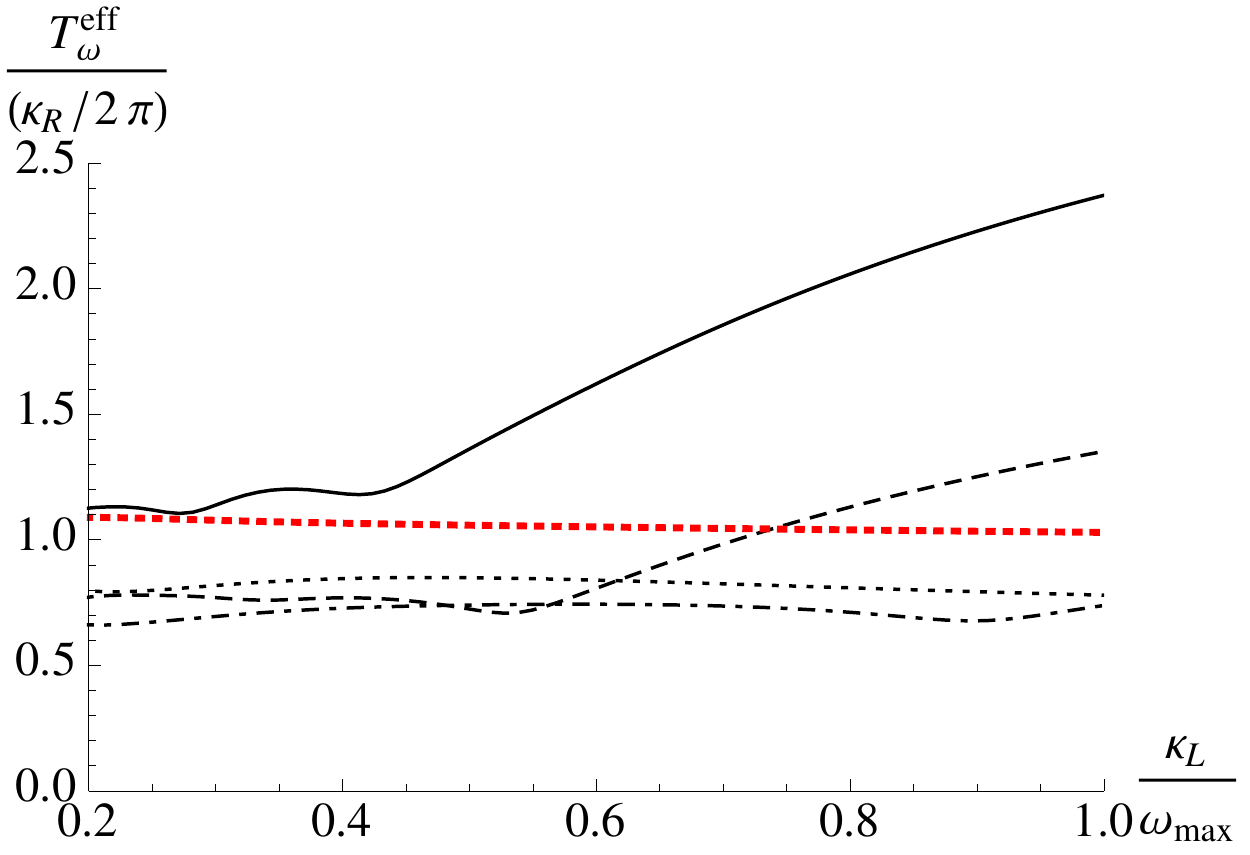}
\caption{Here are shown several plots relating to the effective temperature $T_{\omega}^{\rm eff}$ at the midpoint $(\omega_{\rm min}+\omega_{\rm max})/2$ of the high-frequency regime. In the top row, we have fixed $F_{\rm as}=0.16$, $L/h_{\rm as} = 2.5$ and $\kappa_{L}/\omega_{\rm max} = 0.25$, whilst allowing the downstream slope $\kappa_{R}$ to vary.  The various styles of curve correspond to different values of $F_{\mathrm{max}}$: $0.6$ (solid), $0.8$ (dashed), $1.0$ (dot-dashed) and $1.2$ (dotted). The upper left plot shows $T_\om^{\rm eff}$ normalized by $\kappa_{R}/2\pi$, except for the dotted red curve, which shows the temperature of the transcritical flow ($F_{\rm max} = 1.2$) normalized instead by $T_H = \kappa/2\pi$ evaluated at the horizon (so this curve being equal to $1$ is precisely the Hawking prediction). The upper right plot shows the derivative of $T_{\omega}^{\rm eff}$ with respect to frequency, thus giving an indication of the deviations from the thermality of the spectrum. In the lower left plot, we use the same parameters as in the top row except that the value of the upstream slope $\kappa_{L}/\omega_{\rm max}$ has been increased to $0.75$, so that it is the dominant slope for most of the curve.  We thus see significant deviations for the subcritical flows when $\kappa_{R}$ is sufficiently small.  In the lower right plot, we instead fix the downstream slope $\kappa_{R}/\omega_{\rm max} = 0.25$ and vary $\kappa_{L}$, and once again the subcritical flows show strong deviations when $\kappa_{L}$ is sufficiently larger than $\kappa_{R}$.
\label{fig:Teff}}
\end{figure}

Illustrative examples of these two quantities are shown in Figure \ref{fig:Teff}. Note that $T_{\omega}^{\rm eff}$ has been normalized by $\kappa_{R}/2\pi$, a generalized version of the Hawking temperature, so that what is plotted in all but the upper right plot is in effect the ``Hawkingness'' of the spectrum at the midpoint frequency. In the top row, $\kappa_{L}/\omega_{\rm max}$ is held fixed at $0.25$ and $\kappa_{R}$ is varied, so that the flow has a small upstream slope. The normalized effective temperature is shown on the left and the derivative of the temperature is shown on the right. In the bottom row, the normalized effective temperature is shown for two 
series of flows which exhibit significant deviations from the Hawking-like prediction. In all plots, the parameter $L/h_{\rm as}$ is held fixed at $2.5$, a value close to that of the obstacle used in the Vancouver experiment~\cite{Weinfurtner:2010nu} and which allows the upstream slope to affect the scattering.\footnote{Note that it is $L$ rather than $L_{\rm eff}$ that is held fixed here, since holding both $L_{\rm eff}$ and one of the slopes at a small value can force the remaining slope to be large.  We thus allow $L_{\rm eff}$ to vary a bit, though we expect this variation to have a subdominant effect on the temperature.} The variously styled curves correspond to different values of $F_{\mathrm{max}}$, ranging from $0.6$ to $1.2$ and hence crossing the criticality condition.

When examining the upper left plot, we first note that, independently of the value of $F_{\mathrm{max}}$,  $\kappa_{R}/2\pi$ can generally be said to give a good indication of the effective temperature. Indeed, for all values of $F_{\mathrm{max}}$, the ratio $2 \pi T_{\omega}^{\rm eff}/\kappa_{R}$ is of order 1, and stays approximately constant when $\kappa_{R}$ is multiplied by a factor of $5$.~\footnote{It should also be noticed that, when the flow is sufficiently subcritical (i.e. $F_{\rm max} \lesssim 0.8$), increasing $F_{\mathrm{max}}$ slightly {\it decreases} the effective temperature. Comparing with Figure \ref{fig:2temp}, we see that this is indeed possible at the upper end of the high-frequency regime, but it should be noted that this depends on the choice of frequency at which $T_{\omega}^{\rm eff}$ is calculated, and that had we chosen a frequency significantly lower than $(\omega_{\rm min}+\omega_{\rm max})/2$ we may well have observed the opposite behavior.}  Considering the upper right plot which gives the derivative of $T_{\omega}^{\rm eff}$ with respect to frequency for the same flows, we see that this derivative is always positive, and that it has the clear tendency to increase when decreasing $F_{\mathrm{max}}$. (Only the transcritical flows display a small derivative which is less than $0.03$ for the series here considered.) This indicates that the spectrum in subcritical flows does not follow a Planck law, even approximately. This is in agreement with \cite{Finazzi:2011jd,Leonhardt2012,Michel:2014zsa,Michel:2015aga}, where a temperature increasing with $\omega$ was observed for flows which are not symmetric with respect to the position of the horizon. So, while the effective temperature {\it at any one frequency} is Hawking-like in being approximately proportional to $\kappa_{R}$, the constant of proportionality varies with $\omega$ so that the spectrum as a whole is not a thermal one. 

Consider now the lower panels of Figure \ref{fig:Teff}. In the lower left plot, $\kappa_{L}/\omega_{\rm max}$ is increased to $0.75$, while in the lower right plot, it is $\kappa_{R}/\omega_{\rm max}$ that is fixed at $0.25$ while $\kappa_{L}$ is varied. As expected, we verify that for the critical and transcritical flows $T_{\omega}^{\rm eff}$ remains largely unaffected by $\kappa_{L}$.  We also see that, for the transcritical flows, the good agreement between $T_{\omega}^{\rm eff}$ and $\kappa_R/2\pi$ is well maintained (within $20\%$ relative deviations here).  When considering the subcritical flows, we notice that $T_{\omega}^{\rm eff}$ significantly increases when $\kappa_{L}$ becomes significantly larger than $\kappa_{R}$. This must be due to the residual transmission across the obstacle: although the incoming waves are essentially blocked for $\omega > \omega_{\mathrm{min}}$, there is an evanescent wave on the left of the turning point $x_r(\omega)$ which ``probes'' the gradient on the upstream slope. We thus conjecture that the contribution to $T_{\om}^{\rm eff}$ coming from the upstream slope should be suppressed by the damping factor 
\begin{equation}
D_{L} = \exp \lp - \int_{x_l}^{x_r} dx \, \vert \Im(k^d_\omega(x')) \vert dx' \rp \,,  
\label{eq:damping_factor}
\end{equation}
where $\Im(k^d_\omega(x)) < 0$ is the imaginary part of the complex wave vector of the mode decaying to the left of the downstream turning point $x_r$, and where $x_l$ is the would-be turning point on the upstream side.  (For sufficiently long obstacles, which is the regime of interest to us, the integral can be approximated by $\vert \Im(k^d_\omega(0)) \vert L $.) 

\begin{figure}
\begin{center}
\includegraphics[scale=0.7]{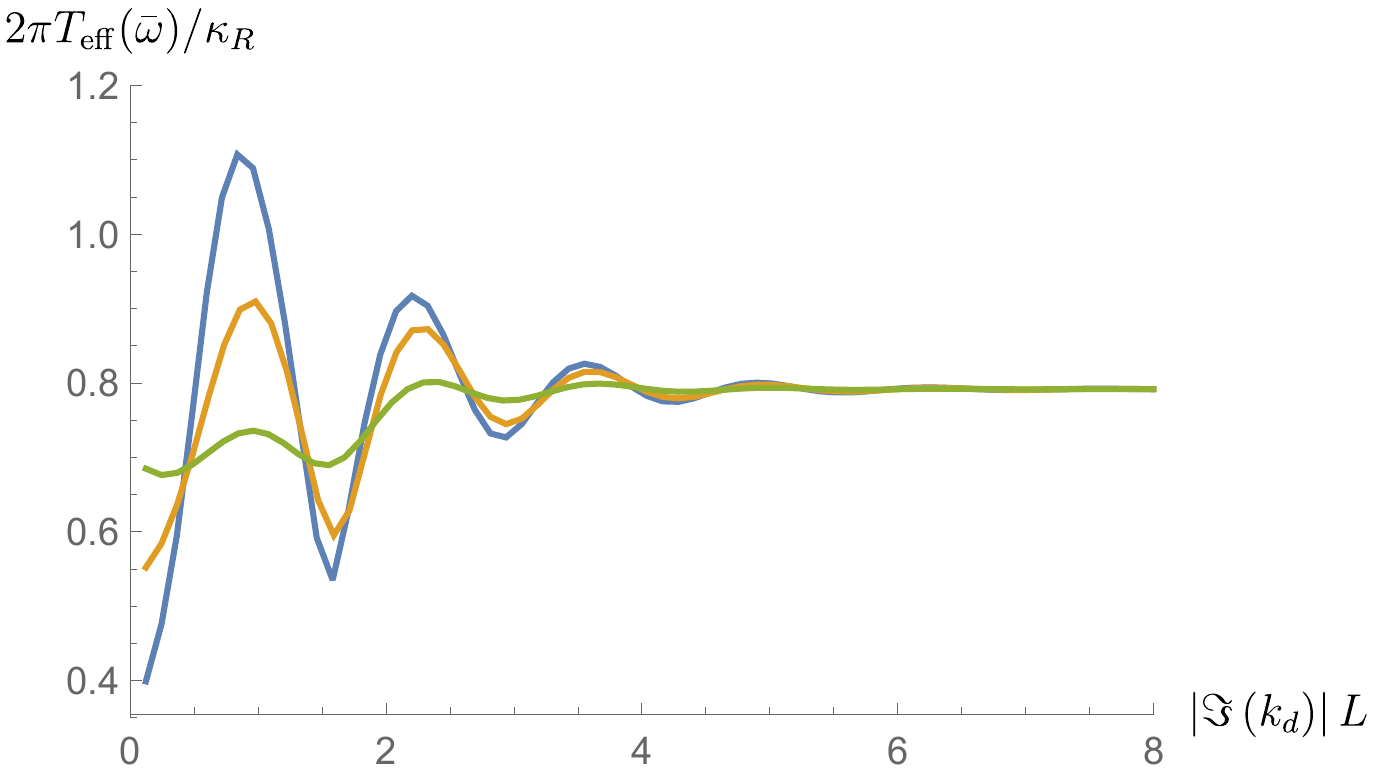}
\caption{The adimensionalized effective temperature $2 \pi T_{\rm eff} / \kappa_R$ evaluated at $\bar{\om} = (\om_{\rm min} + \om_{\rm max}) / 2$ as a function of $L$ for three different upstream slopes. The common parameters of the flows are $F_{\rm as} = 0.4$, $F_{\rm max} = 0.8$, and $a_R = 2 h_{\rm as}$. $a_L / h_{\rm as}$ takes the values $1$ (green curve), $2$ (orange), and $4$ (blue). We notice that the amplitude of the oscillations increases with $a_L$, while they are exponentially damped for increasing $\vert \Im(k_\om^d) \vert L$.
} 
\label{fig:Le}
\end{center}
\end{figure}

The conjecture is confirmed by results shown in \fig{fig:Le}, where we represent $T_\om^{\rm eff}/(\kappa_R/2\pi)$ for 3 different values of the upstream slope $a_L$ while holding fixed the downstream slope $a_R$. Considering first the case with the lowest value of $a_L$, and ignoring the small oscillations, we notice that there is a minimum length ($\vert \Im(k_d) \vert L \approx 2.5$) at which the effective temperature becomes essentially $L$-independent.  This can be understood from the fact that, for $L \lesssim 1/a_R$, the length affects the typical gradient of the obstacle, as was discussed in Sec.~\ref{Sts}. When $a_L \geq a_R$, a larger value of $L$ is required for the oscillations engendered by the upstream slope to be significantly reduced. This larger value of $L$ is such that $\vert \Im(k^d_\om(0)) \vert L \approx 5$, so that the reduction factor of the evanescent wave on reaching the upstream slope is around $e^{-5} \approx 0.01$.  Analyzing further the various curves, we verified that the differences in $T_{\omega}^{\rm eff}$ due to changes of $a_L$ (and thus $\kappa_L$) are proportional to $\exp \lp -\vert \Im(k^d_\omega(0)) \vert L \rp$.  Importantly we also verified that this remains true when considering frequencies other than $(\omega_{\rm min} + \omega_{\rm max})/2$.  For all the frequencies we probed, the difference in $T_{\omega}^{\rm eff}$ becomes insignificant when $\vert \Im(k^d_\omega(0)) \vert L \gtrsim 4$. 

\begin{figure}
\begin{center}
\includegraphics[scale=0.7]{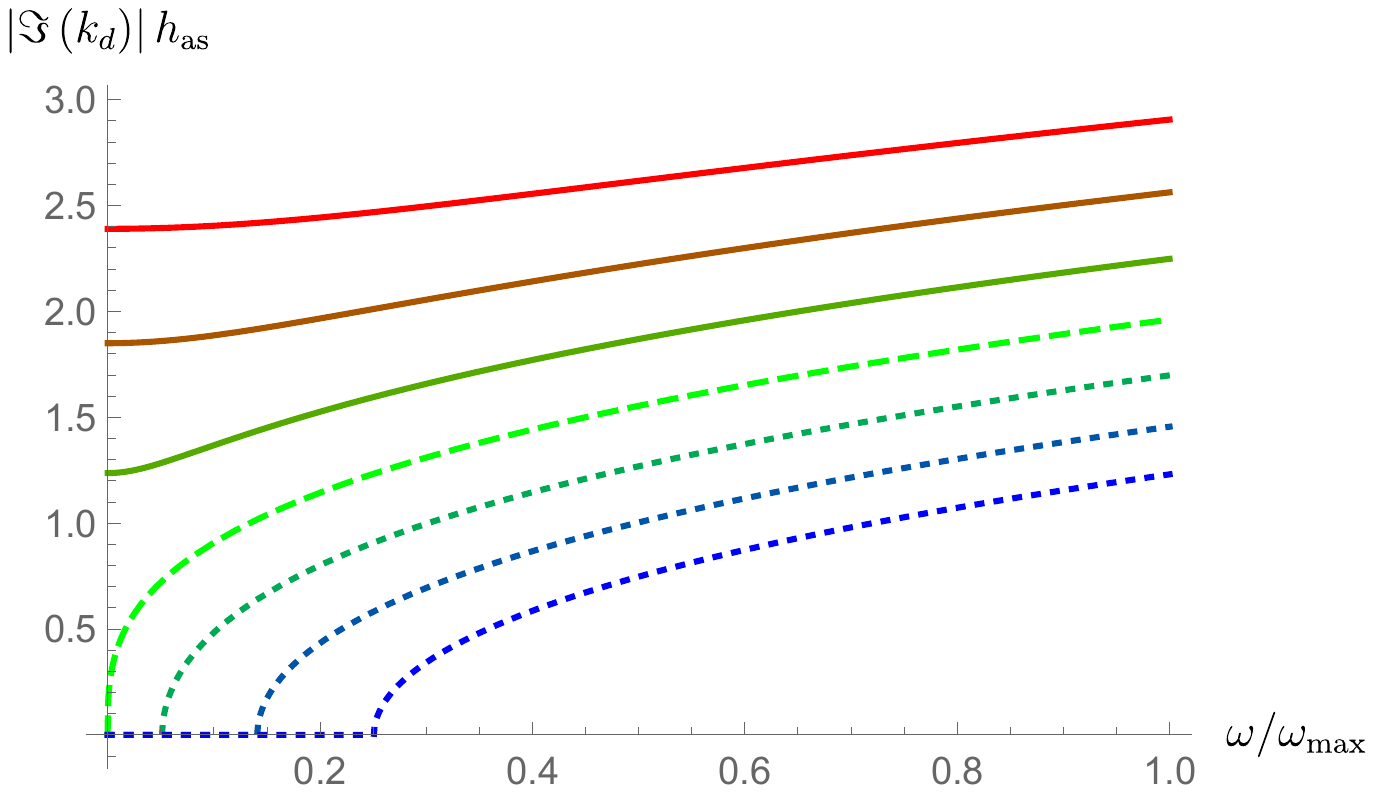}
\caption{Here is plotted $\vert \Im(k_\om^d(0)) \vert h_{\rm as}$, the imaginary part of the evanescent wavevector at $F=F_{\rm max}$. The various curves correspond to the flows used in Figs.~\ref{fig:flow4} and~\ref{fig:2temp}, i.e., $F_{\rm max}$ takes on $7$ equally spaced values between $0.8$ (blue curve) and $1.2$ (red curve) while $F_{\rm min}$ is held fixed. We see that the critical flow (dashed curve) clearly separates the transcritical flows for which $\vert \Im(k_\om^d) \vert$ has a finite value at $\om = 0$ from the subcritical flows where $\vert \Im(k_\om^d) \vert$ differs from zero only for $\om > \om_{\rm min}$.}
\label{fig:Imkd}
\end{center}
\end{figure}

To complete the analysis, in \fig{fig:Imkd} we plot how $\vert \Im(k^d_\omega(0)) \vert h_{\rm as}$ varies with $\om/\om_{\rm max}$ for the 7 flows considered in Figures \ref{fig:flow4} and \ref{fig:2temp}. Considering a fixed frequency, it is clear that there is a steady increase of $\vert \Im(k_\om^d(0)) \vert$ with increasing $F_{\rm max}$. We also note that $\vert \Im(k_\om^d(0)) \vert$ is fixed at zero for $\omega < \omega_{\rm min}$ in subcritical flows, and only begins to increase once $F_{\rm max}$ has reached a value at which $\om=\om_{\rm min}$. For a given value of $\vert \Im(k_\om^d(0)) \vert h_{\rm as}$, there is a frequency window in the high-$\om$ part of the spectrum where this value is exceeded, and the lower limit of this window steadily decreases with $F_{\rm max}$.  In fact, there exists some minimum value of $F_{\rm max} > 1$ above which this frequency window covers the entire spectrum. Therefore, given the result of \fig{fig:Le} that there exists a given value of $\vert \Im(k_\om^d(0)) \vert L$ above which $T_\om^{\rm eff}$ no longer depends on $L$, \fig{fig:Imkd} tells us that that this will be true of high frequencies before low frequencies, and that above a certain value of $F_{\rm max}$ it will be true of the whole spectrum (except for the very small frequencies $\om < \om_c$ of \eq{eq:omc}, where the divergence of some scattering coefficients at the black hole horizon compensates the exponential decay).

In brief, what we learn here is that the emission spectrum of subcritical flows is more sensitive to the properties of the flow on its upstream side, since for a given frequency and length of the obstacle, $\vert \Im(k^d_\omega(0)) \vert L$ is considerably smaller than in transcritical flows. This sensitivity of the scattering coefficients is further studied in App.~\ref{app:slope_asymmetry} for obstacles similar to that used in the Vancouver experiment.

\section{Conclusion}
\label{Conc}

In this paper, we numerically studied the behavior of the 16 coefficients which enter in the $S$-matrix governing the scattering of surface waves on a stationary flow above a localized obstacle.  For simplicity, we assumed that the downstream flow was subcritical and asymptotically homogeneous, i.e., that it was not modulated by an extended zero-frequency wave (as is generally the case in practice, the undulation occurring on the downstream side). 

In the first part of the work, we compared the 16 coefficients of a typical transcritical flow to those of a subcritical one. The main difference concerns the magnitude of the mode amplification: in transcritical flows some coefficients (relating unit norm modes) are substantially larger than 1, thereby revealing that the wave energy measured in the lab frame of some mode is significantly increased by the scattering. This large increase is made possible because of a correspondingly large emission of negative energy waves. In addition, when the flow is significantly transcritical, i.e., when $F_{\mathrm{max}}$ (the maximal value of the Froude number) is larger than $1.2$, the amplification factors closely follow the standard Hawking predictions. Namely, in a wide frequency regime, $|\beta_\om|^2$ (the squared absolute value of the scattering coefficient mixing modes of opposite energy) follows a Planck law at a temperature in close agreement with $\kappa/2\pi$, where $\kappa$ is the analogue surface gravity evaluated where the local value of the Froude number $F(x)$ crosses 1. By contrast, for subcritical flows no coefficient significantly surpasses 1, which means that there are no significant super-radiant effects. 

We then focussed on the coefficients which describe the scattering of counter-propagating long wavelength modes, when gradually decreasing $F_{\rm max}$ from a supercritical to a subcritical value. The effect on $|\beta_\om|^2$ is the most dramatic. While in transcritical flows it behaves as $1/\om$ in a wide domain of low $\om$, in subcritical flows it behaves as $\om$ in a similarly large frequency domain.  As a result, the maximal value of $|\beta_\om|^2$ stays well below 1 for subcritical flows.  Even in the transcritical case, however, there exists an ultra low frequency regime where $|\beta_\om|^2$ is proportional to $\om$, because ultra low frequency modes are essentially transmitted across the obstacle. Interestingly, whenever $\vert \beta_\om \vert^{2}$ scales as $\om$ for $\om \to 0$, $\vert \alpha_\om \vert^{2}$ (the squared absolute value of the coefficient which relates incoming counterpropagating long wavelength modes to reflected short wavelength modes) follows $\vert \beta_\om \vert^{2}$. In fact, their ratio goes to 1 for $\om \to 0$. 

In the second part, we analyzed the detailed properties of the same set of scattering coefficients in sub- and near-critical flows. We have shown the existence of high- and low-frequency behaviors separated by a transitionary regime around the critical frequency $\omega_{\rm min}$.  Above this frequency the counterpropagating incoming long wavelength modes are essentially reflected, while below it they are essentially transmitted. As expected, the width of the frequency domain characterizing the transition decreases when increasing the length of the obstacle (at least for sufficiently narrow obstacles). We have also shown that this width tends to increase with increasing $F_{\mathrm{max}}$, and that it is largely independent of the slopes of the obstacle.  In the low-frequency domain, we observed that  $|\alpha_\om|^2 \sim |\beta_\om|^2 \sim \om/\sigma_\beta$ for $\om \to 0$ both in sub- and transcritical flows. We then showed that $\sigma_\beta$ radically diminishes with increasing $F_{\mathrm{max}}$, see Fig.~\ref{fig:sb_kappaR}.  In transcritical flows, this can be understood from the fact that $\sigma_\beta$, through its relationship to $\omega_c$ of Eq. (\ref{eq:omc}), scales as the square of the damping factor $D_L$ of Eq. (\ref{eq:damping_factor}) associated with the evanescent mode, see also Fig.~\ref{fig:Imkd}.

In the high-frequency regime of subcritical flows, the incoming long wavelength modes are essentially reflected, as is the case for transcritical flows. We could thus expect that the high-frequency scattering coefficients in trans- and subcritical flows behave in the same manner. However, our numerical observations indicate that this is only partially true.  In particular, the scattering coefficients in subcritical flows are seen to be more sensitive to the upstream properties of the flow because there is a larger transmission across the obstacle. This larger sensitivity can be easily understood, and rather well characterized, by evaluating the residual amplitude of the evanescent wave on the upstream side of the obstacle. In addition, we have shown that the effective temperature characterizing the emitted flux significantly depends on the frequency at which it is measured. This means that the emitted flux in general does not follow the Planck law.
  
In Appendix~\ref{3reg}, as a function of the upstream and downstream slopes, we showed the behavior of the effective temperature evaluated in the low, the intermediate and the high frequency regimes. The existence of three different patterns demonstrates that the spectral properties radically differ in each regime. One should thus study each regime separately. In Appendix~\ref{app:slope_asymmetry} we further studied the respective roles of the upstream and downstream slopes for asymmetrical obstacles which are similar to that used in Refs.~\cite{Weinfurtner:2010nu,Euve:2014aga}.  For such narrow obstacles, i.e., obstacles such that the ratio of their effective length to the asymptotic water depth $L_{\rm eff}/h_{\rm as}\lesssim 4$, our analysis reveals that the upstream slope, which is about 4 times larger than the dowstream slope, plays a dominant role in determining the scattering coefficients. Therefore, in future experiments, if one wishes to test the scattering on the downstream slope, it would be necessary to use either longer obstacles, or obstacles with a lower upstream slope.

\section*{Acknowledgments}

SR would like to thank the University of Poitiers, and in particular Germain Rousseaux and L\'{e}o-Paul Euv\'{e}, for their welcome and hospitality while this work was being completed.  This work was supported by the French National Research Agency by the Grants No. ANR-11-IDEX-0003-02 and ANR-15-CE30-0017-04 associated respectively with the project QEAGE (Quantum Effects in Analogue Gravity Experiments) and HARALAB.  We also received support from a FQXi grant of the Silicon Valley Community Foundation.

\begin{appendices}

\begin{figure}[h]
\includegraphics[width=0.35\columnwidth]{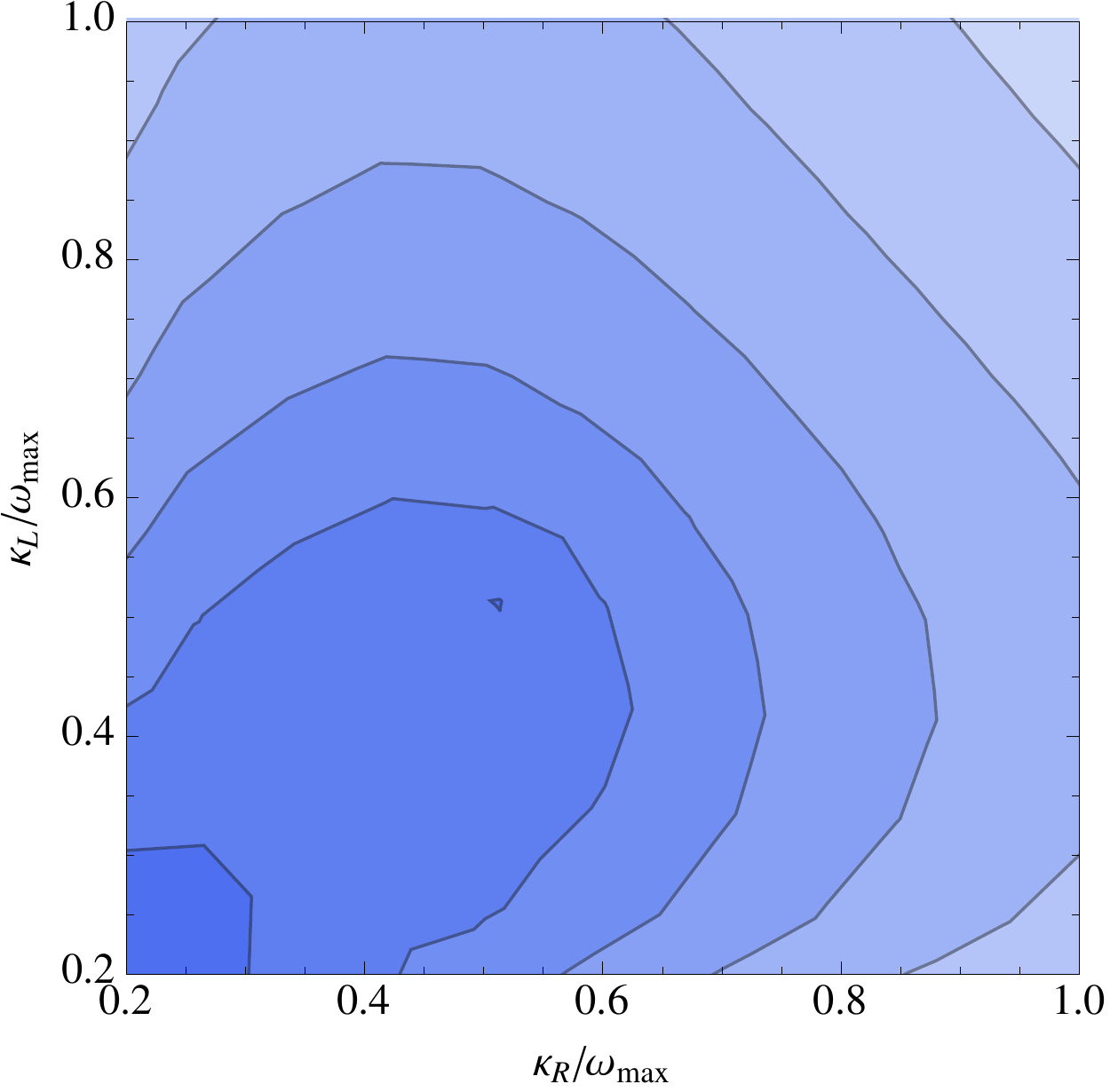} \,\,\,\,\, \includegraphics[scale=0.35]{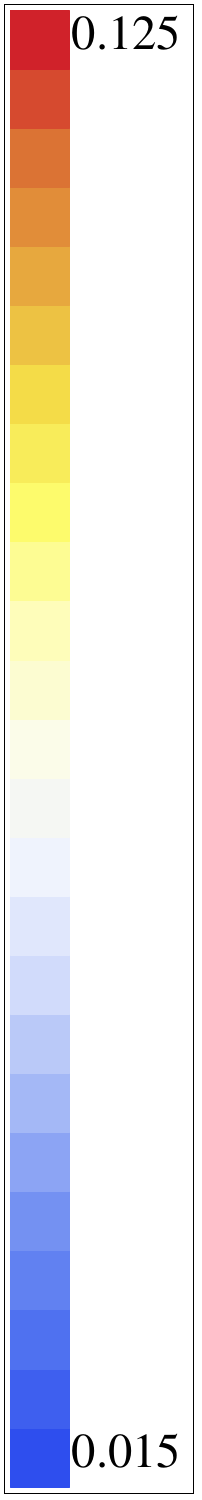} \\
\includegraphics[width=0.35\columnwidth]{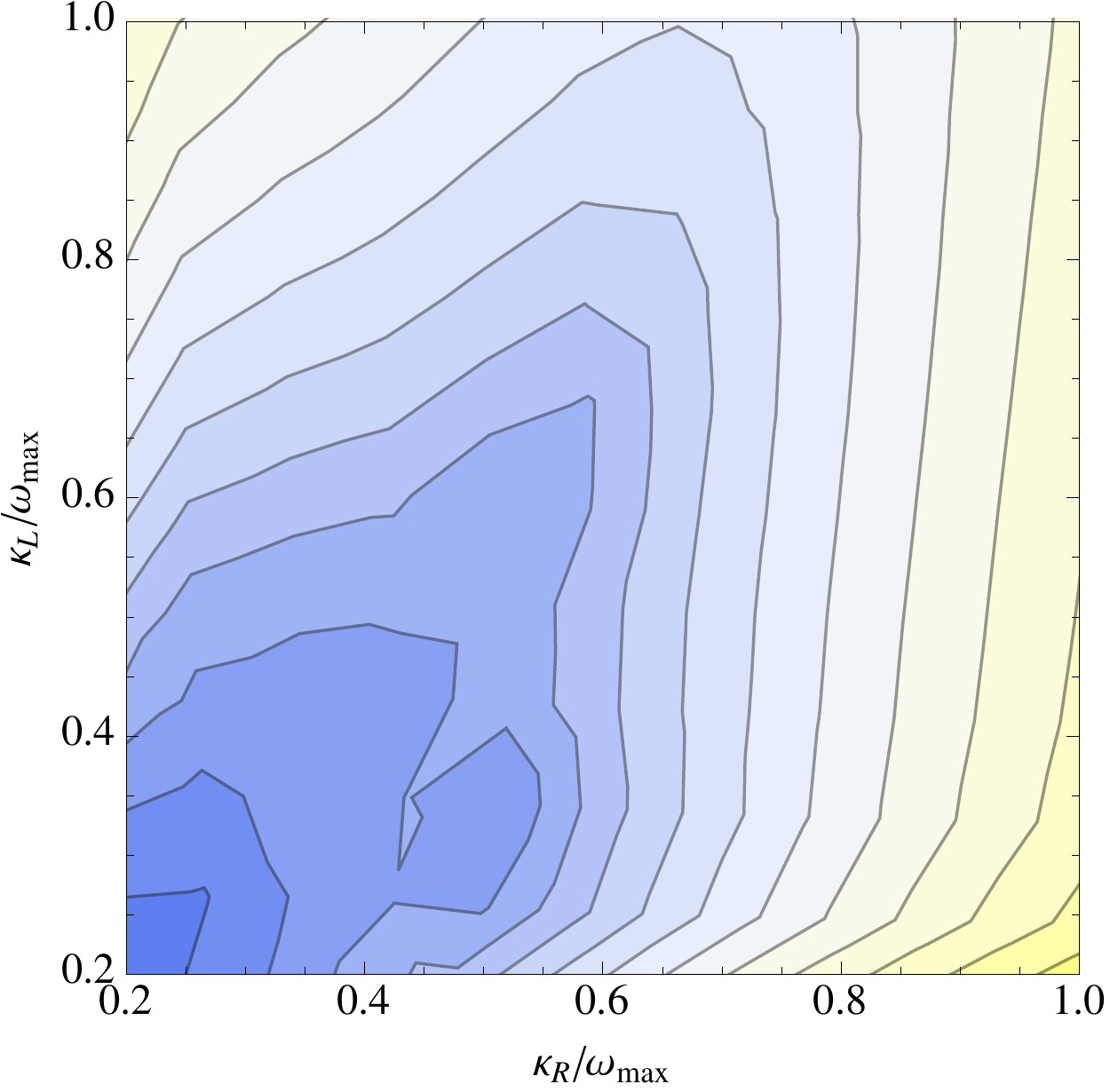} \, \includegraphics[width=0.35\columnwidth]{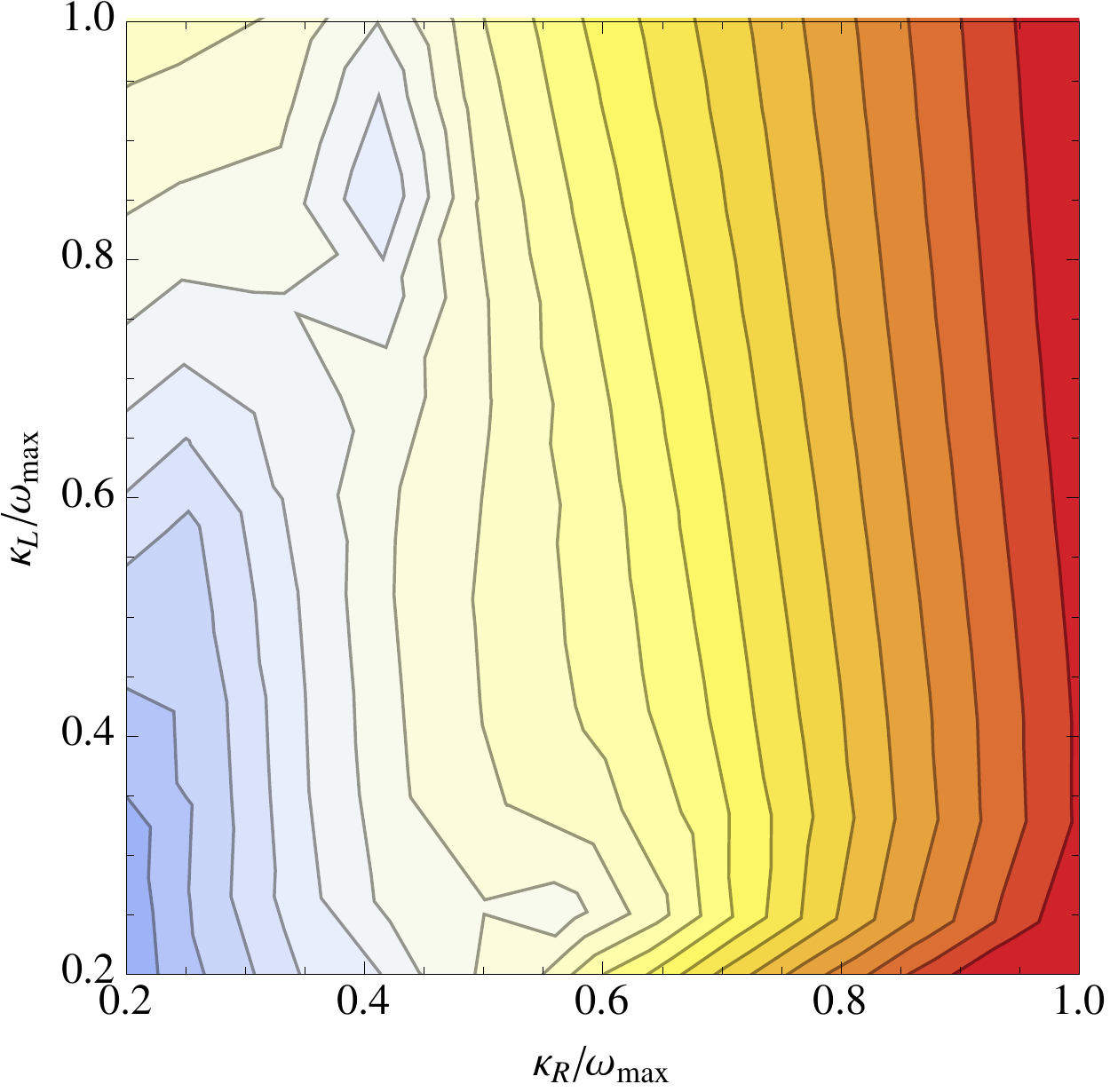}
\caption{Here is plotted, in the $\left(\kappa_{R},\kappa_{L}\right)$-plane with $F_{\mathrm{max}}=0.7$ and $L/h_{\rm as} = 2.5$, the value of $T_{\omega}^{\rm eff}$ of Eq. (\ref{eq:Teff}) (normalized by $\omega_{\rm max}$) for three different values of $\omega$: $\omega_{\mathrm{min}}/2$ (top), $\omega_{\mathrm{min}}$ (bottom left) and $2\,\omega_{\mathrm{min}}$ (bottom right).  The contour shading is the same for each panel, and corresponds to that indicated by the color legend. We clearly see that the effective temperature $T_{\omega}^{\rm eff}$ varies a lot with $\omega$, which means that the spectum is not Planckian, and also significantly depends on the upstream slope $\kappa_L$. 
\label{fig:beta}}
\end{figure}

\section{The 3 different behaviors of the spectrum}
\label{3reg}

As a direct illustration of the existence of three different regimes, we show in Figure \ref{fig:beta} contour plots of the effective temperature $T_{\omega}^{\rm eff}$ of Eq. (\ref{eq:Teff}) in the $(\kappa_{R},\kappa_{L})$-plane (all quantities being adimensionalized by $\omega_{\rm max}$), for frequencies $\omega_{\mathrm{min}}/2$, $\omega_{\mathrm{min}}$ and $2\,\omega_{\mathrm{min}}$.  We clearly see that the shape of the contours radically differs for each plot. In particular, for $\omega = \omega_{\mathrm{min}}/2$ the contours are symmetric about the diagonal $\kappa_{R}=\kappa_{L}$, indicating that in the low-frequency regime the effective temperature is insensitive to the directionality of the flow; on the other hand, for $\omega = 2 \omega_{\rm min}$ the contours are more parallel to the $\kappa_{L}$-axis, indicating that the flow properties on the downstream side are more relevant in this regime.  Much of the residual dependence on $\kappa_{L}$ is due to our use of a narrow obstacle (we have used $L/h_{\rm as} = 2.5$); increasing $L/h_{\rm as}$, the contours for $2\omega_{\rm min}$ are more vertically aligned. We notice that the contours for $\omega = \omega_{\mathrm{min}}$ are somehow in between the two we have just described. What we learn here is that it is inappropriate to look for a (global) description of the scattering that would be valid in the three regimes. This is why we study separately each regime in the main text.

\section{Effects of slope and asymmetry
\label{app:slope_asymmetry}}

\begin{figure}[h]
\begin{center}
\includegraphics[width=0.5\columnwidth]{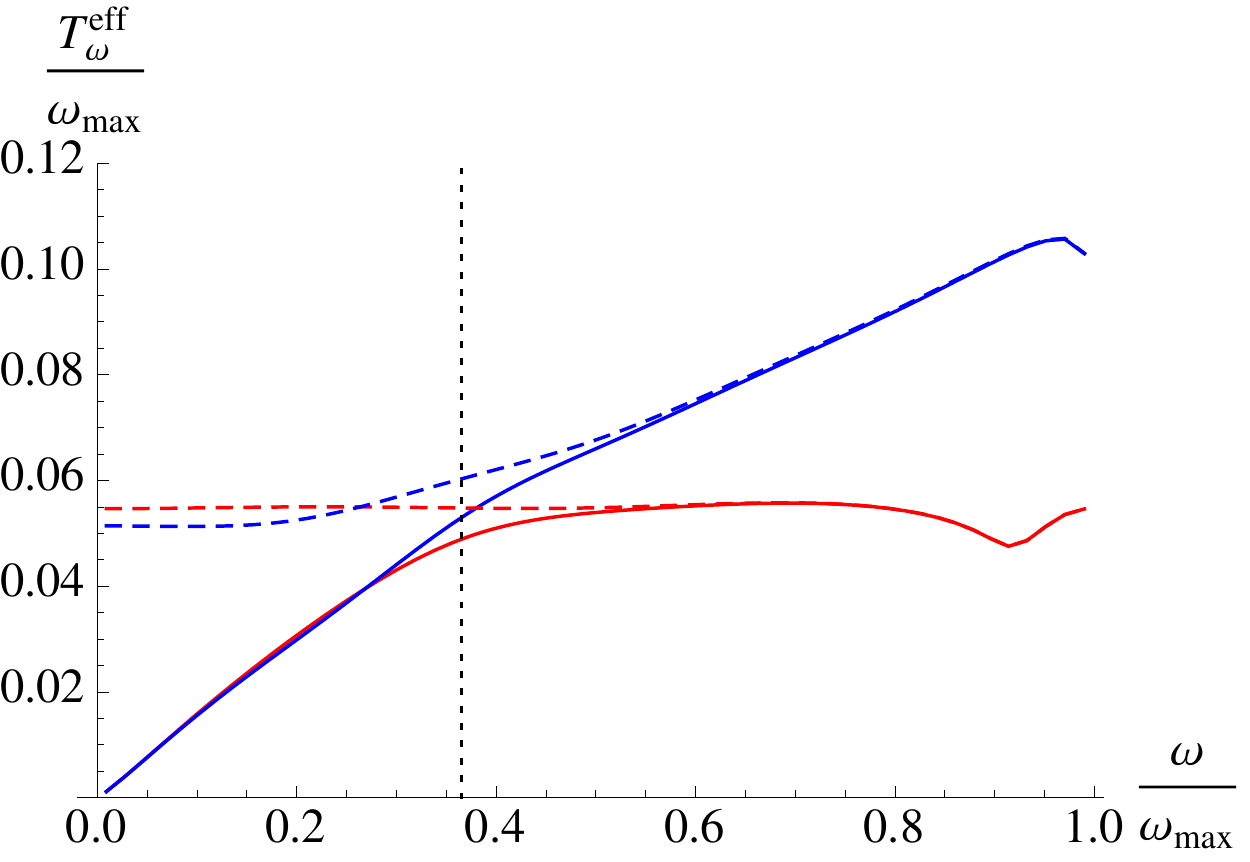} 
\caption{Here are plotted two effective temperatures for two different flows.  The red curves correspond to the flow which resembles that used in the Vancouver experiment, while the blue curves correspond to the same flow with reversed orientation (i.e. with $\kappa_{R}$ and $\kappa_{L}$ swapped).  The solid curves show $T_\om^{\rm eff}$ of Eq. (\ref{eq:Teff}), determined by $|\beta_{\omega}|^{2}$ only, while the dashed curves plot $T_{\omega}^{V}$ of Eq. (\ref{eq:TR}), completely determined by the ratio $|\beta_{\omega}/\alpha_{\omega}|^{2}$. The vertical dotted lines show the critical frequency $\omega_{\rm min}$. 
\label{fig:Teff_Van_Vanrev}}
\end{center}
\end{figure}

\begin{figure}[h]
\includegraphics[width=0.45\columnwidth]{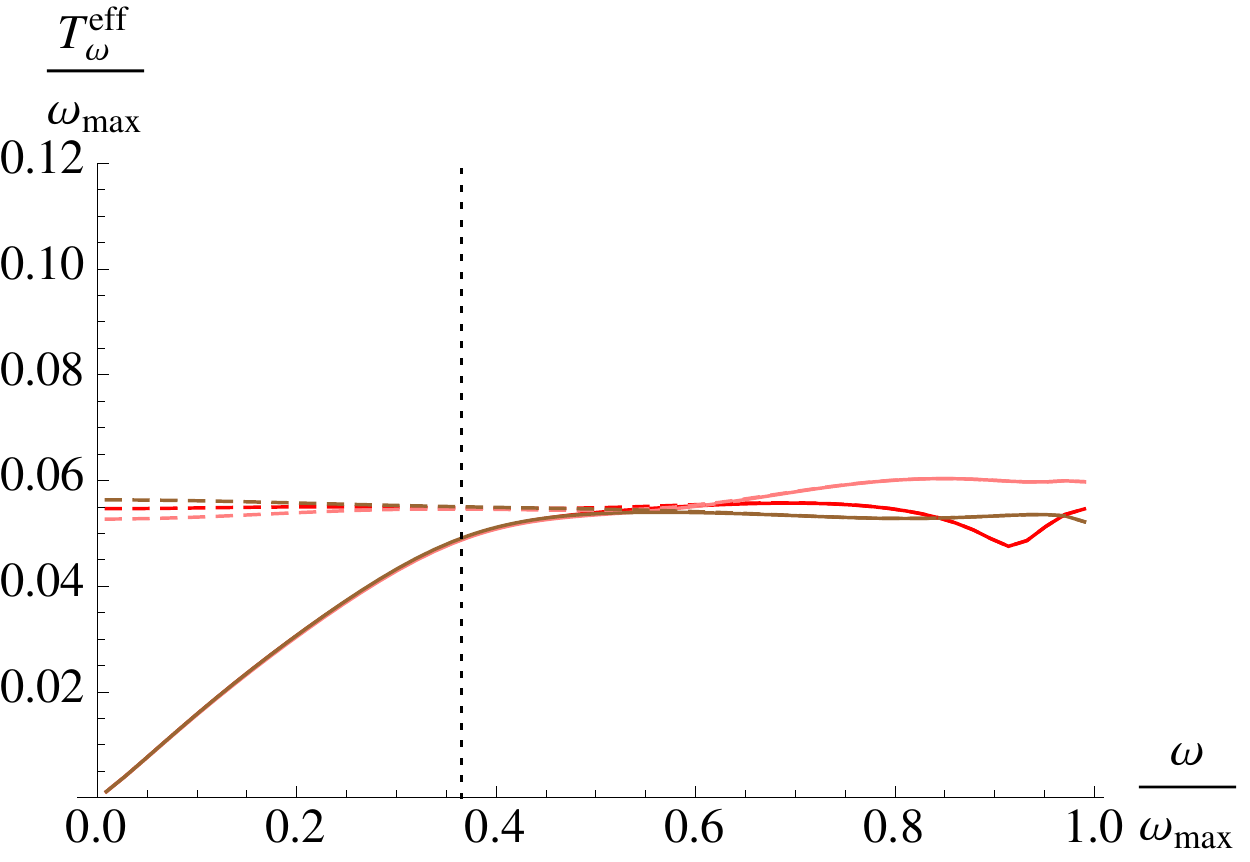} \, \includegraphics[width=0.45\columnwidth]{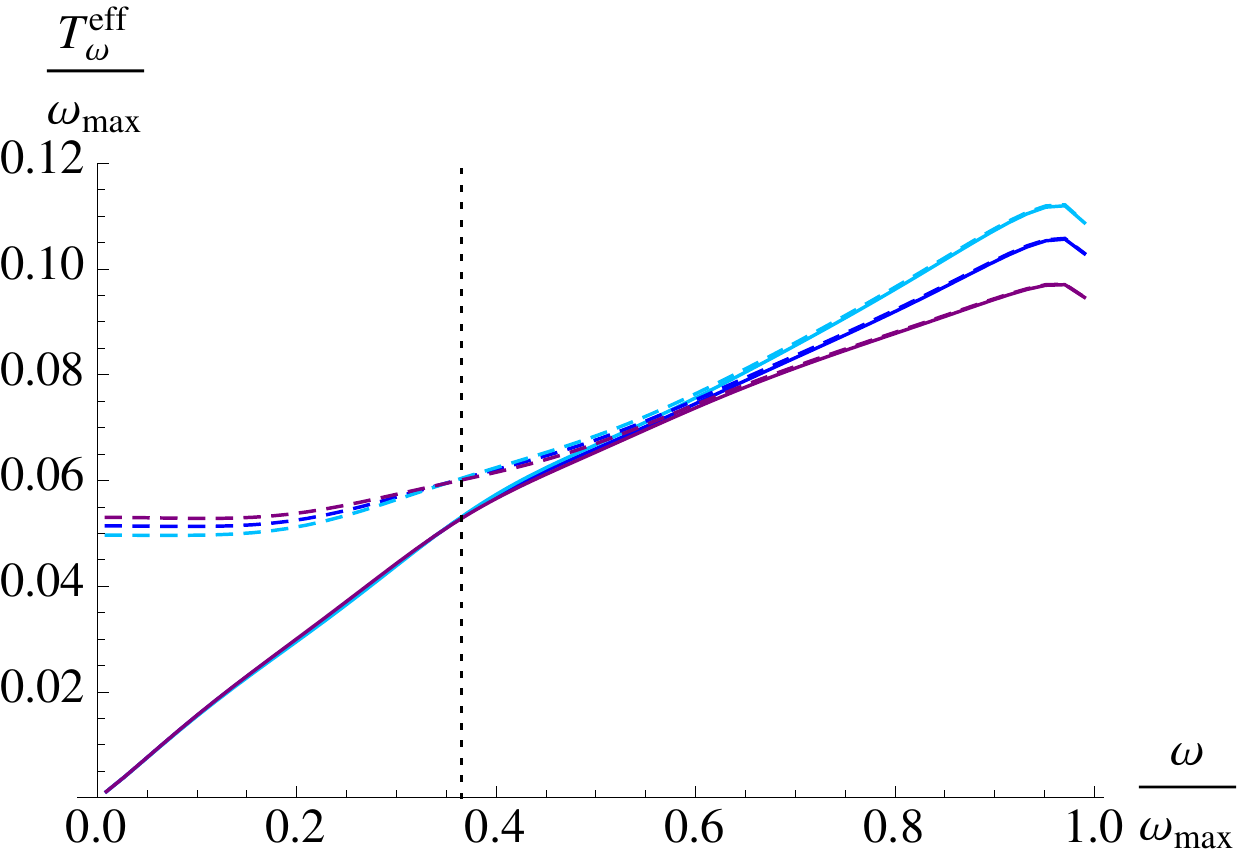} \\
\includegraphics[width=0.45\columnwidth]{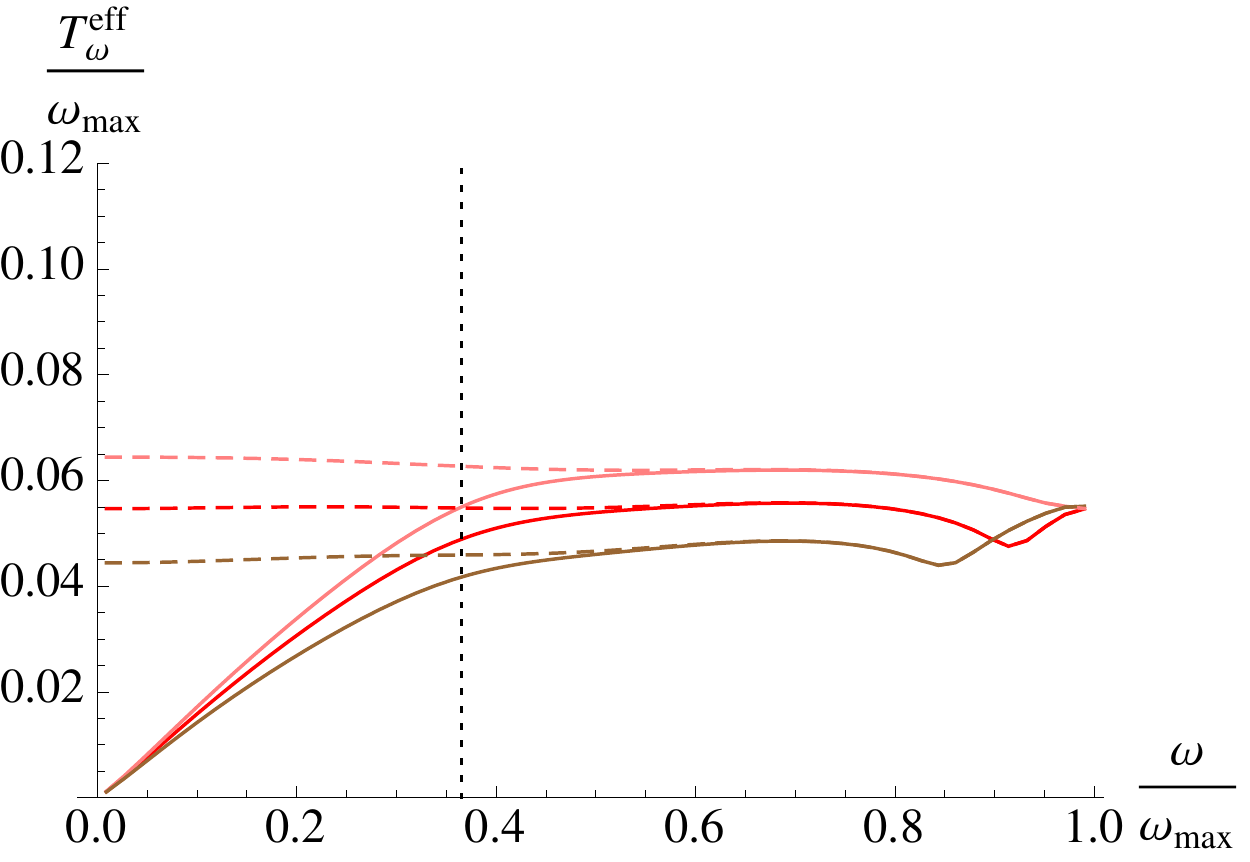} \, \includegraphics[width=0.45\columnwidth]{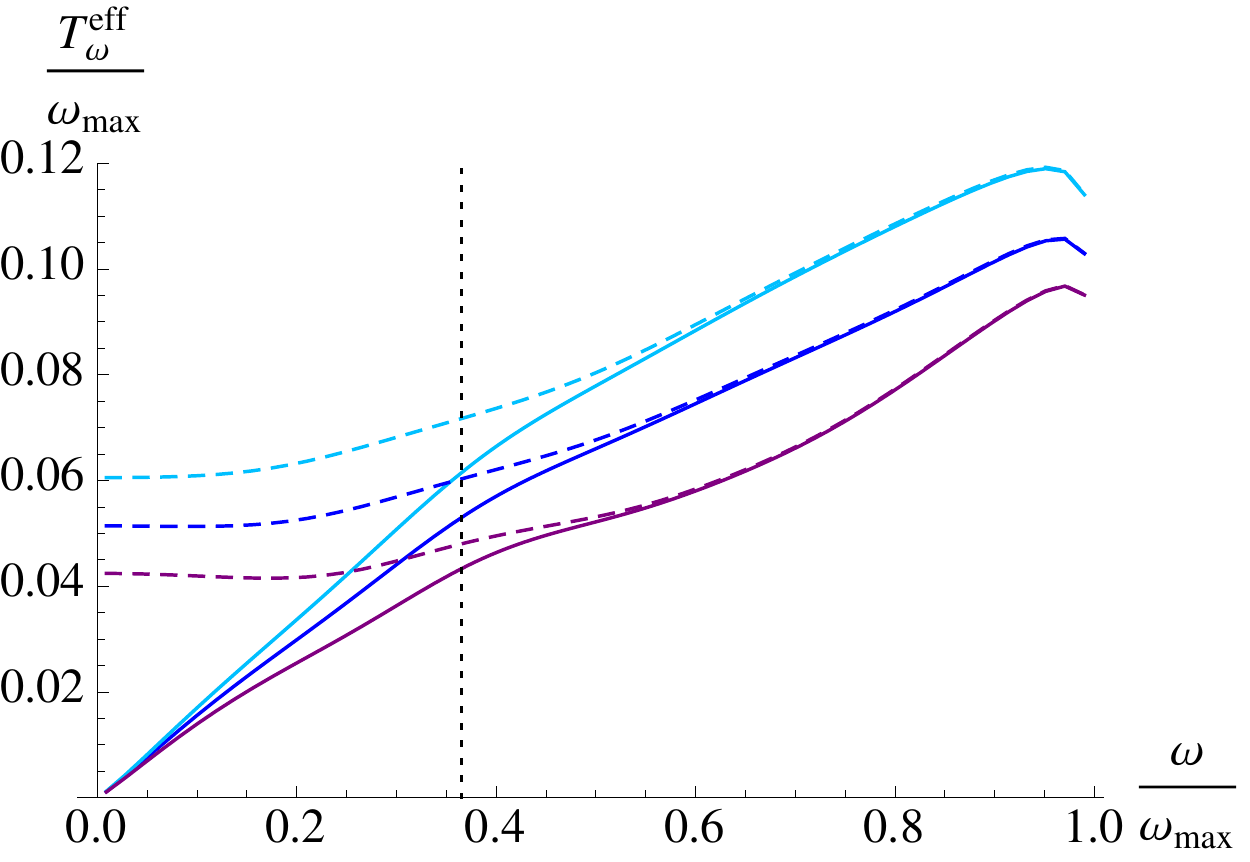}
\caption{As in Figure \ref{fig:Teff_Van_Vanrev}, we plot here the two 
effective temperatures ($T_{\omega}^{\rm eff}$ in solid curve, $T_{\omega}^{V}$ in dashed curve) for two different flows, but we have separated the two flows for clarity.  The red curves on the left correspond to a flow closed to that used in the Vancouver experiment, while the pink and brown curves correspond, respectively, to an increase and decrease in one of the slopes by $15\%$.  Similarly, the blue curves on the right correspond to the reversed flow, 
while the light blue and purple curves correspond, respectively, to an increase and decrease in one of the slopes by $15\%$.  In the upper plots, it is the smaller of these two slopes (i.e. $\kappa_{R}$ in the left column, $\kappa_{L}$ in the right column) that is varied, while in the lower plots it is the larger of the two slopes (i.e. $\kappa_{L}$ in the left column, $\kappa_{R}$ in the right column). 
\label{fig:Teff_varying_slope}}
\end{figure}

When considering asymmetrical obstacles, there arises the interesting question of the respective roles of the upstream and downstream slopes in determining the scattering coefficients. To address this issue, we consider an obstacle described by \eq{eq:f} with properties similar to the one used in the Vancouver experiment.  In particular, we take the upstream slope $a_{L} h_{\rm as} = 1.6$ to be much larger than the downstream slope $a_{R} h_{\rm as} = 0.5$. The length parameter $L/h_{\rm as} = 2.5$, corresponding to an effective length $L_{\rm eff}/h_{\rm as} = 3.5$, is relatively short (compare with Figs. \ref{fig:TransWmin} and \ref{fig:S_kappaR}), a crucial property in that it allows the upstream slope to affect the scattering via tunnelling effects. The maximum and asymptotic Froude numbers are $0.7$ and $0.16$, respectively. 

To illustrate the role of the asymmetry, we first compare the scattering on this flow to that on the reversed flow, i.e., the flow obtained by sending $x \to -x$ while keeping the orientation of the flow (from left to right) unchanged. Two important lessons can be drawn from \fig{fig:Teff_Van_Vanrev}. For frequencies larger than $\omega_{\rm min}$, the temperatures $T_{\omega}^{\rm eff}$ and $T_{\omega}^{V}$ agree for any one flow, indicating that the unitarity condition (\ref{eq:unit1}) in the high-frequency domain is $|\alpha_{\omega}|^{2}-|\beta_{\omega}|^{2}\approx 1$.  However, there is a significant difference between the two orientations of the flow, as can be seen by comparing the red and blue curves.  On the other hand, for frequencies smaller than $\omega_{\rm min}$, the situation is reversed: $T_{\omega}^{\rm eff}$ and $T_{\omega}^{V}$ become independent of the orientation of the flow, but are now in disagreement with each other.  As already noted, $T_{\omega}^{\rm eff}$ vanishes for $\om \to 0$, while $T_{\omega}^{V}$ goes to a constant in this limit.

To further investigate the respective roles of $\kappa_{L}$ and $\kappa_{R}$, we vary these quantities separately around the values given above.  The results are shown in Figure \ref{fig:Teff_varying_slope} in terms of the two temperatures $T_{\omega}^{\rm eff}$ and $T_{\omega}^{V}$ of Eqs. (\ref{eq:Teff}) and (\ref{eq:TR}), respectively.  For either effective temperature, and irrespective of the orientation of the flow, one notices that the changes induced by varying the highest slope by $\pm 15\%$, shown in the lower plots, are much more significant than those resulting from a variation of the lowest slope by the same relative amount, shown in the upper plots. Hence, for subcritical flows that are sufficiently short and asymmetrical, the scattering properties are mostly determined by the steepest slope, whether it is on the upstream or downstream side of the flow.

\end{appendices}

\bibliography{biblio}

\end{document}